# Seedling Report

## Advancing Methodology for Social Science Research Using Alternate Reality Games: Proof-of-Concept Through Measuring Individual Differences and Adaptability and their impact on Team Performance

**PI:** Magy Seif El-Nasr

**Co-PIs:** Casper Harteveld, Paul Fombelle

**Senior Personnel:** Truong-Huy Nguyen, Paola Rizzo, Dylan Schouten, Abdelrahman Madkour, Chaima Jemmali, Erica Kleinman, Nithesh Javvaji, Zhaoqing Teng, Extra Ludic Inc: Samuel Liberty, Wade Kimbrough.

## A. Our Goals

We have several overarching goals for this proposal. We will first enumerate these goals and outcomes then delve into more detailed discussion on each of the outcomes discussed.

### A.1. Developing data-driven models to study team processes

While work in fields of CSCW (Computer Supported Collaborative Work), Psychology and Social Sciences have progressed our understanding of team processes and their effect performance and effectiveness, current methods rely on observations or self-report, with little work directed towards studying team processes with quantifiable measures based on behavioral data. There are a few exceptions, e.g., Sociometric badges (Oluguin and Pentland, 2007) and the work by Zhang et al. (2018) who studied team cohesion and coordination through means of wearables augmented with surveys and other instruments. Even with such work an open problem still persists: *how can we make sense of behavioral data collected through various forms in the virtual or physical world?* In addition, behavioral data is limited in how it can infer cognitive aspects of constructs in question. This then opens up a question of: *how can we augment behavioral data with other forms of data to understand team processes at a scalable, quantitative and empirical level?*

Distribution Statement "A" (Approved for Public Release, Distribution Unlimited)
**Acknowledgement:** this research was developed with funding from the Defense Advanced Research Projects Agency (DARPA).
**Disclaimer:** The views, opinions and/or findings expressed are those of the author and should not be interpreted as representing the official views or policies of the Department of Defense or the U.S. Government.

In this report we discuss work tackling this open problem with a focus on **understanding team adaptation and its effect on performance as both an outcome and process**. We specifically discuss our contribution in terms of methods that augment survey data and behavioral data that allow us to gain more insight on team performance as well as develop a method to evaluate adaptation and performance across and within a group.

To make this problem more tractable we chose to focus on specific types of environments, Alternate Reality Games (ARGs), and for several reasons. First, these types of games involve setups that are similar to a real-world setup, e.g., communication through slack or email. Second, they are more controllable than real environments allowing us to embed stimuli if needed. Lastly, they allow us to collect data needed to understand decisions and communications made through the entire duration of the experience, which makes team processes more transparent than otherwise possible.

## A.2. Effect of individual difference on team performance and adaptation

Previous work discussed many factors affecting team performance, including shared mental models, team cohesion, individual adaptability, and team performance adaptation (Matthieu et al., 2000; Salas et al., 1995; Smith-Jentsch et al., 2008). In addition, some work has emphasized the effect of individual differences on team performance and effectiveness (e.g., Pulakos et al., 2006, Giganac and Szodoral, 2016). However, most of the work in this area focused on observational measures, such as (Rosen et al., 2011)'s model. To our knowledge, no work investigated the effect of individual differences on group performance through quantifiable behavioral data or behavioral data augmented with surveys or observations. We believe and illustrate here in this report that our approach of using an ARG is a promising research environment to study this effect with such types of data.

Our second goal was to investigate whether we can reproduce the results of previous work reported on the effect of individual differences on team adaptation through a series of studies we ran on two game environments: *MarketPlace Live*, an existing game developed to teach marketing and business, and *Daedalus*, a game we developed in the spirit of an ARG that contains puzzles and an overarching narrative to engage groups for a period of a week, and where some of the puzzles require players to look for clues in the real world, such as the Boston Commons. Daedalus was built using Slack to be more congruent to communication environments used in the real-world.

Based on previous work, our hypotheses were:

Distribution Statement "A" (Approved for Public Release, Distribution Unlimited)          1
**Acknowledgement:** this research was developed with funding from the Defense Advanced Research Projects Agency (DARPA).
**Disclaimer:** The views, opinions and/or findings expressed are those of the author and should not be interpreted as representing the official views or policies of the Department of Defense or the U.S. Government.

- **H1:** Degree of **openness** will be highly predictive of **adaptability** and **Performance**. *This is based on previous work by (Burke et al., 2006; Elaine D. Pulakos et al., 2002) who postulated that individuals who rate high on openness display high tolerance and curiosity when they confront unpredictable situations.*
- **H2:** Degree of **neuroticism** will be inversely predictive of **adaptability** and **Performance**. *This is based on previous study by (Burke et al., 2006; Huang et al., 2014; Pulakos et al., 2002) who postulated that individuals with high neuroticism do not have the emotional stability to deal with unpredictable situations.*
- **H3:** Degree of **cognitive flexibility** will be highly predictive of **adaptability** and **Performance**. *This is based on previous study (Burke et al., 2006), which postulates that high cognitive flexibility will increase chances that individuals display adaptability in their behavior.*

Given results from previous studies, we expect the effect size to be **r=0.1-0.3**.

In the proposal, we identified a hypothesis (**H4**) which outlines that there will be a positive correlation effect of practical intelligence and performance as well as adaptability. However, as we investigated the instruments used further, we found them very problematic in terms of reliability and validity. First, they are domain specific. Second, we found one for marketing that we used for the *MarketPlace Live* study, but the scoring of the test was very subjective, and according to our experts, almost all answers can be interpreted differently by different experts, which made reliability of such tests questionable. This led us to exclude this test and this hypothesis.

## B. Outcomes

We outline here the major outcomes and contributions resulting from our work. Details are provided in the remainder of the report. In a nutshell, we created an environment to enable studying team processes through an ARG (B.1), developed novel methodologies to study team processes that are quantifiable and scalable (B.2), and aimed to replicate results from previous work, but while finding some evidence for an effect of individual differences on teamwork not all results reported by prior work were confirmed, further we found more effects than reported as well although not consistent within the environments (B.3).

## B.1 Outcome 1: Daedalus

In order for us to investigate the open problems discussed above, we needed an environment that is comparable to real-life work environment *but* controllable. We designed an Alternate Reality Game called *Daedalus* for this purpose. *Daedalus* was launched in June 2018.

**Application and Impact:** *Daedalus* is a product that can be reused for many studies. It can also be easily modified for other experiments. Therefore, it is a contribution to the field.

**Acknowledgement:** this research was developed with funding from the Defense Advanced Research Projects Agency (DARPA).

**Where it is discussed in the report:** In Section 1, we discuss the environments we used for this research. In particular, Section 1.2 discusses the design process of *Daedalus* and its testing results.

## B.2. Outcome 2: Novel methodology to study team processes

In order to understand team processes at both the cognitive and the behavioral levels, we needed instruments that can tackle these different levels. After many investigations, we developed three methods:

- **Human-In-the-Loop Team Behavior Investigation Tool** using two visualization tools, we contribute a methodology to investigate team behaviors at different levels and within different environments. The method is a combination of: visualization and data pre-processing, abstraction, and labeling to drive knowledge from the data. Human raters are involved in the process of labeling and data pre-processing to create an abstraction level that is deemed useful. Once an abstraction level is reached, the output visualization is then validated through expert review of few randomly selected traces. The visualization is then used to derive decisions about the adaptation process and scores are calculated based on an underlying mechanism that inspects sequences and differences between them. The method has been used to score adaptation from *Daedalus* game log traces. It was also used to identify innovative strategies of winning teams within *BoomTown*, a game developed by the Gallup performer under the NGS2 DARPA program.

  **Application and Impact:** This tool delivers a method for inspecting team processes at a level that is not possible with other tools, specifically combining activity data with sequence analysis to allow humans to make sense of contextual, temporal behaviors exhibited in an environment, where behavior is logged.

  **Where it is discussed in the report:** This method is described in detail in Section 2.1.

- A new method for **Behavioral Situation Assessment Scoring (BSAS)**. We also investigated the development of a data driven method to score situation assessment, in particular cue recognition. The question we asked is: *can we develop a method that uses virtual environment activities to identify if people recognized a cue that allows them to assess a situation or understand the problem?* We were able to develop this scoring mechanism for one of the environments and we tested it against a self-report instrument. Our scoring mechanism produced good correlation, and thus we deemed it successful. For the *Daedalus* environment, we used chat data as indicators of cue recognition. Again, this measure showed great success giving us high correlation with performance. In the future, we aim to expand on this method and see if it applies to other environments.

**Acknowledgement:** this research was developed with funding from the Defense Advanced Research Projects Agency (DARPA).
**Disclaimer:** The views, opinions and/or findings expressed are those of the author and should not be interpreted as representing the official views or policies of the Department of Defense or the U.S. Government.

**Application and Impact:** With more experimentation we believe we can develop a more general data-driven method to gauge elements of situation assessment. This will have many applications as it is scalable and produces more quantifiable results regarding situational assessment that is useful for all team work measurement techniques.

**Where it is discussed in the report:** This part will be discussed in detail in Section 2.2.

- **TAG (Teamwork and Adaptation in Games) Survey** was developed as a survey that gauges team work from participants' perspective.

**Application and Impact:** Since to our knowledge, there are no surveys on adaptation and none in the context of a game environment, this survey presents a real contribution to gauge this construct. Additionally, as a byproduct, we found that this instrument can be used to gauge and understand the affordances of the environment design and its effect on enabling, *or not*, adaptation. Through inspection and comparing distributions of participants' responses, we were able to compare the two example environments we used here. We speculate that such an instrument can also allow us to compare a simulation environment is to a real-world situation. This possibility would be of great value to people interested in gauging teamwork in a simulated environment closer to a real-world environment. Additionally, most often the results reported depend on the environment a way to compare environments would be important for the field.

**Where it is discussed in the report:** This instrument is discussed in detail in Section 2.3.

## B.2. Outcome 3: Results showing individual difference effect on performance and adaptation

As discussed above, much of the previous work has shown that there is some influence of individual differences on team performance and adaptation. However, none of that work used data-driven techniques or experimented with that effect in game environments where most of the experience and team work are done leisurely. We tested our hypotheses described above which were generated from previous work's results. **In agreement with previous work, we found significant effect of neuroticism on adaptation in both environments (r=-0.192, p=0.04) in *Daedalus* and (r=-0.26, p=0.06) in *MarketPlace Live*, with an additional effect of neuroticism on performance within *MarketPlace Live* (r=-0.325, p=0.02).** All other hypothesized effects were not confirmed.

As opposed to previous results, our results show the influence of conscientiousness and agreeableness on some of the team scores, but our hypotheses were not all confirmed (only one of them were confirmed), specifically our results showed **no effect** of openness, and cognitive

**Acknowledgement:** this research was developed with funding from the Defense Advanced Research Projects Agency (DARPA).
**Disclaimer:** The views, opinions and/or findings expressed are those of the author and should not be interpreted as representing the official views or policies of the Department of Defense or the U.S. Government.

flexibility on team performance processes. Perhaps the effect size is too small or the influence is happening at a different level that needs further investigation. Future work will look into more extreme group personality differences and investigate the slack chat as well as sequence of decision-making process.

**Application and Impact:** We were not able to confirm all previous work results and this flags such results for further investigation and perhaps replication. Using the environments we used and the methods we developed, it would be easier for someone to replicate the work.

**Where it is discussed in the report:** The discussion of this work is shown in Section 3.

**Acknowledgement:** this research was developed with funding from the Defense Advanced Research Projects Agency (DARPA).
**Disclaimer:** The views, opinions and/or findings expressed are those of the author and should not be interpreted as representing the official views or policies of the Department of Defense or the U.S. Government.

# More Detailed Report

## 1. Environments

Before we discuss the work done within the project, we will first discuss the environments used and developed for this project. For this project we used two environments: a game that was already developed for teaching marketing, called *MarketPlace Live*, and is being used by the Introduction to Marketing classes in the D'Amore McKim School of Business at Northeastern University. The second environment, called *Daedalus*, was developed by our team.

## 1.1. MarketPlace Live (MPL)

*MPL* is a business/marketing simulation game, developed and maintained by Innovative Learning Solutions, designed for higher education. Players work as a team and strategize to become influential first movers in a chosen emerging market (Cadotte and MacGuire, 2013).

### 1.1.1 The Game

The version of the game discussed here deals with 1990's-era personal computers. Players play the game in 'company' teams, within which they can either assign formal roles and divide responsibilities or collaborate on every decision of the project together. Over six 'business quarters', which last a length of time determined by the game administrator, each team attempts to increase their company's net profit and market share. Players design brands to appeal to target consumer markets, create ad campaigns to generate product demand, and manage sales and support personnel in various offices that can be opened around the world. In-game videos and text guide players through the game and contextualize their actions with a basic narrative.

Players make several decisions within this game, including decisions about which markets to target, what brands to develop and what these brands look like, how many ad campaigns to launch and what the ads look like, how many offices to open and where to open them, and how large the sales workforce should be. The game provides the users with requirements to allow them to make the best decisions, see Figure 1.1.

As teams move to engage with the game, they first discuss their strategies or goals before starting actual play. The early quarters of the game require the team to select a single target market, referred to as a test market, and design simple products and ads to appeal to that market. Teams have to discuss their strategies before they make their decisions (regarding ads, brands, offices, and sales staff), while static data and educational lectures are available to all teams to

**Acknowledgement:** this research was developed with funding from the Defense Advanced Research Projects Agency (DARPA).
**Disclaimer:** The views, opinions and/or findings expressed are those of the author and should not be interpreted as representing the official views or policies of the Department of Defense or the U.S. Government.

inform these early game decisions. An example of the brand decisions that need to be made by the team is shown in Figure 1.2. During these early game segments, each team is restricted to only a single target market, and they face limitations on how much money they are allowed to spend on brands, ad campaigns and sales personnel.

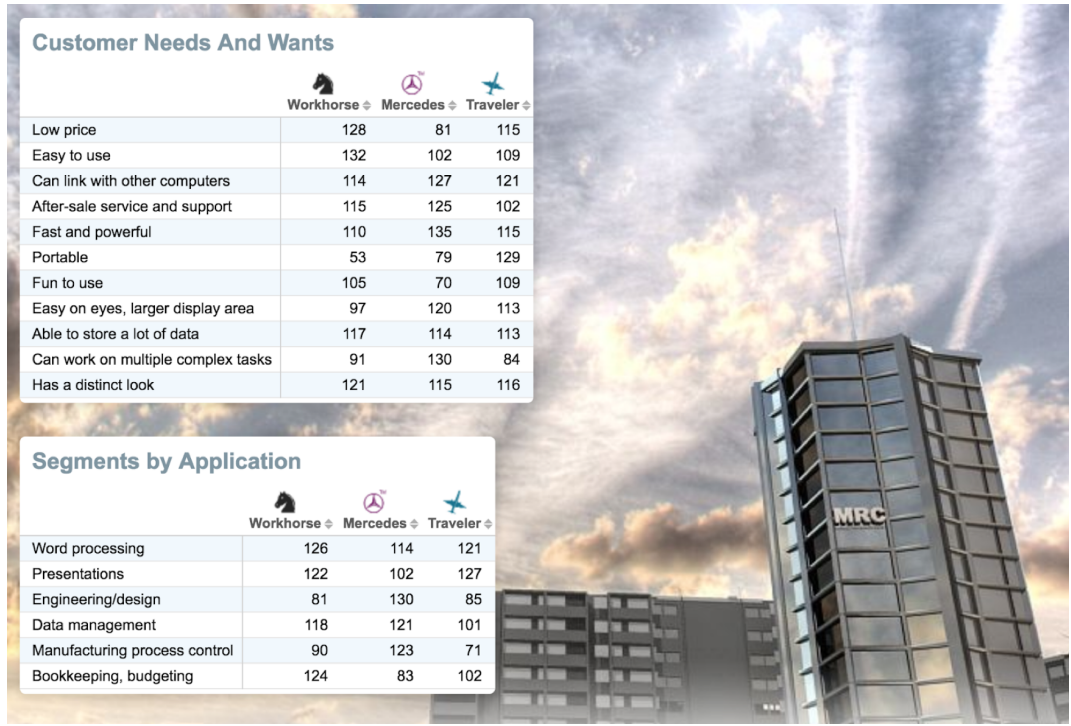

**Figure 1.1.** The game includes screenshots to show the user what the requirements are for the specific decisions they need to make. This figure shows a screenshot showing a portion of the "customer needs" screen detailing what each target segment wants, needs and prioritizes in a product.

Ultimately, around quarters 4 and 5, the game lifts the restrictions and allows players to expand into a new target market as well as lifting the restrictions on brand, ad and personnel costs. This design allows teams to pursue more aggressive marketing strategies later in the game. In addition to the lifted restrictions, new technology will become available for use in brand design and players will find that their results reflect how well they were able to implement this technology. The goal is to adjust the team's brands by incorporating technology that will appeal to their target markets.

Starting from quarter 3, each team gains access to a performance report regarding the previous quarter, as well as a section detailing concerns from the previous quarter according to a fictional

Distribution Statement "A" (Approved for Public Release, Distribution Unlimited)
**Acknowledgement:** this research was developed with funding from the Defense Advanced Research Projects Agency (DARPA).
**Disclaimer:** The views, opinions and/or findings expressed are those of the author and should not be interpreted as representing the official views or policies of the Department of Defense or the U.S. Government.

customer union. The latter explicitly states if any of the brands the team designed fail to meet the minimum requirements for the target segment.

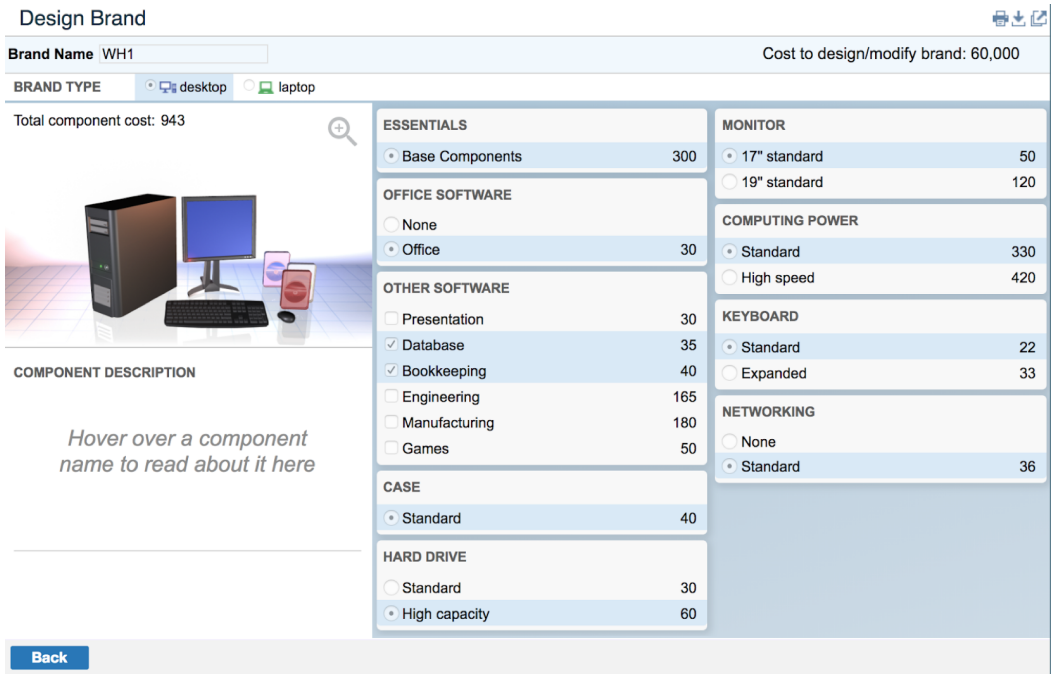

**Figure 1.2.** A screenshot from the game showing the decisions that need to be made by the team around brands.

## 1.1.2 Performance Measures

### 1.1.2.1. Team Scores
The performance report contains a balanced scorecard which displays industry results for the previous quarter based on financial performance (FP), measures how the team is effective at managing their finances, market performance (MP), measures how successful the team was maintaining market share in their target segments, and market effectiveness (ME), measures how the team is effective at targeting their target markets, each scored from 0-100 and used to calculate a total performance score. These are calculated as follows:

- *Market Effectiveness* (ME) = (*Average Brand Judgement + Average Ad Judgement*) / 2, where: *Average Brand Judgment* = average of Highest brand judgment in first and second target segment, and *Average Ad Judgment* = average of Highest ad judgment in first and second target segment.
- *Financial Performance (FP)* = ((*Revenue - Expenses*)/ *20,000,000*) * *100*

Distribution Statement "A" (Approved for Public Release, Distribution Unlimited)    8
<u>Acknowledgement:</u> this research was developed with funding from the Defense Advanced Research Projects Agency (DARPA).
<u>Disclaimer:</u> The views, opinions and/or findings expressed are those of the author and should not be interpreted as representing the official views or policies of the Department of Defense or the U.S. Government.

- *Market Performance (MP)* = Sum of Market share for first and second target segment.
- The final *Balanced Scorecard* is then developed as a combination of these measures:

<div align="center">FP+MP+ME / 3, and</div>

- *Cumulative Balanced Scorecard* (CBS) which is the cumulative scores of the Balanced Scorecard.

Using these results, teams can start comparing themselves to other teams (see Figure 1.3): This includes seeing which teams are targeting which market segments, the relative effectiveness of their brands and ads compared to those of other teams, and a general 'score' ranking teams in regard to market performance, financial performance and marketing effectiveness, which is visualized by which 'floor' each company occupies in a fictional high rise. This information will be available to all teams going forward, allowing teams to inform base decisions and strategies not only on static market data, but also on competitors' performance.

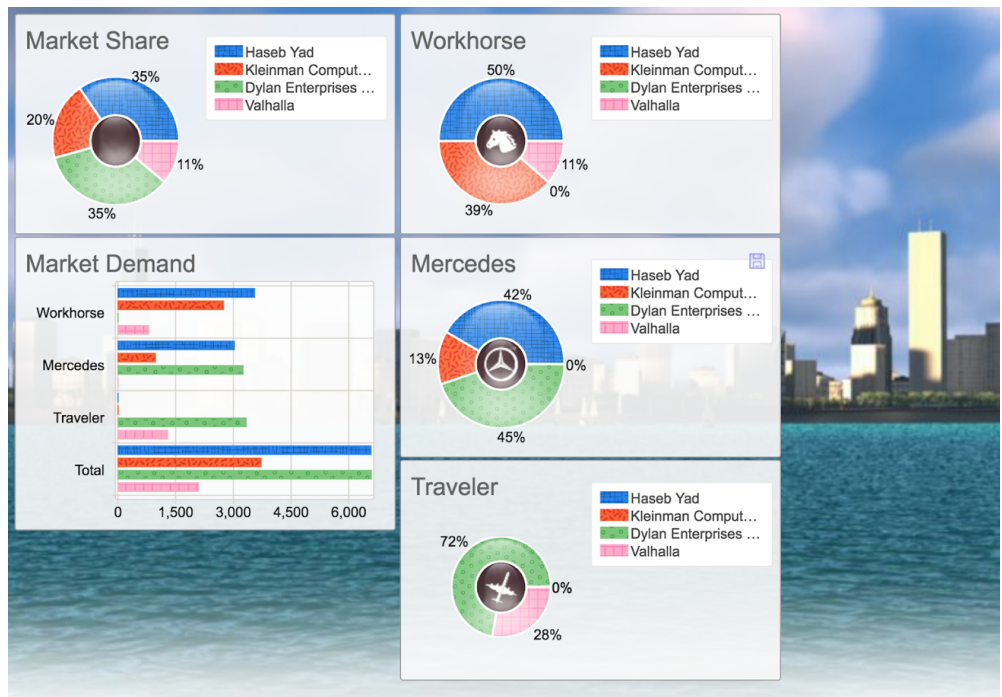

**Figure 1.3.** A screen shot within the game showing Market Share breakdown between teams in a game of Marketplace Live

### 1.1.2.2. Individual Scores

Unfortunately, in MPL all decisions are done by a team and thus there is no way to disambiguate individual contribution. We have collected slack data from our study, but we didn't have time to

**Acknowledgement:** this research was developed with funding from the Defense Advanced Research Projects Agency (DARPA).
**Disclaimer:** The views, opinions and/or findings expressed are those of the author and should not be interpreted as representing the official views or policies of the Department of Defense or the U.S. Government.

triangulate it with behavioral data yet, and thus we left that for future work. The only individual measures we were able to measure are:

- *Screen* (*Scr*), *TimeStamp* (*t*), what screen was pulled up by the player and when
- *Screen time (ScrT)*, how much time an individual spent on the screen.

We also identified specific Screens that are important for specific tasks, such as:

- *competitors' decision screens*: screens that show competitors' decisions
- *competitors' results screens*: screens that show competitors' results
- *previous decisions' screens*: screens that show the teams' decisions for previous quarter
- *current decisions' screens*: screens that show the teams' decisions for the current quarter
- *results screens*: screens that show the teams' results
- *error screens*: screens showing errors when submitting generated by the simulation
- *Performance visualization screens*: screens showing graphs of all teams' results, market conditions over quarters

These measures were used to deduce individual activity level and situation assessment which will be discussed at a later section.

### 1.1.3 Data from MPL

We collaborated with the company that developed the game to get access to data. The data we have access to shows activities per participant: specifically, how much time spent on what screens for all the screens of the game. We also have access to all decisions made on all decision points in the game, such as: all brand configurations, all ad configurations, the amount of money spent on staff and divisions of staff, number of brands made and what target markets they targeted, cost decisions per brand, and locations they decided to expand to. We also have access to all performance scores per quarter. All this data is time stamped and collected per quarter, and thus this gives us a good breakdown to see sequences of adjustments that teams made.

### 1.1.4 Studies

We originally had access to 470 games, 1650 teams and 5793 individuals. These were supplied through the company and we got permissions from instructors who have used the game in their classes to enable us to use the data. However, this data didn't have survey instruments for personality or cognitive flexibility needed for our hypotheses. However, they were a good start for us to start our data modeling process.

**Acknowledgement:** this research was developed with funding from the Defense Advanced Research Projects Agency (DARPA).

We then ran two studies, one in the Fall 2017 and another in the Spring 2018. The studies were setup as part of a class since *MarketPlace Live* is played within the introduction to marketing classes at Northeastern. Students in classes that opt into the study are asked to volunteer in exchange of $40 gift certificate. We also leveraged the business subject pool to disseminate surveys to get more students, since students who take the introduction to marketing are first year students who are targeted by the subject pool.

Through these studies we collected data from 206 Teams 4-5 students each, and individual differences surveys from 228-755 students. We also collected slack data from 110 teams. After data cleaning, we were left with 51 teams of game data and personality surveys and 31 teams with personality and cognitive flexibility surveys. These are the datasets we used for our analyses discussed in the sections below.

## 1.2. Daedalus

Alternate Reality Games (ARGs) are environments where participants engage for a period of time, ranging from few weeks to months, in an alternate reality narrative that seamlessly integrates with participants' lives. Interaction within these games involve methods that participants use in their real-life, such as texts, email, and files shared through text/email, etc. Participants often engage in such an experience through a fictional scenario, such as teams competing to find a drug before a plague infects a city. Clues and information are revealed over time, as authored by researchers to study the phenomenon of interest.

### 1.2.1. Requirements for the game

In this project, our aim was to develop an ARG to allow us to understand adaptability to unpredictable events. Based on our stated goals, we formulated several requirements as follows:

- The ARG should mimic how groups work in real-life tasks, which augments physical and virtual environments.
- The ARG should exhibit unpredictability within different scenarios to enable teams to adapt.
- The ARG should allow participants to interact with one another and coordinate to solve puzzles towards solving a bigger game quest or goal. This requirement was necessary for us to measure teamwork and coordination.
- The ARG should allow participants to formulate their own roles and such roles are easily distinguishable from the data. This requirement was necessary to allow us to determine role assignment and backup behaviors which are important for team adaptation process.

Distribution Statement "A" (Approved for Public Release, Distribution Unlimited)          11
**Acknowledgement:** this research was developed with funding from the Defense Advanced Research Projects Agency (DARPA).
**Disclaimer:** The views, opinions and/or findings expressed are those of the author and should not be interpreted as representing the official views or policies of the Department of Defense or the U.S. Government.

## 1.2.2. Design Process

In order to develop an ARG environment with these requirements, we have to define the constructs we are modeling within the design and also develop measurement techniques to allow us to assess if the game is moving towards the requirements through playtests. The constructs we need to define is **unpredictability** and **team composition and roles**.

To define **unpredictability**, we hosted a workshop using Pulakos et al.'s (2000) dimension of adaptability as a guide. At the workshop we began by trying to recreate Pulakos et al.'s Job Adaptability Inventory (JAI), which they used to create their taxonomy of adaptability. Participants were asked to use participatory design techniques to generate ideas for types of jobs that fit under each of the definitions. These techniques produced several jobs that fit each. Participants then further organized the jobs into clusters based on similarity. Finally, participants formulated questions about each cluster. These questions formed a rubric to assess whether or not  a new proposed job would fit that definition. Just like Pulakos et al.'s JAI, answering yes to a majority of these questions would identify whether that job fit into the definition of the dimension.

Based on these definitions, we then developed some questions (shown in Table 1.1) as a rubric, which we call **Puzzle Adaptability Index (PAI)**. We modified them slightly to apply them to the ARG puzzles. These definitions were then used in the design process to assess whether or not a puzzle contained the definition.

**Table 1.1.** Puzzle Adaptability Inventory, a rubric to evaluate puzzles developed for the ARG

1. Refusing to be paralyzed by uncertainty or ambiguity.
    a. Are players actions affected by environmental changes?
    b. Does this puzzle have volatility or crises?
    c. Do players have to make decisions with limited information?

2. Taking effective action when necessary without having to know the total picture all the facts at hand.
    a. Is there urgency for players to make a decision without the full picture?
    b. Do you need to make decisions based on partial or incomplete info?

3. Imposing Structure for self and others that provide as much focus as possible in dynamic situations.

**Acknowledgement:** this research was developed with funding from the Defense Advanced Research Projects Agency (DARPA).

**Disclaimer:** The views, opinions and/or findings expressed are those of the author and should not be interpreted as representing the official views or policies of the Department of Defense or the U.S. Government.

a. Does one player need to enforce order in this puzzle?
b. Does one player need to manage others?

4. Effectively adjusting plans, goals, actions, or priorities to deal with changing situations.
a. Does the puzzle involve an unpredictable environment that makes players need to adjust their plans?
b. Do players need to adjust to competitors' strategies?

5. Readily and easily changing gears in response to unpredictable or unexpected events and circumstances.
a. Do players need to change tasks at a moment's notice?
b. Are players unable to accurately predict the outcomes of their actions?

6. Not needing things to be black and white.
a. Do you deal with things that can't be quantified? e.g., Are things fuzzy/ambiguous? Or Do players need to make decisions based on emotions, not fact?
b. Do players have to negotiate with other players in this puzzle?
c. Do players make decisions based on dynamic feedback?

Salas et al. (1992) defined a team as "two or more people who interact dynamically, interdependently, and adaptively towards a common goal/object/mission, who have been assigned specific roles or functions to perform and who have a limited life span of membership (p. 4)." This is important as it will allow us to measure the individual tasks and performance as well as team performance. However, this definition raises several design questions. In a typical ARG, many players engage at once, cooperation is emergent, not designed. Teamwork in an ARG is not role-based, but instead is required by the difficulty and volume of the puzzles.

This created an early design problem for us, because we could not create an ARG with the typical difficulty and volume of puzzles that other ARGs have with that constraint. This is specifically problematic because ARGs do not require teams to work synchronously together and this is an important ARG element that we wanted to keep in the design since it mimics real-life and thus fulfills one of our requirements.

To solve this problem, we decided to design puzzles that required multiple team members to solve, e.g., puzzles that require players to be in different geolocations, or to perform simultaneous actions. We then let players self-organize in their roles, like a typical ARG ("imposing structure for self or others," in Pulakos' words). This allows a team to be fluid,

Distribution Statement "A" (Approved for Public Release, Distribution Unlimited)                13
**Acknowledgement:** this research was developed with funding from the Defense Advanced Research Projects Agency (DARPA).
**Disclaimer:** The views, opinions and/or findings expressed are those of the author and should not be interpreted as representing the official views or policies of the Department of Defense or the U.S. Government.

continuing to play even if one player must drop from the game temporarily or permanently. This is extremely important because we predict that player attrition on a team may be as high as 85%, which is comparable to other similar games as reported by *Game Analytics* (WebReport, accessed 2018).

After the inception workshop we took into consideration all the design constraints: timeline and budget, the need for team-based problem solving, the software platforms to deliver game content and collect data, and the need to test all parts of the definitions for "dealing with uncertain and unpredictable work situations." With these constraints clearly defined, we began the agile and iterative process of designing the ARG. To save development time, we elected to use existing platforms, e.g., *Slack* and *ARIS*. *ARIS* is an engine that uses location data to deliver puzzles. In the spirit of ARGs, delivering the game content through these platforms also enables us to take a playful experience and use it to co-opt a serious space that many players maybe familiar with in their everyday lives, i.e., *Slack* is a very popular chatting and collaboration software.

We simultaneously began to design a narrative and puzzles for the ARG. We started with a brainstorming process. The brainstorming meetings were attended by members of the design and development team, project PIs, and members of the data team. This enabled us to conceptualize puzzles that satisfies our design constraints and also the research and theoretical constraints for the study. Based on these brainstorming meetings, we conceptualized 5 key puzzles and a narrative for the ARG.

### 1.2.2.1. Central Narrative
The central narrative is the story that engages the player throughout the entire ARG. Early on in the design process we decided that we wanted to make this story about a sentient AI that was testing players on their ability to solve puzzles. We decided to name the AI 'Daedalus' after the maker of the labyrinth from Greek Mythology. We have not iterated on this idea much because through playtests and focus groups, participants showed much excitement towards such a narrative. The only major consideration that was brought up by some participants was that an evil AI is an overplayed trope in science fiction. Therefore, we decided against making *Daedalus* evil and developed the key plot points around having players fix *Daedalus*' neural net so that it can learn from their puzzle solving abilities and solve humanity's most pressing problems.

### 1.2.1.2. Core Mechanics
The core mechanics of a game are central actions that a player takes over and over again to progress through play. For *Daedalus*, it is important to make the core mechanic and central narrative closely related. This is one way to help scaffold the game onto the real world. Since the interface for our game is *Slack* and *ARIS*, and *Daedalus* is the creator of the Labyrinth in Greek

**Acknowledgement:** this research was developed with funding from the Defense Advanced Research Projects Agency (DARPA).

Mythology, we decided to use early text-based games from the 1980s and examples of Interactive Fiction to inspire our core mechanic.

After some discussion, we settled on having players navigate through a text-based labyrinth. Each part of the labyrinth was described in words and players then need to solve puzzles as a team to get to the end. Some of these puzzles will take players into the real world, but their core action would always be to progress through the labyrinth.

### 1.2.2.3. Prototype 1

In the first iteration, the labyrinth contained 30 rooms. Each room had a text description and a series of 3 doors that lead to other rooms. Players were tasked with making their way through this network from room 1 to 10 and then back to 1 again. To make things even more difficult, each door was locked with a riddle or logic puzzle.

The design was refined through playtesting a paper prototype with two groups of players. Through surveys, focus groups, and observations, we determined that the game was fun and engaging and observed that teams were organizing themselves to accomplish tasks adapting to the game's uncertainty. Because of this assessment we moved onto the next iteration and began to develop a digital prototype.

Our design process for the subsequent set of puzzles was similar to the labyrinth composed of six conceptualized puzzles. We paper prototyped and tested one. Because two of these puzzles were based on popular games, we decided not to prototype and test them and instead opted for using focus groups for them as a way to test their integration within *Daedalus*. After discussion, three puzzles were thrown out and a new one was conceptualized to replace them.

### 1.2.2.4. Prototype 2

We ran several playtests of the digital prototype for the labyrinth. Unfortunately, the results of those playtests were not the same as the paper prototype. First, the game was less engaging for a team playing over Slack asynchronously, as opposed to the paper prototype ,which required the team to be there synchronously solving the puzzles and navigating together. Second, the puzzles proved too easy to solve, so teamwork was not necessary. Third, it was difficult to measure exactly how the team interacted.

We, therefore, pivoted our design and adjusted the labyrinth's format. Specifically, we decided to shed parts of the game that were not highly interactive and engaging and streamlined the ARG by organizing it into five phases, each focusing on different elements of adaptability. The basic labyrinth structure was altered into an "escape the room" style text puzzle. Results of playtesting showed that playtesters were extremely engaged. The puzzles were more complex, discrete, and

**Acknowledgement:** this research was developed with funding from the Defense Advanced Research Projects Agency (DARPA).
**Disclaimer:** The views, opinions and/or findings expressed are those of the author and should not be interpreted as representing the official views or policies of the Department of Defense or the U.S. Government.

varied allowing better measures of adaptation. The puzzles also required teams to work together which was an intended effect.

The new format of the game followed the following outline:

- **Overview**: A Dynamic, Text-based Escape Room
  The entire game takes place over Slack and is contained within an "escape room" environment conducted in text by chatbots. In order to beat the game, players have one week to escape five rooms. Each room will have a series of interconnected puzzles. Some can be solved by analyzing the text descriptions in the room, others will require players to print and manipulate paper items or assemble objects, and some will require players to travel to a real-world location. In each stage there is a Master Puzzle, which is more difficult and measures a specific type of adaptation. Some puzzles are solvable individually, while others require organization and teamwork. *Puzzle Adaptability Inventory:* Refusing to be paralyzed by uncertainty or ambiguity.

- **Master Puzzle Stage 1:** Scavenger Hunt A.
  *Gameplay:* Players must visit three real locations in Boston and take photos of what they see there. The instructions are vague, but the importance of "good" photos is stressed. *Puzzle Adaptability Inventory:* Taking effective action when necessary without having to know the total picture or all the facts at hand.

- **Master Puzzle Stage 2:** Tamagotchi.
  *Gameplay:* Players must hatch an egg and raise a creature to adulthood in a short time. The egg/creature only responds to certain inputs, such as a limited set of emoji, so players will need to puzzle out what actions to take. After the egg is hatched they must feed and grow their animal by moving it around a chess board to pick up food using emoji-coded button clicks that do not overtly state their functions. *Puzzle Adaptability Inventory:* Readily and easily changing gears in response to unpredictable or unexpected events and circumstances.

- **Master Puzzle Stage 3:** ARIS Garden Puzzle.
  *Gameplay:* Players are required to travel to the Public Garden in Boston where they must quickly solve a puzzle that requires the communication of complex ideas between players in different locales. *Puzzle Adaptability Inventory:* Imposing Structure for self and others that provide as much focus as possible in dynamic situations.

- **Master Puzzle Stage 4:** Scavenger Hunt B.
  *Gameplay:* Players must complete a very difficult escape room, which requires them to rely on the photos they took in Stage 1. They must adjust their actions based on feedback,

**Acknowledgement:** this research was developed with funding from the Defense Advanced Research Projects Agency (DARPA).
**Disclaimer:** The views, opinions and/or findings expressed are those of the author and should not be interpreted as representing the official views or policies of the Department of Defense or the U.S. Government.

potentially revisiting locations if their original pictures were not of the right things. *Puzzle Adaptability Inventory:* Not needing things to be black and white.

- **Master Puzzle Stage 5:** Prisoner's Dilemma
  *Gameplay:* After "winning" the game, players must play a modified prisoner's dilemma to determine who will actually receive prize money, and how much. Players will be given the opportunity to cooperate or betray their team. *Puzzle Adaptability Inventory:* Effectively adjusting plans, goals, actions, or priorities to deal with changing situations.

### 1.2.3. Link to Video of Daedalus

A video of the game is available at:
https://drive.google.com/file/d/1QCyxP_RXlwntTzUkK0FXoXJXfxFTQNZj/view?usp=sharing

### 1.2.4. Validation and Testing

All puzzles and narrative were tested based on several criteria, which was done through design inspection:

- **Unpredictability:** Do puzzles satisfy the definitions of the PAI?
- **Team Performance:** How far does the team progress in the allotted time and how quickly do they complete the puzzles?
- **Individual Adaptation:** What actions do players take in the game? How readily do they change tactics? How long does it take a player to act after receiving an instruction? How long and how many attempts before the instruction is completed successfully?
- **Team Adaptation:** Is the team able to self organize to effectively address each task? What roles do team member assign themselves/each other? How much does each team member contribute? Are team members able to negotiate with one another when necessary?
- **Individual Performance:** Who enters the most correct solutions? Who contributes to the most different puzzle solutions? Who is able to secure prize money for themselves at the end of the game? How many messages do they send over Slack?

Furthermore, as prototypes conformed to these inspection criteria, we started to playtest them. We conducted five playtests between March and May of 2018. Iterations were made after each playtest. Each playtest is composed of 5-7 players who were asked to play the game as a team. They then were asked to fill in a survey to gauge their engagement with the game. This survey is in Appendix C.

**Acknowledgement:** this research was developed with funding from the Defense Advanced Research Projects Agency (DARPA).
**Disclaimer:** The views, opinions and/or findings expressed are those of the author and should not be interpreted as representing the official views or policies of the Department of Defense or the U.S. Government.

The final playtest results were overall positive. Some remarks led us to make small adjustments to the game, but we learned that players were engaged and understood play, felt communication was important, and were motivated by a broad cross section of factors.

When asked about the amount of money that would motivate them to play the game. As shown by the data in Figure 1.4 and Table 1.2, factors related to the game system itself and intrinsic motivations were the most common choice, which told us that our game was well designed. The cash prize was important to some people and not to others. According to this data we set the cash prize for the study at $100 for finishing and $300 for finishing in the top five teams.

**Table. 1.2.** Ranking of aspects of the game that would motivate the player to play

| | |
|---|---|
| Cash Prize | 3 |
| Helping my team | 3 |
| Competing against other teams | 3 |
| Challenging myself to complete hard puzzles | 6 |
| Seeing what the next stage looks like | |
| Once I start something I can't stop | 3 |
| Learning more of the story | 1 |
| Other | 1 |

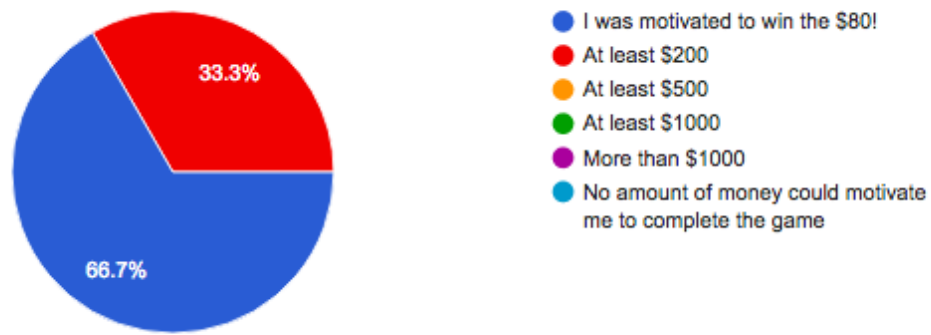

**Acknowledgement:** this research was developed with funding from the Defense Advanced Research Projects Agency (DARPA).
**Disclaimer:** The views, opinions and/or findings expressed are those of the author and should not be interpreted as representing the official views or policies of the Department of Defense or the U.S. Government.

Our final playtest resulted in good scores, as shown in Table 1.3. One player told us "Once our team was more cohesive (and we started responding to each other and working as a group), I enjoyed the fact that technology was actually bringing us together to work towards a common goal. It's easy to say that electronics and technology connect us (and they do at a surface level), but valuable cooperation via technology is much more rare." Other comments included :

- "These have been fun puzzles. I am very interested in reading the report once you are done. This seems like a fascinating concept."
- "The entire process has been fun and engaging, and definitely stimulating. This was the first 'escape room-like' encounter I have had and I will definitely be seeking out more in the future because of Daedalus!"
- "The immersive project has been great thus far, I really like the unique bot interface in Slack - I didn't know you could build essentially an Infocom game into the bots. I don't like that the interface is somewhat slow (delays when you push buttons) but it doesn't really detract from the experience, one just works with it as provided. I have a lot of past ARG and puzzle experience, and while the puzzles tend to be on the easier side, they do seem to be designed to have the solving experiences evaluated, if that's what you're really studying here. Nice job, folks!"

**Table 1.3.** Number of Responses for a Likert scale of questions from final playtest
played on May 2018 (9 out 10 players responded to the survey)

| Question | Disagree / Strongly disagree | Agree / Strongly agree |
|---|---|---|
| Puzzles were fun to solve | 1 | 7 |
| Puzzle were the right difficulty | 2 | 4 |
| Videos engaged me | 1 | 5 |
| Communication was important | 0 | 7 |

## 1.2.5. Designing for better Situation Awareness

The most notable change through our prototype testing was the addition of a *Progress Log* channel, which was made due to player feedback of confusion about the game system and inability for team members to quickly catch up with players who had moved ahead through

**Acknowledgement:** this research was developed with funding from the Defense Advanced Research Projects Agency (DARPA).
**Disclaimer:** The views, opinions and/or findings expressed are those of the author and should not be interpreted as representing the official views or policies of the Department of Defense or the U.S. Government.

multiple puzzles and stages while working asynchronously with the others. In our May Survey, players told us the game "rules were confusing to all team members," that "everyone moved on and [they] felt bad asking questions," and "the interface contributed to the disjointment" felt.

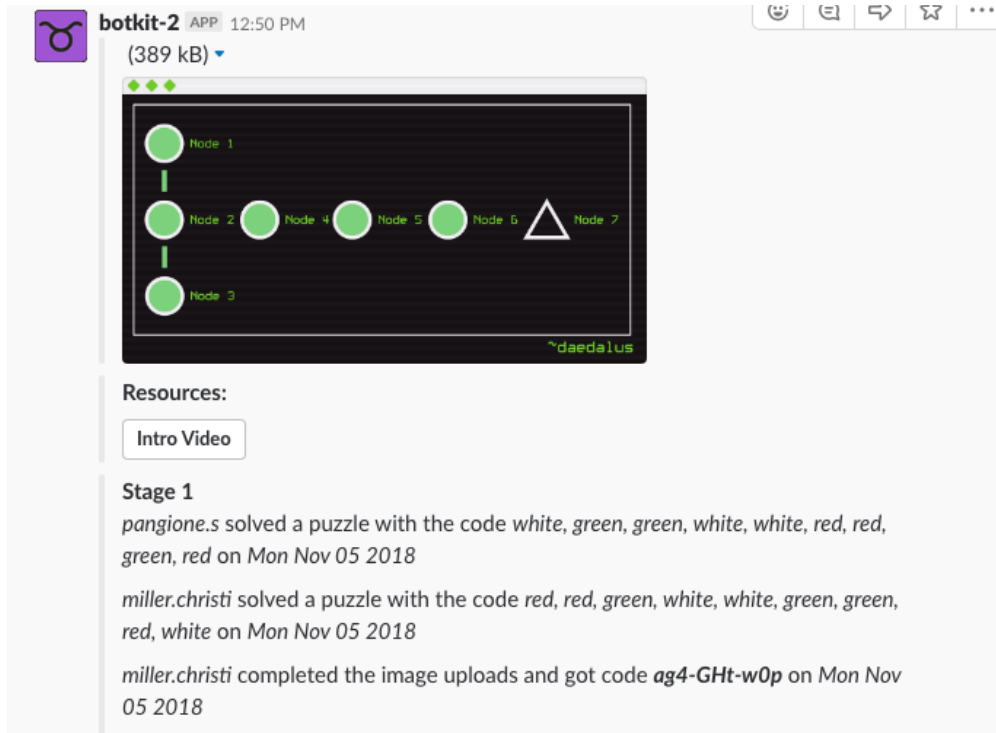

**Figure 1.5.** A Screenshot of the *Progress Log* channel in Slack.

The Progress Log, shown in Figure 1.5, is a channel in Slack that displays a map of the escape room nodes showing all steps of the farthest player's progress. Underneath, it lists a log of each puzzle solved, who solved it, date of the solution, and what the solution was. The addition of this channel significantly lessened player confusion as it created a way to get players up to speed and a shared situation awareness within the group. In addition, to enforce this shared mental model of progression, each individual will need to go through all steps to get from where they are to where the team is, therefore solving all the puzzles that the team has solved to catch up. This also improved clarity (see below).

### 1.2.6. Daedalus Study

Similar to MPL, we also ran a study to collect data through *Daedalus*. The study ran in Summer of 2018. Rather than using the classroom setting as in *MPL*, we recruited through different

Distribution Statement "A" (Approved for Public Release, Distribution Unlimited)                    20
**Acknowledgement:** this research was developed with funding from the Defense Advanced Research Projects Agency (DARPA).
**Disclaimer:** The views, opinions and/or findings expressed are those of the author and should not be interpreted as representing the official views or policies of the Department of Defense or the U.S. Government.

venues, including game clubs, classrooms within MIT, BU and Northeastern, etc. We knew attrition will be high in such a game, and so recruited as many players as possible. The cash prize was a good motivator. We thus collected data from 116 individuals (26 teams) who completed the game and completed the TAG survey. Out of those, 105 individuals completed the personality survey and out of those 103 individuals completed all surveys including the cognitive flexibility. We also have game data from 12 teams who didn't complete the game.

## 1.2.7. Data from Daedalus

We collected all data time stamped from Daedalus. This includes: all chat utterances by each team member stamped. All links used for external videos or images were also captured. We also captured navigation through screens and puzzle activity in the form of button presses.

From these button presses and screen times, we were able to deduce the following information:

- *Errors*: this was formulated as pressing the button with the incorrect sequence for solving a puzzle.
- *Puzzle Solved*: this is formulated based on button press with the right key or sequence to solve the puzzle.
- *Relevant Cues*: Based on screens needed to solve the puzzle, this is similar to MPL where we named specific relevant screens. But since the puzzles in Daedalus all need specific screens we didn't need to use feature selection to understand screen importance as designers already know this information and encoded in the data log, allowing us to formulate this abstraction.
- *Irrelevant Cues*: Similar to relevant cues, screens that are not relevant to solving the puzzle in the stage of the game is named irrelevant.

## 1.2.8. Scoring Performance
### 1.2.8.1. Team Score

The ARG used time to completion as a performance metric by which the teams were assessed. Players were encouraged monetarily to finish before other teams, as well as hatch all the eggs in the tamagotchi stage. Each additional egg that was hatched beyond the required 3 removed 5 hours from the team's final completion time. This adjusted time was then used to figure out which team finished first, and that team was rewarded with the largest prize.

**Acknowledgement:** this research was developed with funding from the Defense Advanced Research Projects Agency (DARPA).
**Disclaimer:** The views, opinions and/or findings expressed are those of the author and should not be interpreted as representing the official views or policies of the Department of Defense or the U.S. Government.

*1.2.8.2. Individual Performance Score*

Based on logs of activity per player, we were able to develop two scores: puzzle activity score and chat activity score, which show the amount of effort the individual did in solving puzzles and coordinating with others to find solutions. These two metrics were calculated as follows:

- **Puzzle Activity Score (PA)** $= \frac{\sum_i (PCR * TF_i)}{N}$, where $N$ is the number of Puzzles, and $P$ are the Puzzles. $PCR$ is computed as: $\frac{\sum completed(P)}{N}$, and $TF_i$ is computed as $\frac{(T_s^i - T_q^i)}{(T_s^i - T_n^i)}$ Where $T_s^i$ is the team's slowest time at completing the puzzle $i$, $T_q^i$ is the team's quickest time at completing the puzzle $i$ and $T_n^i$ is the time that player $n$, current player, completed puzzle $i$. This factor is a value between 0 and 1 and signifies how quickly an individual finished the puzzle after it was solved by the quickest team member.

- **Chat Activity Score (CA)** $= \frac{\sum m_s}{\max_{s \in T} m_s}$ where $m$ are messages and subject is a $s$ in a Team $T$. This measure captures amount of effort afforded by players who are not focused on puzzles but rather on communicating and coordinating with their fellow teammates. This measure is how close each player was in terms of the number of their messages to the most active player. This score then gives us a value from 0 to 1, where 1 is given to the most active player chat-wise in the team.

Given then two scores, we calculate **Individual Performance Score** as:

$$\left( \frac{2}{3} * PA + \frac{1}{3} * CA \right) * Score_T$$

Where $PA$ is Puzzle Activity Score, and $CA$ is Chat Activity Score and $T$ is a team whose member we are determining the performance for. As can be seen from this formula, puzzle activity is two times the chat activity, as we believe it is more reflective of an individual's actual contribution to the team's performance. However, different formula can be substituted here with little effect on our drawn conclusions.

## 2. Novel methodology to study team processes

Studying team processes to understand adaptation, team learning, coordination or innovation requires an in-depth study of team activity. Previous work has already started discussing the different team processes that is captured under such teamwork (Baard, Rench & Kozlowski, 2013). By studying adaptation as a process, previous work identified **cognitive**, **affective**, and **motivational** processes that underlie team adaptation. To further solidify the process of adaptation we adopted Rosen et al.'s work. Figure 2.1 shows the processes identified by the researchers to fall under the team adaptation process. The processes include four stages:

Distribution Statement "A" (Approved for Public Release, Distribution Unlimited)       22
**Acknowledgement:** this research was developed with funding from the Defense Advanced Research Projects Agency (DARPA).
**Disclaimer:** The views, opinions and/or findings expressed are those of the author and should not be interpreted as representing the official views or policies of the Department of Defense or the U.S. Government.

- **Situation Assessment:** The team starts collecting information about their environment, including identifying and assigning meaning to cues and ensuring good communication.
- **Plan Formulation:** The team formulates a plan of action in order to reach the desired outcome of adaptation).
- **Plan Execution:** The team attempts to carry out the plan, adjusting to new information or unexpected changes as necessary.
- **Team Learning/Reflection**: The team reflects on what went well, what went poorly, and what changes to the team's structure and processes should be made in the future.

This model developed a set of **behavior markers** (specific behavioral signatures that can be observed and identified by a researcher or manager). This was then meant to develop an instrument for observing behaviors, and thus allowing researchers to capture adaptation process through these observed behaviors. Examples include a marker for *coordination*: articulated information about their status, needs and objectives as often as necessary (and not more), and a marker for *reactive conflict management*: utilized negotiation or mediation strategies for conflict resolution (Rosen et al., 2011). Such behavior markers are closer to what we envision as an instrument to capture such behaviors directly from data and thus we adopted this model.

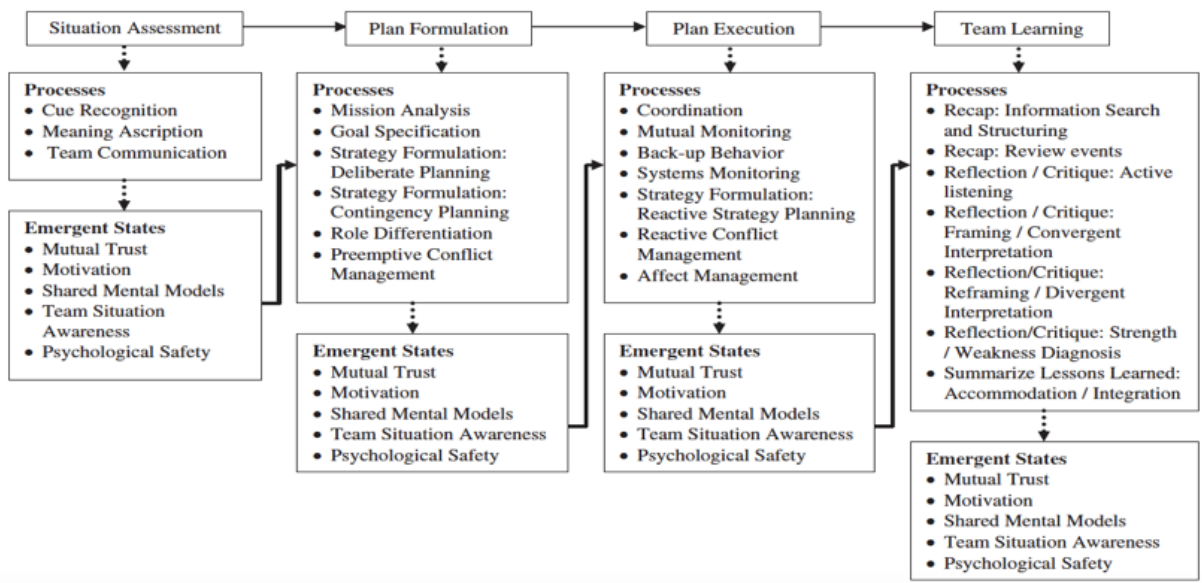

**Figure 2.1.** Rosen et al.'s Adaptation Process

As it can be seen from the list of processes and Figure 2.1., some of the processes are impossible to measure through behavioral data alone. For example, meaning ascription, or many aspects of plan formulation are not available to us through game logs. In such a case, we needed to

**Acknowledgement:** this research was developed with funding from the Defense Advanced Research Projects Agency (DARPA).

triangulate behavioral data with another instrument. Since our goal is to develop a scalable approach to gauging adaptation, we chose to explore two instruments: 1) a self-report measure to capture aspects of adaptation process that we cannot capture through behavioral data, which we called TAG (Teamwork and Adaptation in Games) survey, and 2) *Slack* data qualitative coding. Since slack has been used by many teams to coordinate activity, effectively coding it to gauge adaptation process may give us a seamless approach that can scale well to measure adaptation for many teams. We discuss these two instruments and methods next. We also discuss their reliability and validity.

In addition to the two instruments to capture the non-behavioral aspects of adaptation, we also developed two new data-driven methods and tools to gauge different parts of the adaptation process through game logs: 1) Human in the loop team behavior investigation tool and 2) Behavior Situation Assessment Scoring method.

## 2.1. Human-In-the-Loop Team Behavior Investigation Tool

To understand teamwork and adaptation process we developed a new tool to allow us to gauge this level of analysis. Unfortunately, many of the adaptation processes discussed above are context dependent and also task dependent and are therefore hard to generalize from one scenario to another. Therefore, we adopted a *human-in-the-loop* approach whereby we leverage our visualization system, *Glyph*, to visualize sequential behavioral patterns in the data. The human can then interpret the sequences. In order to get to the right level of abstraction, more data pre-processing is needed to visualize the patterns at a level that is interpretable for the human involved. To fully understand this process, we will first discuss our visualization system Glyph and then discuss the approach we took to abstract the data to visualize it in an interpretable level.

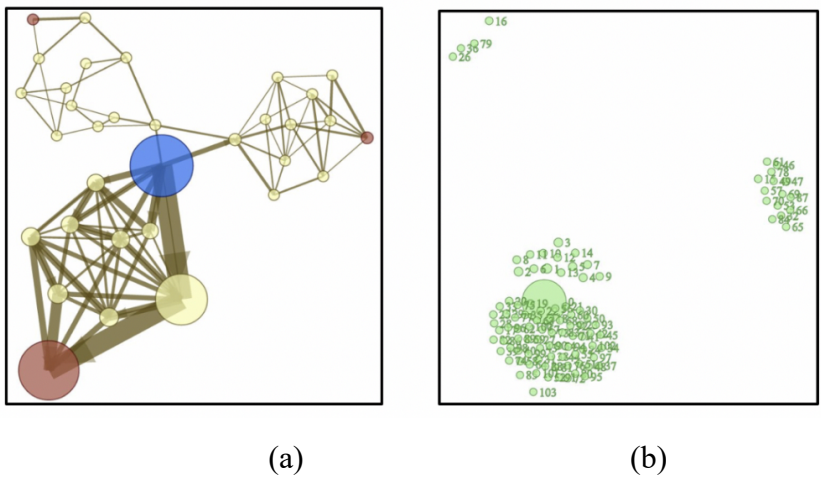

(a)                    (b)

**Acknowledgement:** this research was developed with funding from the Defense Advanced Research Projects Agency (DARPA).

**Disclaimer:** The views, opinions and/or findings expressed are those of the author and should not be interpreted as representing the official views or policies of the Department of Defense or the U.S. Government.

**Figure 2.2.** Glyph 2-window visualization of action. (a) state graph and (b) is sequence graph. Data visualized is from a puzzle-based game called *Wuzzit Trouble* developed by *BrainQuake* to teach algebraic math to middle school kids.

### 2.1.1. Glyph

*Glyph* (Nguyen et al., 2015) is a visualization system composed of a dual-view interface that shows data from two related perspectives: a state graph and a sequence graph. A state graph shows a sequence of abstract behaviors and abstracted states (Figure 2.2a). A state-graph in Glyph is a node-link graph, where nodes represent different game states, at which players make decisions and execute actions, and directed links are actions that players took to get from one state to another. For example, Figure 2.2(a) shows a state graph with one start state (in blue) and three end states (in red). In general, a game state captures all information associated with in-game entities that affect players' choice of actions. To facilitate comparison of individual action sequences, we augmented the state graph with a **synchronized** sequence graph showing the popularity and similarity of sequence patterns exhibited by users (Figures 2.2b and 2.3). Each node in the sequence graph represents a full play trace, the size is an indication of popularity. Further, the distance between each node provides a visual representation of similarity/dissimilarity.

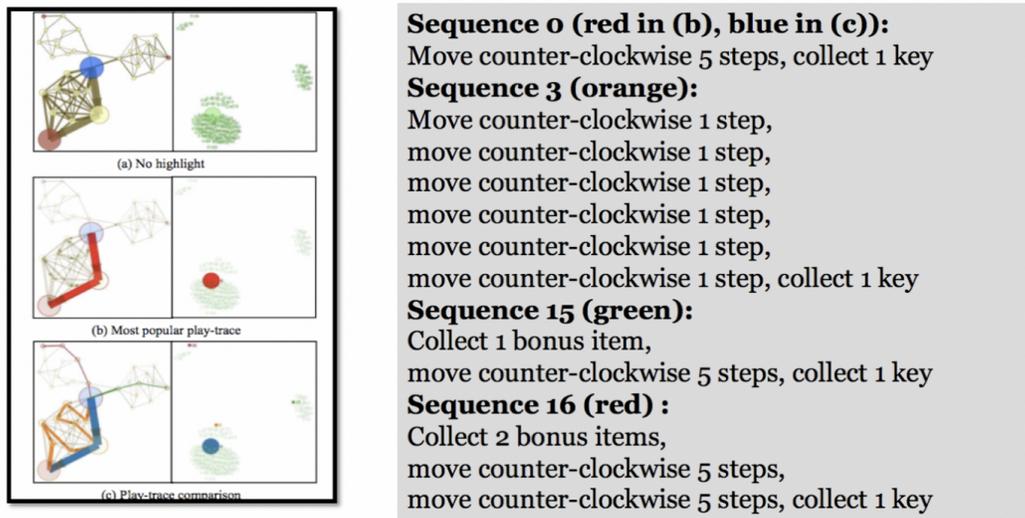

**Sequence 0 (red in (b), blue in (c)):**
Move counter-clockwise 5 steps, collect 1 key
**Sequence 3 (orange):**
Move counter-clockwise 1 step,
move counter-clockwise 1 step,
move counter-clockwise 1 step,
move counter-clockwise 1 step,
move counter-clockwise 1 step,
move counter-clockwise 1 step, collect 1 key
**Sequence 15 (green):**
Collect 1 bonus item,
move counter-clockwise 5 steps, collect 1 key
**Sequence 16 (red) :**
Collect 2 bonus items,
move counter-clockwise 5 steps,
move counter-clockwise 5 steps, collect 1 key

**Figure 2.3.** Glyph 2-window visualization of action. To the right are example sequences of actions performed by selected players (shown by the larger red, blue, green, orange dots on the left).

**Acknowledgement:** this research was developed with funding from the Defense Advanced Research Projects Agency (DARPA).

**Disclaimer:** The views, opinions and/or findings expressed are those of the author and should not be interpreted as representing the official views or policies of the Department of Defense or the U.S. Government.

Utilizing synchronized visual information presentation, the interface provides multiple perspectives on the same data at different levels of granularity, allowing instant examination of play data at different scales. Figure 2.3 shows traces from many players from *Wuzzit Trouble* (BrainQuake, 2014), and a comparison between different ways that different players have solved a particular puzzle.

## *2.1.2. Scoring and Modeling Adaptation Process*

Using this visualization system, we can inspect all the sequences of behavioral data to gain more insight on the team processes at a deep level. Quantitatively, we can also develop a scoring mechanism based on the sequence graph (shown in the right window). The sequence graph developed uses a metric to compute distances between sequences to show how different the sequences are. This metric is determined using *Dynamic Time Warping* (Berndt and Clifford, 1994) -- a technique developed to determine difference measure between two time series or sequence data.

To develop a score for each pattern or individual trace in the data, we needed to identify an *optimal strategy*. In some games, this optimal strategy is easy to define based on expert or designer knowledge. In some other games, due to their dynamic nature, we may have to identify an expert game player and then use his trace as optimal but match the situational variables to identify what traces are applicable given the situation. Fortunately, for the two environments we are using here, optimal strategies are easily developed by experts, and thus we used that trace as a comparative trace. The visualization system then allowed us to inspect all other traces in opposition to that trace and verify the use of such a trace as optimal trace. Once that is verified a distance function based on Dynamic Time Warping is used to quantify the adaptation scores. Below we will discuss how we derived these scores and also all insights we gained by inspecting the visualized sequences about adaptation and individual patterns.

### *2.1.2.1 Applying Human-In-the-Loop Team Behavior Investigation Tool to MPL*

We used this approach to understand adaptation of teams within MPL. We specifically looked at decisions made in regard to brands and how such decisions affect the affecting Market Effectiveness (ME) performance score, see Section 1.1.2 for more information on performance scores within MPL. As discussed above, in order to develop a reasonable visualization to interpret the information needed, we had to develop a good state representation and sequence representation.

**State Representation.** In the game, Market Effectiveness is calculated based on decisions made on brands designed for target markets, therefore to capture how teams adapted to this performance metric, we need to capture how they adjusted and developed brands and how these

**Acknowledgement:** this research was developed with funding from the Defense Advanced Research Projects Agency (DARPA).

brands are assessed by target markets and non-target markets. Target markets are market segments selected to be the primary and secondary markets by the team, and non-targets are those not selected. We then computed if the teams were able to improve the judgments of their brands from one quarter to another or not. Thus, there are three possible values for each segment type:

- **Increase**: The team are able to add new lines of products that increase the brand judgment in that segment. This **positively** affects ME.
- **Decrease**: The team have adjusted their products in ways that decrease the brand judgment in that segment. This **negatively** affects ME.
- **Unchanged**: The team's new brand designs, if any, do not change the judgment in that segment. ME remains the **same**.

For example, let's say in Quarter 4, a team's state is **"target: increase_unchanged, non-target: decrease"**. This means that as compared to the products in Quarter 3, in this quarter, the team was able to create new products to improve the brand judgment in one target segment, while keeping the judgment the same for the other. Meanwhile, the brand judgment in the remaining segment (i.e., their non-target) had reduced. While this has no effect on ME, it has indirect effects on *Financial Performance*, as products in the non-target segment(s) are now less appealing to customers. Note that at some point in the game, every team is requested to choose up to two target segments among the three: Mercedes, Traveler, Workhorse.

**Sequence Representation.** Each team's data can now be represented as a sequence of five states, representing decision data from Q2 to Q6.

We also made one adjustment to the original Glyph visualization system. Since all sequences are equally long (i.e., 6 quarters), the sequence distances are computed as the sum of the pair-wise state distances, instead of using DTW. More specifically, the distance between two sequences seq1 and seq2 is computed by adding the differences between seq1's Q2 decisions and seq2's Q2 decisions, seq1's Q3 decisions and seq2's Q3 decisions, etc., together. DTW can handle sequences of varying lengths, by disregarding the absolute position of each state in the sequence (i.e., warping). In this case, there is a semantic meaning to the absolute position of each state, as each of them is ordered by the quarter, so DTW may not be appropriate.

Differences between states were calculated as follows: for two states, we compare targets first, if they are different and one of them is "decrease", diff is increased by 10, otherwise, diff is increased by 4 (higher weight on decrease to accentuate failures in targets). Next, we compare non-targets, the same logic applies, except that the respective values are 5 and 2. Again, decrease is heavily penalized.

**Acknowledgement:** this research was developed with funding from the Defense Advanced Research Projects Agency (DARPA).

It should be noted that the decision of the representation and the weighting of the different nodes in the distance function was done through the human-in-the-loop process described above, whereby we visualized the sequences and inspected the differences between the ideal pattern and the other patterns and adjusted the distance metric to capture the right distance function. Iterations were done to arrive at this metric and validation was done through expert review.

The visualization of the traces for 50 groups from Spring 2018 data (see Figure 2.4). A link for this visualization is at: https://truonghuy.github.io/mpl_rule_generation/index.html for more interactive experience. In the visualization we also noted the *Market Effectiveness* scores, which are the numbers on the nodes in the sequence graphs.

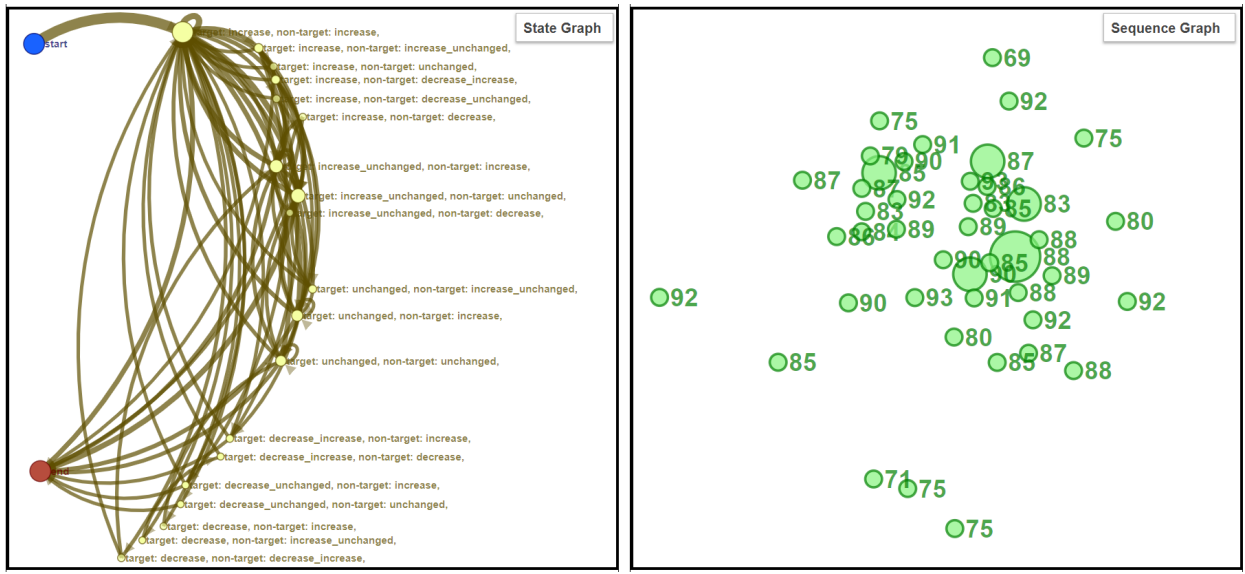

**Figure 2.4.** The Glyph visualization. We arrange the nodes in the state graph in decreasing order of the effects on target markets (increase → unchanged → decrease). Distances between nodes in the sequence graph represents how dissimilar the sequences are. The node labels in the sequence graph are the average ME score of trajectories belonging to that respective node (the higher the better, with maximum being 100).

Some notable observations:

**Convergence of high market effectiveness scores:** From the sequence graph (on the right), we can tell that at the end of the game, most teams are able to design brands with high judgments (most are close to 90, with the highest ME score being 93 – done through inspection of the sequence graph on the right where nodes are labeled with ME scores). This is a sign that the

**Acknowledgement:** this research was developed with funding from the Defense Advanced Research Projects Agency (DARPA).

**Disclaimer:** The views, opinions and/or findings expressed are those of the author and should not be interpreted as representing the official views or policies of the Department of Defense or the U.S. Government.

game did well in teaching students important marketing concepts, including market effectiveness.

**Clusters are closer to Optimal Strategy:** From interaction with the visualization (which is not available from the screenshot), we were able to deduce that the big cluster (see the sequence graph on the right) in the sequence graph comprise of sequences containing of mostly "target: increase" or "target: increase_unchanged" nodes, which mean they are able to improve ME as the game progresses. This is good, as these would be closer to an optimal strategy, which we identified as increase -> increase -> increase -> increase -> increase.

**Diverse Strategies (innovation or failure to adapt):** Through the visualization, we can inspect strategies that are different. These are at the perimeter of the sequence graph and they depict teams that have different (and potentially more interesting) decisions in the game.

For example: the three nodes labeled as 71, 75, 75 at the bottom of the graph (see Figure 2.5) are teams who performed below average and thus failed to adapt. A closer look at them reveals something in common: They stumbled in Q3, resulting a decrease on ME. They tried to bounce back with varying degrees of success (see Table 2.1), which shows the three sequences in different columns, since it doesn't matter which is which we didn't label the columns. From the table we can see that the attempts to adapt were not good enough, resulting in low final ME scores of 71, 75, and 75.

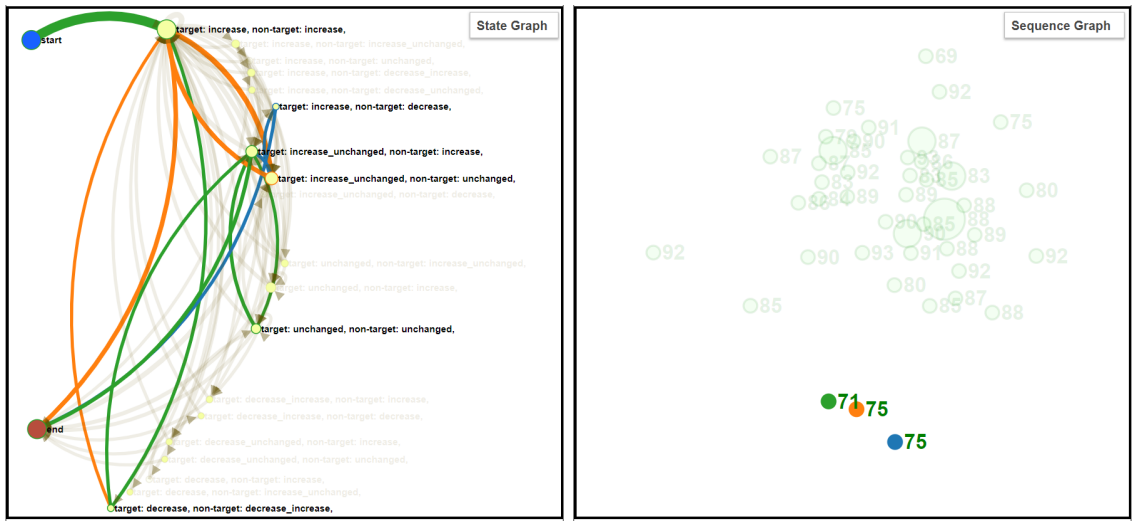

**Figure 2.5.** Figure shows the use of glyph to highlight the three underperforming teams.

**Table 2.1.** Failure to Adapt

| Q2. target: increase, non-target: | Q2. target: increase, non-target: | Q2. target: increase, non-target: |
|---|---|---|

**Acknowledgement:** this research was developed with funding from the Defense Advanced Research Projects Agency (DARPA).

**Disclaimer:** The views, opinions and/or findings expressed are those of the author and should not be interpreted as representing the official views or policies of the Department of Defense or the U.S. Government.

| | | |
|---|---|---|
| increase<br>**Q3. target: decrease, non-target: decrease_increase**<br>Q4. target: **increase_unchanged**, non-target: increase<br>Q5. target: **unchanged**, non-target: unchanged<br>Q6. target: **increase_unchanged**, non-target: **increase** | increase<br>**Q3. target: decrease, non-target: decrease_increase**<br>Q4. target: **increase**, non-target: increase<br>Q5. target: **increase_unchanged**, non-target: unchanged<br>Q6. target: **increase**, non-target: **increase** | increase<br>**Q3. target: decrease, non-target: decrease_increase**<br>Q4. target: **increase_unchanged**, non-target: increase<br>Q5. target: **increase_unchanged**, non-target: unchanged<br>Q6. target: **increase**, non-target: **decrease** |

Another interesting example is at the top of the sequence graph (see Figure 2.6), there are two nodes with similar decisions, but resulted in vastly different results: one ending with ME score 92, and the other 69 (Figure 2.6).

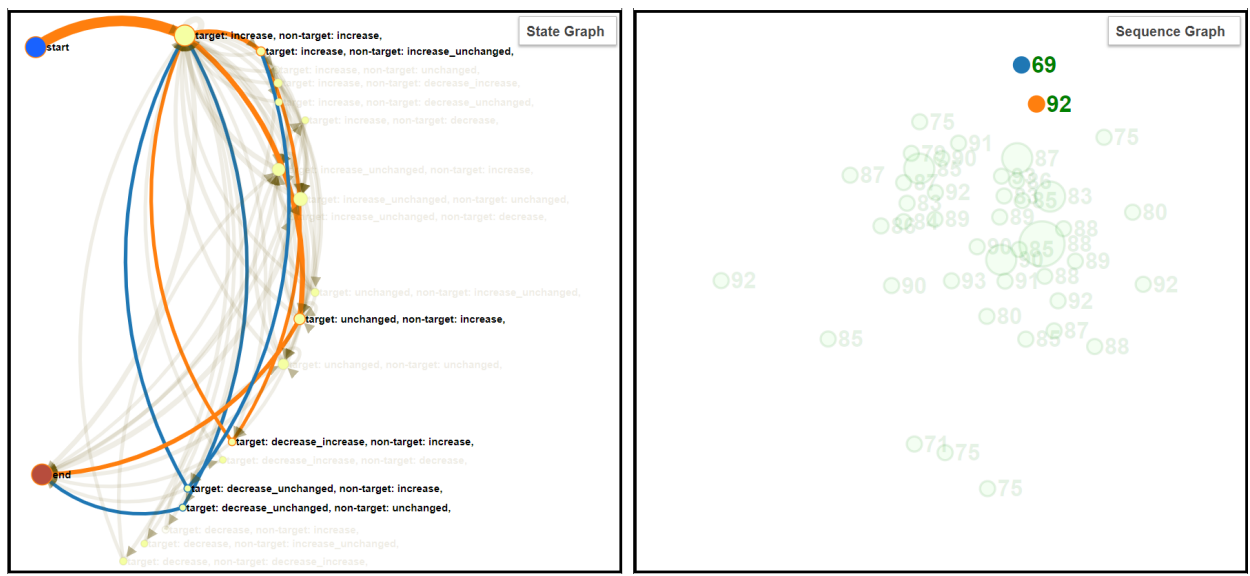

**Figure 2.6.** Two seemingly similar struggles in brand design that result in vastly different ME scores at end-game. They were moving back and forth between states with increasing and decreasing brand judgments.

Both of them stumbled in Q4. However, despite the decrease in one target segment in Q4 (right column in Table 2.2) that team managed to improve one of the two target segments in Q4 and Q6, while the other team failed to do so. Furthermore, the 69 team made another major mistake at the last quarter (Q6) as it got worse judgment in one target segment. This probably dealt the final blow to its final performance. The other team on the other hand seemed to have learned a good lesson in Q4 (which resulted in the decrease in a target market) and were able to reach the end of the game with success. As you can see from the visualization, both are far away from the

**Acknowledgement:** this research was developed with funding from the Defense Advanced Research Projects Agency (DARPA).
**Disclaimer:** The views, opinions and/or findings expressed are those of the author and should not be interpreted as representing the official views or policies of the Department of Defense or the U.S. Government.

ideal in the center of the sequence diagram, but one is closer than the other. Therefore, this distance metric we devised seem to capture well how well teams adapt and further inspection allowed us to adjust the metric towards what we feel is more robust measure of adaptation.

**Table 2.2.** Different degrees of adaptation

| Final ME Score: 69 | Final ME Score: 92 |
|---|---|
| Q2. target: increase, non-target: increase, | Q2. target: increase, non-target: increase, |
| Q3. target: increase, non-target: increase_unchanged, | Q3. target: increase, non-target: increase_unchanged, |
| **Q4. target: decrease_unchanged, non-target: increase,** | **Q4. target: decrease_increase, non-target: increase,** |
| Q5. target: increase, non-target: increase, | Q5. target: increase, non-target: increase, |
| Q6. target: **decrease**_unchanged, non-target: **unchanged** | Q6. target: unchanged, non-target: **increase** |

*2.1.2.2 Applying Human-In-the-Loop Team Behavior Investigation Tool to Daedalus*
Similar to the process with MPL, we first visualized the raw data into Glyph and then developed methods to abstract the data to arrive at the right distance function and state representation for easier interpretation by experts and also inspection of behavioral patterns.

**State Representation.**

- **Navigational Nodes:** we distinguished between screens that just served as **navigational state** which is basically a state where the user is passing through a navigational screen.
- **Cue Nodes:** To identify cues that are important to solving a puzzle (cue recognition part of the situational assessment phase of Rosen et al.'s model discussed above), we identified the sequence in which the puzzles must be solved. For stage 1, a player can complete the safari puzzle at any point in time, but must complete few dots, many dots, the cypher puzzle and glyph puzzle in that order. After completing all of these puzzles, the player gets the code for the safe to proceed to the next stage. We use this information to figure out if the player is visiting a relevant cue to a puzzle they can solve or not and use that to abstract the cue screens into two states: **irrelevant cue** and **relevant cue**.
- **Failure nodes:** We also observed that players often attempt a puzzle multiple times incorrectly. To capture this behavior more compactly we collapsed sequences of failures. We do this by looking at the number actions between two failures and if they are below a certain threshold, we collapse the series of failures into the state **failed_many_times**.
- **Help from a Friend Node:** Another thing we observed is that some players solve puzzles without visiting relevant cues, this is likely due to them getting the solution form their team-mates or through the progress screen, as discussed earlier. Since we wanted to capture this behavior we added a state **no_relevant** to represent it.

**Acknowledgement:** this research was developed with funding from the Defense Advanced Research Projects Agency (DARPA).
**Disclaimer:** The views, opinions and/or findings expressed are those of the author and should not be interpreted as representing the official views or policies of the Department of Defense or the U.S. Government.

- **Gave up Node:** Finally, we wanted to show explicitly the number of puzzles an individual solved before abandoning the game so that the differences of the traces can be more meaningfully extenuated. And thus, we added a state **gave_up_x** at the end of traces of those who did not solve all of the puzzles, where *x* is the number of puzzles they solved before abandoning the game. Completely empty traces were given another state to **gave_up_without_trying** to further penalize them over other traces.

At the end we arrived at these states for the traces:
- relevant_cue
- irrelevant_cue
- failed_once
- failed_many_times
- solved [name of puzzle]
- navigation
- no_relevant
- gave_up_[number of puzzles solved before abandoning the game]
- gave_up_without_trying

We then attempted to differentiate traces in terms of their performance using the Dynamic Time Warping algorithm as a clustering mechanism. This algorithm compares individual states of the traces and if there are not same a distance of 01 unit is added between them. To differentiate the players who completed the puzzles a distance of 01 unit is added between the traces every time when one of them contains the 'solved_[name of puzzle]' and other trace doesn't. An additional 01 unit distance is attributed for solving the final puzzle, i.e. 'solved_safe', the final puzzle of stage 01 when only one of the traces has it. Figure 2.7 shows the trace for some game traces where individuals or groups have solved the puzzle.

**Acknowledgement:** this research was developed with funding from the Defense Advanced Research Projects Agency (DARPA).

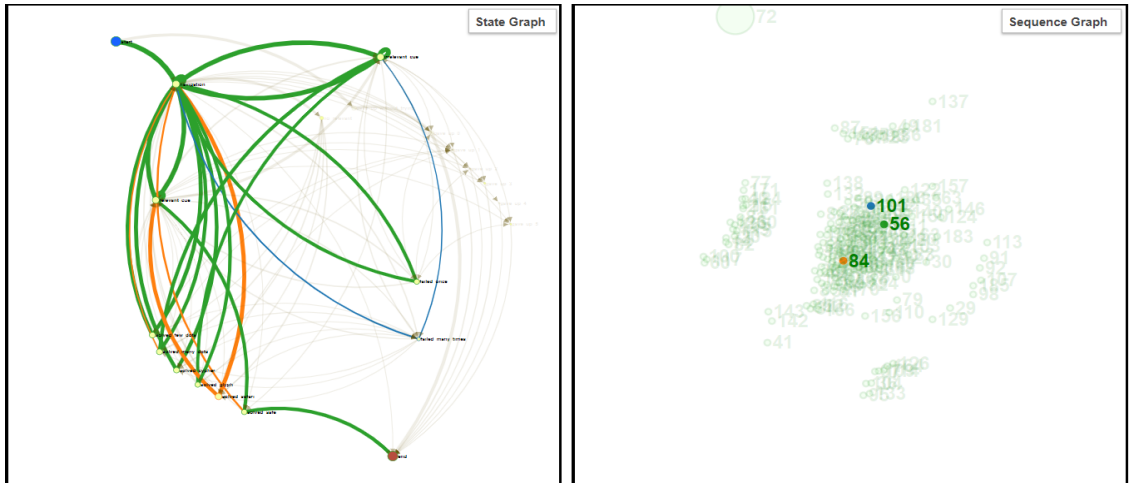

**Figure 2.7.** Example traces of completed game traces

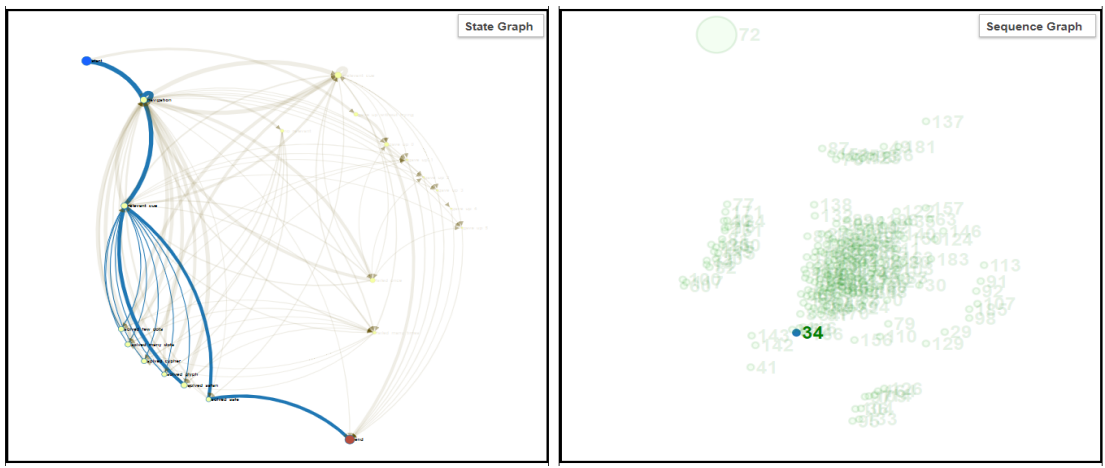

**Figure 2.8.** Visualization Ideal solution

We visualized the ideal pattern, see Figure 2.8.

Distribution Statement "A" (Approved for Public Release, Distribution Unlimited)          33
**Acknowledgement:** this research was developed with funding from the Defense Advanced Research Projects Agency (DARPA).
**Disclaimer:** The views, opinions and/or findings expressed are those of the author and should not be interpreted as representing the official views or policies of the Department of Defense or the U.S. Government.

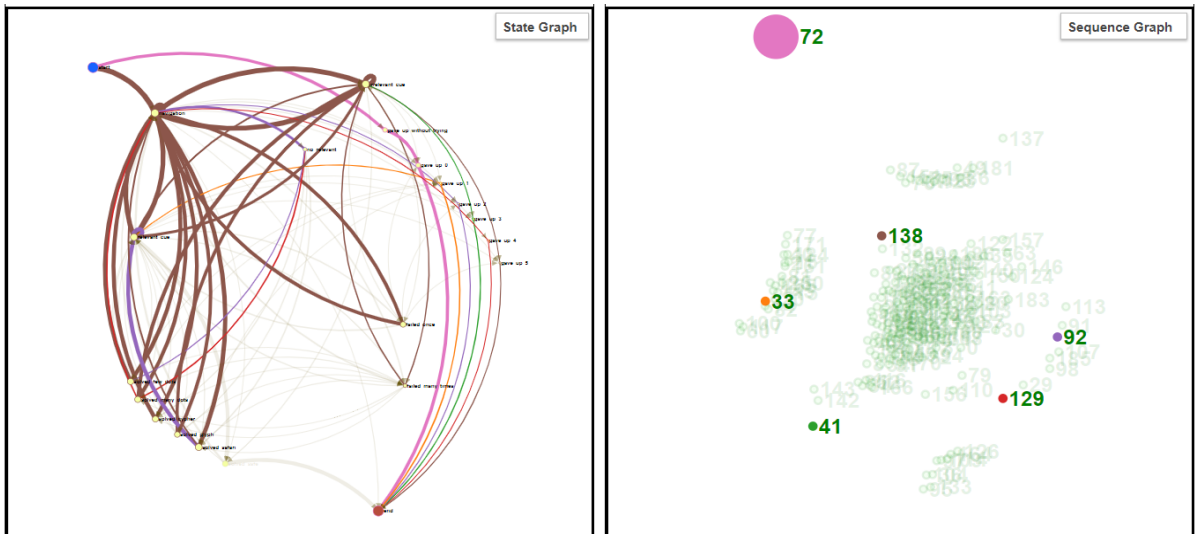

**Figure 2.9.** Example of clusters with incomplete game traces who left the game but some left after completing some puzzles. Gave_up_0 (seq no.72), left without completing any puzzles. Gave_up_1 (seq no.33) left after completing 1 puzzle. Gave_up_2(seq no.92) left after completing 2 puzzles. Gave_up_3 (seq no.41) left after completing 3 puzzles. Gave_up_4 (seq no. 129) left after completing 4 puzzles. And, finally, gave_up_5 (seq no.138) left after completing 5 puzzles.

Further, to separate the players who left before reaching the end of the first stage from the rest a heavy penalty distance of 300 units has been added if only one the them has this state. Such a high distance would make the trace very far from the traces which completed stage 01. Further, a penalty distance of lesser value has be added to these incomplete game traces when they leave early in the stage, i.e., the earlier they leave the game the higher penalty distance is attributed to the trace. This would enable us to distinguish the players that left early from the players that left in the later phase of the game on top of the distinction from the players who completed the game. Figure 2.9 shows such traces.

In addition to this high-level discrimination, we attempted to differentiate both these clusters further based on the players attempts/failures in solving puzzles, picking relevant or irrelevant or no cue at all before solving the puzzles. In the context of incomplete game cluster, we heavily penalized the play traces who left the game without even trying, i.e. 'gave_up_without_trying' with a distance of 400. We did a similar thing in completed game cluster, when we observed play traces who solved without effort i.e. from the solutions of teammates. We identified these traces when they solved the puzzles without visiting the relevant cue in eight previous states i.e. 'no_relevant_cue'. This threshold of eight previous steps is decided arbitrarily by considering the general memory and can be adjusted.

**Acknowledgement:** this research was developed with funding from the Defense Advanced Research Projects Agency (DARPA).

**Disclaimer:** The views, opinions and/or findings expressed are those of the author and should not be interpreted as representing the official views or policies of the Department of Defense or the U.S. Government.

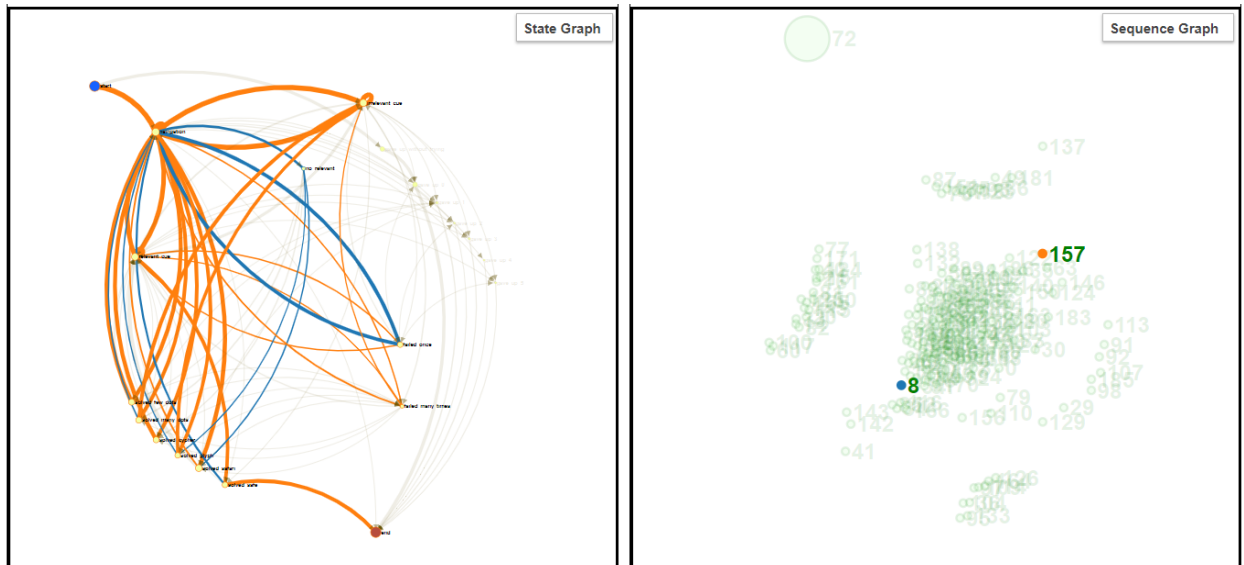

**Figure 2.10.** Example of traces that completed game with struggle (seq no. 157) and without much struggle (seq no.8).

There are also traces where the players made several failed attempts by visiting irrelevant cues before eventually finishing stage 01 or leaving the game. These traces can be identified with the occurrence of states such as **failed_once**, **failed_many_times**, and **irrelevant_cue**. Every occurrence of these states in a trace is penalized with a distance of 01, 03 and 02 units respectively. It should be noted that the state **failed_many_times** is not much different from **failed_once** except for the fact that it is an abstraction of **failed_once**, when a player made more than 3 unsuccessful attempts to solve a puzzle by visiting irrelevant cues or misinterpreting the relevant cue. Having these states enabled us to differentiate players who failed several times before finishing (mid-high adaptable) or leaving (mid-low adaptable) the game from the players that finished (high adaptable) or left the game without much struggle (low adaptable). See Figure 2.10 for an example of one player who struggled a lot and other who finished without much of a struggle.

Using this visualization, we can also look at team interaction. Unlike MPL, with Daedalus we have a clear record of each individual's contribution and activity log and thus can look at activities with the teams to see if there is a sole contributor or if there are slackers or people who struggled in the team. Figure 2.11 shows such a team. As you can see from the figure, this team had a person who gave up without trying (Seq no.72), but there were members who tried and solved several puzzles, but then gave up (e.g., Seq no. 129, 109, 104). Note also that the ideal solution is in the center cluster (as was the case with MPL). For more inspection of the teams and traces, please use the link at: https://a3madkour.github.io/.

Distribution Statement "A" (Approved for Public Release, Distribution Unlimited)   35
**Acknowledgement:** this research was developed with funding from the Defense Advanced Research Projects Agency (DARPA).
**Disclaimer:** The views, opinions and/or findings expressed are those of the author and should not be interpreted as representing the official views or policies of the Department of Defense or the U.S. Government.

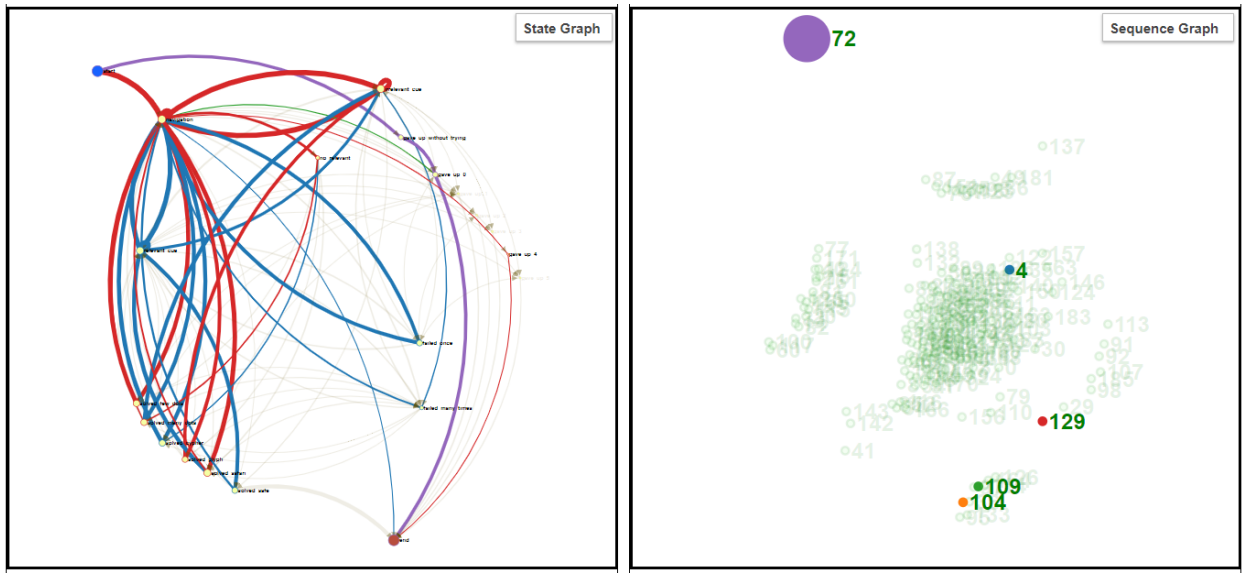

**Figure 2.11.** Visualization of a group with many people who gave_up

*2.1.2.3 Applying Human-In-the-Loop Team Behavior Investigation Tool to BoomTown*

As part of the project we also analyzed and used this approach to understand innovative patterns in a game called *BoomTown* developed by another performer working under the NGS2 DARPA program. Due to the length of this report, we will add the report on this part in another document called Gallup Report.

### 2.1.3. Validation

Since Glyph produces a visualized sequence of actions that is interpretable by a human, it is easy for a human expert to see the scores and the patterns and how they deviate from the optimal strategy to give us a qualitative indication of the validity of the score over groups within the game. This process was then done through inspection of experts for both environments. It is also the same method used for Gallup work, which was inspected by the Gallup team.

## 2.2. Behavioral Situation Assessment Scoring (BSAS) Method

As established earlier *Situation assessment*, according to Rosen et al.'s model, often occurs in the first phase of team work process during team adaptation. It is an individual level process, where one or more team members search the environment for cues for decision-making. This information collection itself is not sufficient to trigger team adaptation: the raw information collected has to be assigned meaning in relation to the existing knowledge, which is referred to in Rosen et al.'s model as 'Meaning Ascription'. Rosen et al's. model divides the process of

Distribution Statement "A" (Approved for Public Release, Distribution Unlimited)                36
**Acknowledgement:** this research was developed with funding from the Defense Advanced Research Projects Agency (DARPA).
**Disclaimer:** The views, opinions and/or findings expressed are those of the author and should not be interpreted as representing the official views or policies of the Department of Defense or the U.S. Government.

situation assessment into three important steps: information collection/cue recognition, meaning ascription and team communication. The section describes the process of quantitatively modeling and validating situation assessment in both *MarketPlace Live* and *Daedalus*.

## 2.2.1. Measuring Situation Assessment in MPL

Information collection process of Situational Assessment is directly observable from the behavioral data obtained from MarketPlace Live studies. The meaning ascription part of the process is cognitive, and thus is not observable from behavioral data. Unfortunately, we also don't have enough communication data to allow us to infer this part. In fact, for Spring 2018, we were able to get slack data from some teams and thus future studies can analyze this data to infer meaning ascription, for now we will focus on *Information Collection* part of this process.

Information collection by each member of the team can be measured from the time spent by that member on screens, which provide information about their own team performance and their competitors decisions and performance. The time spent on each screen is the consolidation of the activities the member performed on that screen: view, modify, calculate, save etc. Each of these activities indicate that the team member is collecting information on that screen. Table 2.3 illustrates the list of screens available for players to assess their situation.

**Table 2.3.** Screens providing information about own and competitor performance

| Competitors Screens | Own Performance Screens |
|---|---|
| Competitors Profiles | Balanced Scorecard |
| Competitors Salesforce | Cumulative Balanced Scorecard |
| Competitors Local Advertising | Market Share |
| Competitors Brands | Sales |
| Competitors Ads | Division Profitability |
| Competitors Prices | Brand Profitability |
|  | Detailed Brands |
|  | Demand report |
|  | Strategic Graphs |

Our approach is to investigate the relationship between information collection and performance to obtain the important features in each quarter that can predict performance. Identifying these features will enable us to score the teams for their situation assessment in that quarter. To get at these features, we first used machine learning technique to predict performance from time spent

Distribution Statement "A" (Approved for Public Release, Distribution Unlimited)
**Acknowledgement:** this research was developed with funding from the Defense Advanced Research Projects Agency (DARPA).
**Disclaimer:** The views, opinions and/or findings expressed are those of the author and should not be interpreted as representing the official views or policies of the Department of Defense or the U.S. Government.

on the screens (Table 2.3). The performance metric used in this case is the balanced scorecard score generated by the simulation (as discussed in Section 1) at the end of each quarter.

The balanced scorecard is a measure used to evaluate the team's overall performance in finances, markets and marketing effectiveness, with a value between 1 and 100. Initial exploratory analysis, such as scatter plots, suggested that this relationship between information collection from the above screens and performance is non-linear. We opted to use classification as a technique to predict performance, since some classification methods, such as Random Forest can work well with small datasets and can give us the importance of features as an entropy measure. In order to do that we needed to turn the performance value from continuous to discrete, i.e. classes that we can then use classification to predict. For this, we used equal frequency binning to divide teams into 5 classes: "Very High", "High", "Medium", "Low","Very Low" in each quarter. The decisions of how to split the continuous scale was done through analysis of the distribution to find best cut offs.

We then trained a Random Forest Classifier, using an 80%-20% split of the data, where 80% was used for training and the rest for testing. We used entropy as a measure of information gain from features for the decision tree feature selection. We used Out-Of-Bag (OOB) score as a measure of accuracy evaluating each tree in the forest. We used 10-fold cross-validation to hypertune the parameters of the random forest model.

We then calculated the F1-score for Q6, which resulted in 0.55. While this number may be low, it is significantly better than the baseline = 0.19 obtained from a random/dummy classifier. The confusion matrix of test data obtained from random forest classifier and dummy classifier are presented in Figure 2.12. Table 2.4 indicates the precision, recall, F1-score and support of each class from classifier of Q6. From the table we can see that the precision and recall for the very high and very low groups is a lot higher than the other classes, it may be that the number of data points for the classes in between is smaller affecting the rate of our prediction model.

Distribution Statement "A" (Approved for Public Release, Distribution Unlimited)          38
Acknowledgement: this research was developed with funding from the Defense Advanced Research Projects Agency (DARPA).
Disclaimer: The views, opinions and/or findings expressed are those of the author and should not be interpreted as representing the official views or policies of the Department of Defense or the U.S. Government.

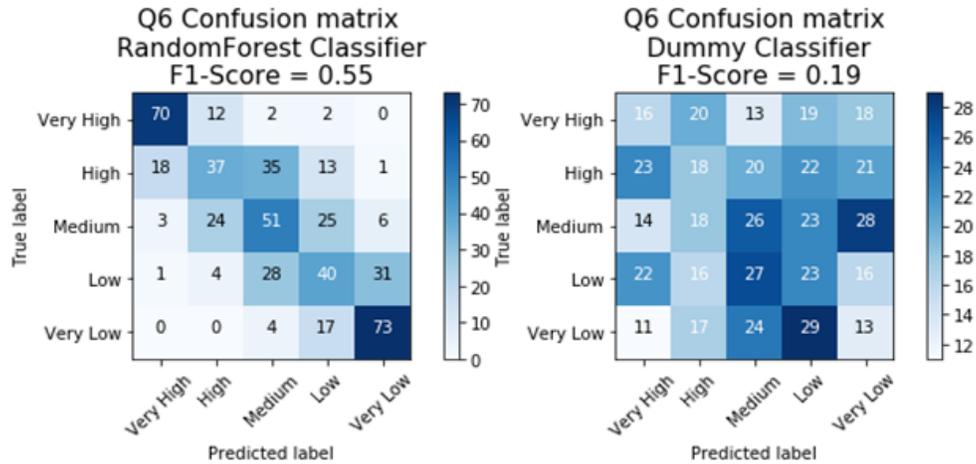

**Figure 2.12.** Confusion Matrix for the Random Forest Model vs.. Random Classifier

**Table 2.4.** Model Evaluation Metrics for Q6 Random Forest Classifier

| Class | Precision | Recall | F-Score | Support (GT) |
|---|---|---|---|---|
| Very High | 0.76 | 0.81 | 0.79 | 86 |
| High | 0.48 | 0.36 | 0.41 | 104 |
| Medium | 0.42 | 0.47 | 0.45 | 109 |
| Low | 0.41 | 0.38 | 0.40 | 104 |
| Very Low | 0.66 | 0.78 | 0.71 | 94 |

**Acknowledgement:** this research was developed with funding from the Defense Advanced Research Projects Agency (DARPA).

**Disclaimer:** The views, opinions and/or findings expressed are those of the author and should not be interpreted as representing the official views or policies of the Department of Defense or the U.S. Government.

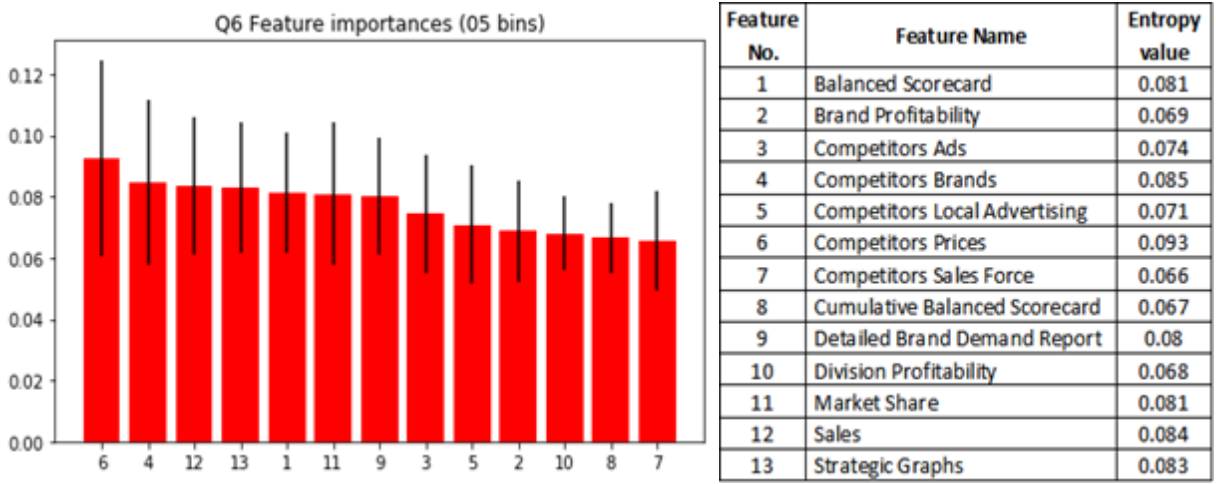

**Figure 2.13.** Feature Importance from the Random Forest Model

Features importance values of the Q6 random classifier model are shown in Figure 2.13. Similarly, importance scores are derived for all other quarters. Such feature importance measures can then be used to deduct information collection part of the situation assessment measure. We derived a formula to denote Information Collection of a team in a quarter, given feature importance $X$, where $x_1, x_2, ..., x_n$ are entropy values in that quarter and $y_1, y_2, ..., y_n$ are time spent by a team on different screens *1-n* in that quarter as follows:

$$InfoColl = \sum_{i=\{1,2,...,n\}} \frac{x_i}{\sum_n x} f(y_i - \mu_i), \text{ where}$$

$$f(y) = \frac{1}{1 + e^{-y}}$$

Here, $\mu_i$ is the mean time spent of all the teams on the screen *i* in the study and *f(y),*is the sigmoid function which takes the time spent on a time and returns the probability of team recognizing the cue from that screen.

The formula above calculates the weighted average of feature importance, i.e. weighted average time spent on specific screens with the probability of cue recognition on that screen in a quarter. This implies time spent on more important screens would be rewarded more than others. The probability of cue recognition in that screen is calculated using the *f(y)* shown above given the mean adjusted time spent on each screen by corresponding team. So, the closer to 1 (or 100) the more time they spent into cue recognition, or we can say that the higher probability they have recognized the cues.

**Acknowledgement:** this research was developed with funding from the Defense Advanced Research Projects Agency (DARPA).

**Disclaimer:** The views, opinions and/or findings expressed are those of the author and should not be interpreted as representing the official views or policies of the Department of Defense or the U.S. Government.

## 2.2.2. Measuring Situation Assessment in Daedalus

Due to the structure of the Daedalus it was hard to know which cues are important and thus needed expert coding. Therefore, we relied on Slack data qualitative coding to identify important screens or cues from Slack data. As with MPL, meaning ascription and team communication were not apparent from the Slack behavioral data, therefore we focused on information collection or cue recognition.

As discussed in Section 1.2.6, the Slack data coding allowed us to identify:

- *Relevant Cues*: Based on screens needed to solve the puzzle, this is similar to MPL where we named specific relevant screens. But since the puzzles in Daedalus all need specific screens we didn't need to use feature selection to understand screen importance as designers already know this information and encoded in the data log, allowing us to formulate this abstraction.
- *Irrelevant Cues*: Similar to relevant cues, screens that are not relevant to solving the puzzle in the stage of the game is named irrelevant.

Given these variables from the qualitative coding, we define cue recognition as a function of whether or not the players looked at the cues. Therefore, the formula for Cue Recognition is defined as:

$$\frac{\sum_j relevant(i)}{\sum_i relevant(i) + \sum_i irrelevant(i)}$$

Where *i* is a cue in the game and *j* are cues that are coded for that participant.

## 2.3. TAG (Teamwork and Adaptation in Games) Survey Instrument

The TAG survey is one of the two new measurement approaches created to measure those aspects of the Rosen et al.'s (2011) model that were *not* easily measurable via behavioral data. We elected to create a self-report survey measure for three reasons. First, a survey affords mixed-method measurement by naturally combining open and closed questions. This allows us to combine the benefits of both and alleviate the drawbacks: Closed questions can be used to assign numerical scores to participant attributes, while open questions can be used to gauge how players understand and experience the idea of adaptation. Second, related, self-report measures are a great way to tap into the subjective experiences of participants. While this can be a drawback in cases where objective data is required, we position it here as a plus: We can attempt to correlate survey scores that reflect players' subjective experiences of adaptation with our extant objective game data measures, letting us evaluate the degree to which players *think* they

Distribution Statement "A" (Approved for Public Release, Distribution Unlimited)                41
**Acknowledgement:** this research was developed with funding from the Defense Advanced Research Projects Agency (DARPA).
**Disclaimer:** The views, opinions and/or findings expressed are those of the author and should not be interpreted as representing the official views or policies of the Department of Defense or the U.S. Government.

are adaptable matches the degree to which the game says they *actually are*. Third, surveys are easy to scale up to large numbers (Evans & Mathur, 2005), and small surveys can be applied at almost any point in an experiment without being overly intrusive.

Below, we describe the iterative development and application of TAG. Three versions were made: Versions 1 and 2 were applied to the *MarketPlace Live* study, while version 3 was applied to the *Daedalus* study.

## 2.3.1 Results of using TAG with MPL

### 2.3.1.1. Fall 2017

Version 1 of the TAG was created by converting the behavioral markers for each Rosen et al.'s model phase into questions. The instrument was composed of 28 closed 5-point likert scale questions and 5 open questions, intended to solicit higher-level feedback about each player's perception of team strategy, team adaptation, team performance, and team communication. Table B.1 in Appendix B shows this version of the TAG.

The questions mirrored Rosen et al.'s model but with some adjustments. For example, the *Situation Assessment* phase includes *meaning ascription*, which has as a marker 'assigning meaning or relevance to cues.' In *MarketPlace Live*, an important set of cues is the market success of created products, as products that perform poorly must be improved or pulled. We translated this into the question 'My team quickly noticed problems that could occur in the market'.

TAG version 1 was used in the Fall 2017 *MarketPlace Live* study and was administered twice through this experience, after the fourth quarter of play (Q4), and after the sixth quarter of play (Q6). A total of 82 participants completed this survey. We analyzed the closed and open questions in several steps, described below. The data from this study as well as the results of other parts of this study is described in more detail in a later section.

The first quantitative analysis was a high-level overview of the raw question answers. The data were tested for internal consistency (data correctness and obvious outliers), and we compared medians and first and third quartiles. The main finding of this analysis is that most players reported being active *MarketPlace Live* players: Most items were either centrally distributed or skewed rightward (towards 'strongly agree'), with modes of 3-5. In all four phases of the Rosen et al. model, players were actively engaged with adaptation processes. This also means that *MarketPlace Live* as a game supports and requires adaptation processes for all four phases.

Other findings of this analysis include:

Distribution Statement "A" (Approved for Public Release, Distribution Unlimited)                    42
**Acknowledgement:** this research was developed with funding from the Defense Advanced Research Projects Agency (DARPA).
**Disclaimer:** The views, opinions and/or findings expressed are those of the author and should not be interpreted as representing the official views or policies of the Department of Defense or the U.S. Government.

- Question TAGC.17 ("One or more of my teammates showed negative feelings such as stress, anger, depression, frustration or similar emotions that negatively impacted the team") was the only question that was strongly skewed leftward. However, that question was the only negatively-written one; that this question is skewed far leftward suggests that emotional intra-team issues did not occur in most teams. This result also suggests that participants paid attention filling out the survey.
- Survey items scaled to SA, Situation Assessment, show that most players thought they were good at identifying and appraising cues.
- Survey items scaled to PF, Plan Formulation, show that most players thought they were good at setting objectives and developing strategies. A minor outlier here is TAGC.19 ("*My team came up with a list of additional tasks in case a strategy did not work*"). This suggests that a higher number of players lacked contingency planning skills.
- Survey items scaled to PE, Plan Execution, were more centrally distributed. This suggests that many players were more skilled at making plans than in following through on them, and that communicating with and giving feedback on teammates was a low priority.
- Finally, question TAGC.7 ("*My team identified the root causes of conflicts between its members*") was very centrally distributed. This suggests either that most players were not very capable of identifying internal conflicts, or that these conflicts did not occur.

We intended to use this survey to measure adaptability scoring, and thus decided to develop a factor model for the purpose of grouping the scores for adaptation. We conducted a Principal Components (PCA) analysis to understand the factors that affect this in a game like MarketPlace Live. Note that Rosen et al.'s factors were based on observations in work environments. We expected that the importance of different elements of the planning process will be different for other environments and methods, in our case a game environment and a self-report instrument, respectively. Therefore, the factors that come out of this factor analysis could show different factors than Rosen et al.'s model. PCA analysis works on the assumption that each question in a survey linearly and uniquely contributes to one composite 'factor'; this sets it apart from other types of factor analysis, which assume that one or more unidentified factors influence the (patterns of) answers given on the survey. Because our survey is designed along the same assumptions as PCA analysis, we used this approach. Eigenvalues and scree plots suggested a solution with 2, 3, or 4 factors. We found a 3-factor structure that encompassed 24 of the 28 questions:

- **TAG Factor 1: Game Strategy.** This factor contained questions that deal with aspects of playing *the game* effectively, such as 'making plans', 'articulating objectives', and 'identifying reasons for success and failure'. This factor includes the following questions:
  - TAGC.1: My team explicitly articulated its objectives.

**Acknowledgement:** this research was developed with funding from the Defense Advanced Research Projects Agency (DARPA).
**Disclaimer:** The views, opinions and/or findings expressed are those of the author and should not be interpreted as representing the official views or policies of the Department of Defense or the U.S. Government.

- TAGC.2: My team brainstormed strategies on how to play MarketPlace Live.
- TAGC.3: My team members passed information to one another in a timely and efficient manner.
- TAGC.4: My team defined potential problems it could encounter.
- TAGC.5: My team made plans for improving our performance in future quarters/encounters.
- TAGC.10: My team brainstormed new plans and strategies when it encountered problems.
- TAGC.12: My team anticipated problems that could occur in the market.
- TAGC.15: My team tried to resolve team conflicts by negotiating
- TAGC.16: My team articulated the best and worst strategies
- TAGC.17: One or more of my teammates showed negative feelings such as stress, anger, depression, frustration or similar emotions that negatively impacted the team
- TAGC.22: My team was able to see why our strategies were or were not successful thus far
- TAGC.25: My team identified the key issues for improving its performance in Marketplace Live
- TAGC.26: My team discussed errors and their causes
- **TAG Factor 2: Team Planning.** This factor contained questions that deal with team interaction, teamwork, and team planning.
  - TAGC.6: My team came up with a list of tasks to accomplish its objectives
  - TAGC.11: My team maintained and updated a list of to-dos, needs, and objectives
  - TAGC.14: I was assigned specific tasks according to my abilities
  - TAGC.19: My team came up with a list of additional ideas in case a strategy did not work
  - TAGC.23: My team explicitly outlined everybody's responsibilities
- **TAG Factor 3: Feedback.** This factor contained questions that describe or contain the concept of feedback.
  - TAGC.8: I asked my teammates for feedback regarding my performance
  - TAGC.13: I received feedback from my teammates about my performance
  - TAGC.18: I analyzed team member contributions to identify their errors
  - TAGC.20: I provided feedback to teammates about their performance
  - TAGC.24: I accepted suggestions from teammates about how I could improve my performance
  - TAGC.28: When offered feedback, I was able to identify what would help me improve my performance

**Acknowledgement:** this research was developed with funding from the Defense Advanced Research Projects Agency (DARPA).

**Disclaimer:** The views, opinions and/or findings expressed are those of the author and should not be interpreted as representing the official views or policies of the Department of Defense or the U.S. Government.

- **Questions kept separate.** One question was not included in any factors, but still considered interesting enough to include as a stand-alone question.
  - TAGC.9: My team referred to past strategies when making decisions

The open questions were analyzed using an open coding approach. Four researchers developed a code book by first open-ended coding on a test set of 10 participants, independently creating new codes as needed. Brainstorming sessions were held to combine these independent code lists, condensing similar codes and scrapping codes that researchers did not agree on. A two-level code hierarchy was created and filled out, resulting in a code book of **51 codes** sorted in three categories: **Strategy**, **Adaptation**, and **Teamwork**. Three coders then used this code book to code the open question answers from 78 participants (4 less than the number of quantitative answers, since four participants did not answer the open questions). Before full coding, inter-rater reliability was determined using the Fleiss-Kappa measure. We found an inter-rater reliability of $k=.53$ ($SEk=.367$). While is this still relatively low (values of .7 or higher would generally be considered acceptable), we decided to proceed due to time limits and the exploratory nature of our work. Given more time / future work, we can revisit this coding process with the aim of obtaining higher IRR ratings, and then re-validating results.

We then used heuristic analysis to reduce the code base of 51 codes to the 16 'most interesting' ones, by looking at which codes were used often, which codes were used in concert, and which codes formed interesting logical groupings. This resulted in X sets of codes:

- **"No Strategy".** This code (literally "No Strategy") was deemed interesting because a surprising number of players flat-out mentioned 'not having a strategy' -- usually right before describing their strategy. It suggests that directly asking players what their 'strategy' was might be a flawed approach, as some players might not be able to answer the question.
- **Adjustment Targets:** Adjust Advertising, Adjust Product, Adjust Price(s), Adjust Target Segment(s). These codes describe the most important elements in *MarketPlace Live* when it comes to adaptation. Interesting here is that the codes Adjust Sales Force and Adjust Location did not make the cut, as they were not used often enough. This tells us that while *MarketPlace Live* puts several gameplay elements on equal footing, players quickly decide / learn that some of them are worth focusing on more than others. Alternatively, it could be that there simply is not any interesting adaptation in these two elements -- or that players just forgot about them.
- **Adjustment Reasons:** Adjust Based On Own Results, Adjust Based On Competitors' Results, Adjust Based On Competitors' Actions, Adjust Based On Market Reaction, Copy Competitors. These codes describe the reasons players gave for making changes to

**Acknowledgement:** this research was developed with funding from the Defense Advanced Research Projects Agency (DARPA).
**Disclaimer:** The views, opinions and/or findings expressed are those of the author and should not be interpreted as representing the official views or policies of the Department of Defense or the U.S. Government.

their strategy, abstracted into higher levels. Of these, Copy Competitors is the only 'clear' strategy: This code was used in conjunction with Adjust Based On Competitors' Actions often to indicate that teams would literally copy what they saw their competitors do. Other strategies were generally kept to the higher levels: Players would that *that* they adjusted their products or looked at their competitors' results, but not *how*. This might be a drawback in our style of question asking.

- **Organization Style:** Face-To-Face Meetings, Group Discussion, Group In Agreement, Making Decisions As A Group, Mediated Communication, Voting. These codes show that the vast majority of teams preferred an equitable style of decision making. Codes like Leader And Followers were coded, but during the analysis only showed up a handful of times. This might be due to the fact that our *MarketPlace Live* play sessions were incorporated into student classes, where all participants were (assumed to be) on more-or-less equal skill levels. Perhaps in different social contexts, other styles of organization would be more prevalent.

In summary, we see that the open coding approach presented here can be used both to specifically characterize the game context of choice, and to provide more generalizable insights into the way the game is played. We also believe that it will be possible to do follow-up, perhaps mixed-method analyses to compare the distribution or application of certain codes to behavioral data or factor scores. An early attempt at this kind of analysis was started, but low participant counts, and lack of devoted time has halted this.

*2.3.1.2. Spring 2018*
Based on the Fall 2017 factor analysis results, we decided to change the TAG survey to bring it in line with the new factors we found. Questions TAGC.7 (*My team identified the root causes of conflicts between its members*), TAGC.20 (*My team conversed without using sarcastic or non-constructive comments*), and TAGC.27 (*My team attempted to provide opportunities for socializing*) were deleted because they were either better interpreted from our behavioral data or did not apply to a game, such as TAGC.27 which targets socialization. Gallup further reviewed the survey and some of the wording was changed as a result. We also added one additional open question to better capture an individual's role within their team. This resulted in version 2 of the TAG, shown in Table B.2 in the Appendix.

A total of 210 participants filled out this version of the TAG, which was applied after Q4 and after Q6, as done in the previous study. Data were again tested for consistency.

*2.3.1.3 Take-aways from using TAG for MarketPlace Live (2017 and 2018)*
Most questions got a mode score of 4-5, with the exception to the following questions:

**Acknowledgement:** this research was developed with funding from the Defense Advanced Research Projects Agency (DARPA).
**Disclaimer:** The views, opinions and/or findings expressed are those of the author and should not be interpreted as representing the official views or policies of the Department of Defense or the U.S. Government.

- TAGC.8 - I asked my teammates for feedback regarding my performance: received mode of 3
- TAGC.13 & TAGC.20 - I received feedback from my teammates for self-correction & I provided feedback to teammates to facilitate self-correction: received mode of 2 and 2, respectively.
- TAGC.17 - One or more of my teammates showed negative feelings such as stress, anger, depression, frustration or similar emotions that negatively impacted the team: received mode of 1.
- TAGC.18 - I analyzed team member contributions to identify their errors: received mode of 3.

These questions concern feedback of performance and self-correction and errors, indicating that playing the game lacked reflection and feedback on individual performance. The answers also showed that aspects of negative feeling were not expressed through the game experience.

Due to time constraints, we did not complete the coding of qualitative data from Slack or open questions and we will keep that for future work.

## 2.3.2. Applying TAG to ARG

We used version 2 of TAG for the Summer/Fall 2018 *Daedalus* study. The closed questions in this version were functionally identical to version 2 (as seen in Appendix B), except with all references to *MarketPlace Live* replaced with *Daedalus* (and assorted grammar changes). The open questions were removed from this version, as we wanted to focus on Slack chat data for qualitative analysis instead.

A total of 124 participants filled out version 3 of the TAG. This version was only applied once, at the end of the *Daedalus* game: Since players played *Daedalus* at different paces, we could not put the questionnaire at a set point in time (i.e. 'after two days'), since players could be at any point in the game by then, or even done. And since *Daedalus* is time-driven, we were worried that putting the questionnaire at certain game milestones (i.e. 'when players reach phase three') would incentivize players to rush through the questions instead of giving accurate answers. Correspondingly, only the game milestone 'end of game' was seen as a clear, reliable measurement moment. The results for Daedalus is different from the *MarketPlace Live* results. In particular, the results are as follows:

- Questions related to feedback, self-correction and error (TAGC.8, TAGC.13, TAGC.17, TAGC.18, and TAGC.20) all had low scores similar to the MPL, all receiving a mode 1.

**Acknowledgement:** this research was developed with funding from the Defense Advanced Research Projects Agency (DARPA).

**Disclaimer:** The views, opinions and/or findings expressed are those of the author and should not be interpreted as representing the official views or policies of the Department of Defense or the U.S. Government.

- TAGC.5 - My team made plans for improving our performance in future puzzles, receiving a mode of 2.
- TAGC.11 - My team brainstormed new plans and strategies when it encountered problems, receiving a mode of 1
- TAGC.14 - I was assigned specific tasks according to my abilities, receiving a mode of 1
- TAGC.15 - My team tried to resolve team conflicts by negotiating receiving a mode of 3
- TAGC.16 - My team articulated the best and worst strategies, receiving a mode of 3
- TAGC.23 - My team explicitly outlined everybody's responsibilities, receiving a mode of 1
- TAGC.24 - I accepted suggestions from teammates about how I could improve my performance, receiving a mode of 3
- TAGC.28 - When offered feedback, I was able to identify what would help me improve my performance, receiving a mode of 3

This shows that in addition to feedback, reflection and error correction, which were also lacking in the *Daedalus* game, participants in the *Daedalus* game did not express that they planned from puzzle to puzzle (TAGC.5), brainstormed problems together (TAGC.11), or resolved conflict (TAGC.15). They agreed much less with how the game allowed them to improve through feedback or through teammates (TAGC.24 and TAGC.28). Furthermore, the game does not require or assign roles, and, as a result, the participants did not respond positively to that (TAGC.14 or TAGC.23). This does indicate that participants paid attention to filling out the survey as this difference was expected.

## 2.3.3. Histogram Investigations for Questions Scores between Daedalus and MPL

We further investigated the score distributions with a series of statistical tests. For this analysis we used the results from version 2 of the TAG survey (MarketPlace Live Spring 2018 *study*) and Daedalus *study* to make sure the questions and answers are comparable. **Mann-Whitney tests show that all results are statistically significant, except for TAGC.17** ("One or more of my teammates showed negative feelings such as stress, anger, depression, frustration or similar emotions that negatively impacted the team.")

We discuss the meaning of these differences underneath each histogram *below*. Several questions particularly show major differences between Daedalus and MarketPlace Live, which we believe demonstrates the significant difference in task environment between these two games. That we can find these differences here indicates that the TAG is a useful instrument for characterizing these different task environments, but also that this needs to be taken into account when applying

**Acknowledgement:** this research was developed with funding from the Defense Advanced Research Projects Agency (DARPA).

**Disclaimer:** The views, opinions and/or findings expressed are those of the author and should not be interpreted as representing the official views or policies of the Department of Defense or the U.S. Government.

the instrument. Orange histogram bars show results for the Spring 2018 MarketPlace Live study; blue histogram bars show results for the Daedalus study; and brown histogram bars indicate overlap.

Non-significant results:

**TAGC.17**: One or more of my teammates showed negative feelings such as stress, anger, depression, frustration or similar emotions that negatively impacted the team. *<Game Strategy>*

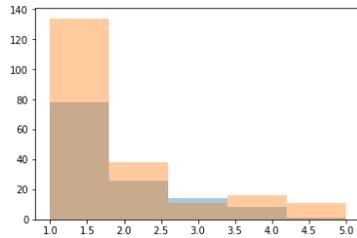

The non-significant, almost identical distributions for this question suggest that all players across all games were able to work together without letting tensions run high.

Significant results showing similar answer patterns:

**TAGC.1**: My team explicitly articulated its objectives. *<Game Strategy>*

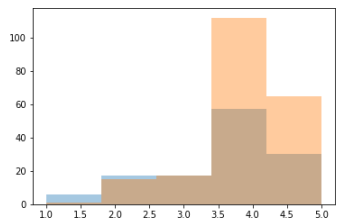

The difference between 'players who scored 1-3' and 'players who scored 4-5' in this question is sharper for *MarketPlace Live* than for *Daedalus*. This indicates that *MarketPlace Live* players as a group were much more likely to explicitly set objectives, but that a large section of *Daedalus* players did also (try to) do this.

**TAGC.2**: My team brainstormed strategies on how to play MarketPlace Live/Daedalus. *<Game Strategy>*

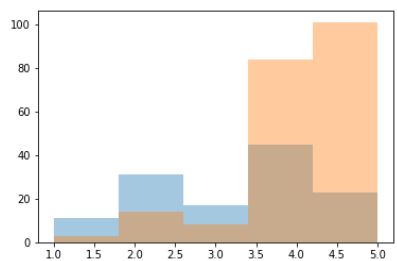

**Acknowledgement:** this research was developed with funding from the Defense Advanced Research Projects Agency (DARPA).

**Disclaimer:** The views, opinions and/or findings expressed are those of the author and should not be interpreted as representing the official views or policies of the Department of Defense or the U.S. Government.

Results here show that *MarketPlace Live* players were overwhelmingly more likely to brainstorm than *Daedalus* players. Since *MarketPlace Live* revolves around players interpreting complete-but-uncertain market data, while *Daedalus* revolves around adapting to new and uncertain puzzle scenarios, this is not unexpected; while certain *Daedalus* puzzles (like the escape rooms) likely afford brainstorming, other puzzles (like taking pictures around Boston) do not.

**TAGC.3**: My team members passed information to one another in an efficient manner. *<Game Strategy>*

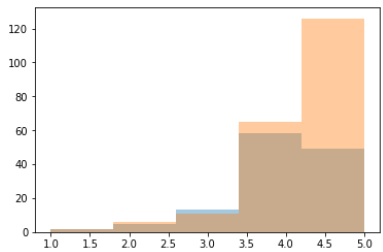

*MarketPlace Live* scores mode=5, while *Daedalus* scores mode=4; both groups of players overwhelmingly reported being efficient at information-passing, but *MarketPlace Live* players did so more.

**TAGC.4**: My team identified potential problems it could encounter while playing Marketplace Live/Daedalus. *<Game Strategy>*

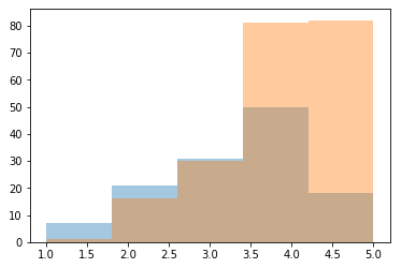

*MarketPlace Live* players reported a significantly higher degree of preemptive problem identification. As with TAGC.2, this is likely because *MarketPlace Live* both affords and centers this kind of preparatory problem solving, while *Daedalus* centers reacting to the unknown.

**TAGC.10**: My team brainstormed new plans and strategies when it encountered problems. *<Game Strategy>*

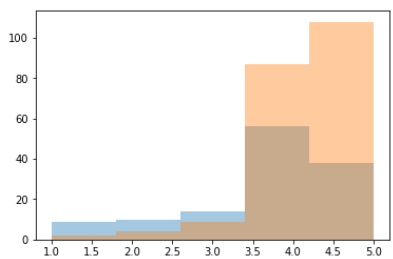

**Acknowledgement:** this research was developed with funding from the Defense Advanced Research Projects Agency (DARPA).

**Disclaimer:** The views, opinions and/or findings expressed are those of the author and should not be interpreted as representing the official views or policies of the Department of Defense or the U.S. Government.

Similar to TACG.2, this score discrepancy reflects how *MarketPlace Live* affords and encourages brainstorming throughout the length of gameplay, while *Daedalus* has several sections that do not really afford brainstorming at all.

**TAGC.12**: My team anticipated problems that could occur in the market/while playing Daedalus. *<Game Strategy>*

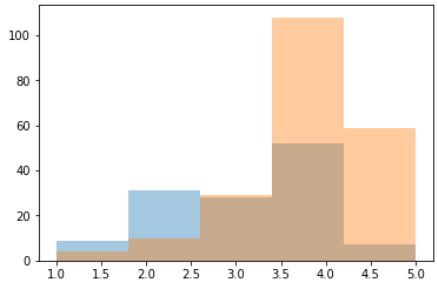

Again, we see here that *MarketPlace Live* affords and encourages preemptive planning, while *Daedalus* does not. Predicting future problems is a core skill set of being good at *MarketPlace Live*, while in *Daedalus* this skill set is only occasionally tested; it is functionally impossible to 'predict' what the next set of puzzles will be like.

**TAGC.19**: My team came up with a list of additional ideas in case a strategy did not work. *<Team Planning>*

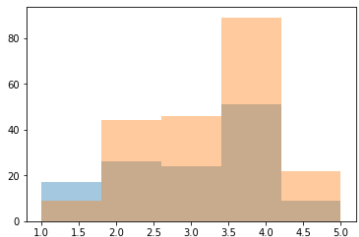

*MarketPlace Live* players were a little more likely to engage in contingency planning than *Daedalus* players, but both groups of players were similarly ambiguous about the idea.

**TAGC.20**: I provided feedback to teammates about their performance. *<Feedback>*

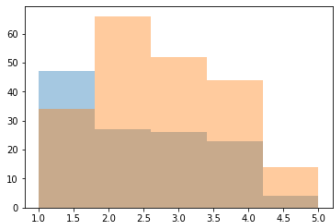

Similarly, both *MarketPlace Live* and *Daedalus* players were mostly ambiguous to the idea of providing feedback to teammates. This is likely because neither game has strong explicit markers for individual player performance. Since *MarketPlace Live* does have several clear measurement/'taking stock' moments (the lulls between quarters), it is not surprising to see a

Distribution Statement "A" (Approved for Public Release, Distribution Unlimited)     51

**Acknowledgement:** this research was developed with funding from the Defense Advanced Research Projects Agency (DARPA).

**Disclaimer:** The views, opinions and/or findings expressed are those of the author and should not be interpreted as representing the official views or policies of the Department of Defense or the U.S. Government.

slightly stronger trend in favor of providing feedback in this game.

**TAGC.22**: My team was able to see why our strategies were or were not successful thus far. *<Game Strategy>*

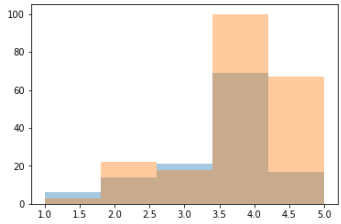

In both games, success and failure are obvious and explicit states -- scoring low in *MarketPlace Live* or failing to solve a puzzle in *Daedalus* -- which explains the similar pattern. *MarketPlace Live* does give more explicit information about *why* an approach was unsuccessful, which is why the percentage of 5 scores is higher than *Daedalus*; as a puzzle game, *Daedalus* by design does not give this feedback.

**TAGC.26**: My team discussed errors and their causes. *<Game Strategy>*

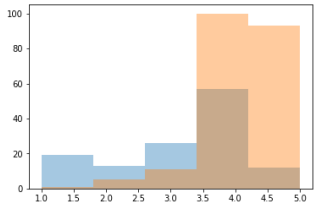

Related to the above TAGC.22, *Daedalus* does not have 'errors' in strategy so much as it has 'failures to proceed'. *MarketPlace Live* allows players to make *errors*, which can be discussed.

Significant results showing different answer patterns:

**TAGC.5**: My team created strategies for improving our performance in future quarters/later challenges. *<Game Strategy>*

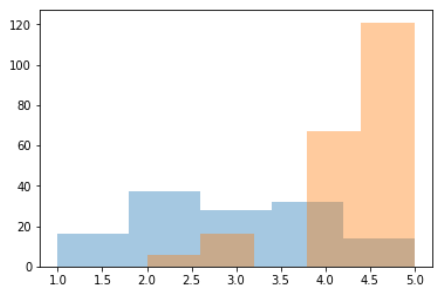

The significant difference (MarketPlace Live mode = 5, Daedalus mode = 2) here is a reflection of the intrinsically different nature of the two game environments. As discussed, *MarketPlace Live* allows players space between quarters to reflect, and functionally demands that players do so in order to be good at the game -- players need to take in all their own information, the

**Acknowledgement:** this research was developed with funding from the Defense Advanced Research Projects Agency (DARPA).
**Disclaimer:** The views, opinions and/or findings expressed are those of the author and should not be interpreted as representing the official views or policies of the Department of Defense or the U.S. Government.

information of all competitors, and possible market changes. This is explicitly the intended approach. In contrast, *Daedalus* encourages players to solve the game as quickly as possible. There are no game-mandated break moments, time is at a premium, and there is not even all that much consistent information between challenges that players can reflect on. While certain teams were observed to try and 'take everything in' between challenge sets, it is not surprising that the *Daedalus* answers to this question are scattershot, while the *MarketPlace Live* answers skew heavily to the right.

To characterize this even further, we can look at the open question TAGO.2: "*How did your team's strategies change during play? And why?*". Answers to this question included:

- *"We focused on where we did poorly and how our competitors did in comparison. Then adjusted accordingly."*
- *"Yes, because in a new quarter, our performance level isn't great at all, so we have to come up with a new brand, ad campaigns, and so on."*
- *"We copied some of competitors products to pass the consumer union."*
- *"We changed our strategies based on results. We first were concerned about having high costs, so we were cautious about how many ads we placed and whether we re-designed products. However, this hurt us. We instead decided to spend as much as we needed to improve our products and advertising."*
- *"No, since our strategy was results-driven."*

We see from the above that *MarketPlace Live* play revolves around this adaptive strategy planning. In contrast, the *Daedalus* Slack chats contain almost no long-term planning; those messages that were coded as 'strategy planning' were almost always about short-term tactics.

- *"I'm free between 6:30 to 8:30 pacific to help troubleshoot the cyber garden if you need it."*
- *"Yep. Just let us know if those don't work, otherwise when you find one that does keep typing it."*
- *"If we get stuck it could be worth pulling those out."*

**TAGC.6**: My team came up with a list of tasks to accomplish its objectives. <*Team Planning*>

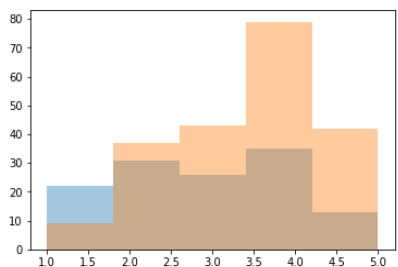

Difference here is MarketPlace Live mode = 5, Daedalus mode = 4. This question again reflects that *MarketPlace Live* affords long-term planning in a way that *Daedalus* does not. Both games

**Acknowledgement:** this research was developed with funding from the Defense Advanced Research Projects Agency (DARPA).

**Disclaimer:** The views, opinions and/or findings expressed are those of the author and should not be interpreted as representing the official views or policies of the Department of Defense or the U.S. Government.

do support task planning, but *MarketPlace Live* places it front and center while *Daedalus* encourages players to be fast.Neutral for the ARG vs. left skewed for MPL indicating that the ARK does not really have much of lists of tasks or planning.

**TAGC.8**: I asked my teammates for feedback regarding my performance. *<Feedback>*

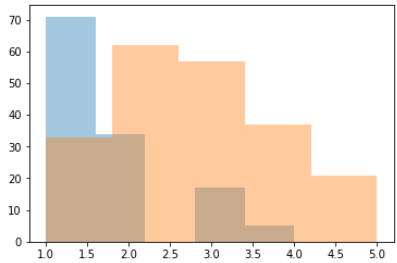

These answer patterns (MarketPlace Live mode = 3, Daedalus mode = 1) reflect an intrinsic difference in the two game environments: *MarketPlace Live* is about handling complete-but-uncertain data to achieve fuzzy outcomes, while *Daedalus* is about solving puzzles. The only 'performance' metrics that *Daedalus* has are puzzle completion (which feedback does not apply to) and speed (to which the only feedback can be 'do it faster'). In contrast, it is possible for players to make wrong calls in *MarketPlace Live*, misinterpret data, or hold different ideas about the importance of certain mechanics to other players. All of these are things players could consider requesting feedback about. An example of feedback-requesting behavior from the Spring 2018 Slack chats follows: Underlined text is message identifiers, italicized text is chat messages, bolded text is actions.

TeamMember-1
*open new office in paris? could capture the entire market*
*additional advertisements also*

TeamMember-2
**uploaded this image: Screen Shot 2018-02-08 at 1.54.32 PM.png**

TeamMember-2
*use this to create new ads*

TeamMember-1
*changing advertisement: replacing pictures of engineers with features of engineering apps*

TeamMember-1
*also added "highest rated brand: mercedes"*

The only similar behavior we can see in *Daedalus* is players asking for help when getting stuck on puzzles, which is only tangentially related to the intent of this question:

Distribution Statement "A" (Approved for Public Release, Distribution Unlimited)                    54
**Acknowledgement:** this research was developed with funding from the Defense Advanced Research Projects Agency (DARPA).
**Disclaimer:** The views, opinions and/or findings expressed are those of the author and should not be interpreted as representing the official views or policies of the Department of Defense or the U.S. Government.

- *New ideas? Or should I just cycle around the monuments?*
- *How / where to make for quicker texts?*
- *Yes, entered the pattern and also got the cypher wheel, all of you already have this?*

This also indicates that the ARG compared to MarketPlace Live does not emphasize taking feedback and reflection.

**TAGC.9**: My team referred to past strategies when making decisions. *<no factor>*

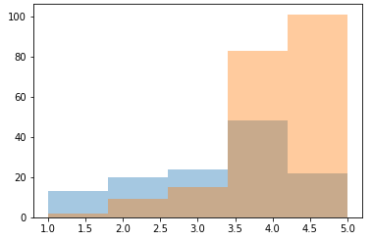

The differences here are (MarketPlace Live mode = 5, Daedalus mode = 4). *MarketPlace Live* is explicitly designed as one long running simulation: Decisions made in earlier quarters persist and reverberate through later quarters. In contrast, four of *Daedalus*' five phases are entirely different, with only phases 1 and 4 having mechanical similarities (i.e,, both are 'escape room' puzzle sections). That *Daedalus* ranks as highly in 4 as it does seems to indicate that players were definitely *looking* to use past information in future challenges, even if this was barely supported.<text>

**TAGC.12**: My team periodically updated a list of to-dos, needs, and objectives. *<Team Planning>*

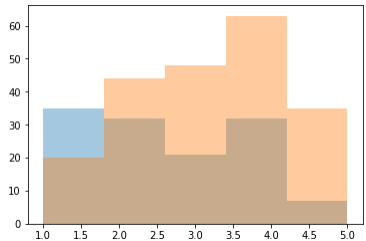

The differences here are (MarketPlace Live mode = 4, Daedalus mode = 1). The spread of scores here seems to indicate that players of both teams were not uniformly interested in formal task planning. *MarketPlace Live* players can draw on past activities and performance during downtime to plan for the future, while *Daedalus* players must be fast and adaptable to unknown new circumstances.

**TAGC.14**: I received feedback from my teammates about my performance. *<Feedback>*

**Acknowledgement:** this research was developed with funding from the Defense Advanced Research Projects Agency (DARPA).

**Disclaimer:** The views, opinions and/or findings expressed are those of the author and should not be interpreted as representing the official views or policies of the Department of Defense or the U.S. Government.

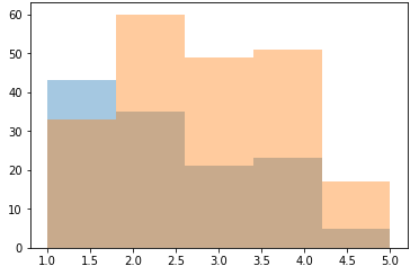

The differences here are (MarketPlace Live mode = 2, Daedalus mode = 1). This question text and answer pattern is similar to TAGC.8, another question about feedback, and the conclusions drawn there apply here as well. The most significant difference is the wider range of scores for *Daedalus*, which in TAGC.8 was almost exclusively 1-2; this is likely because *giving feedback to others* is a natural extension of talking over Slack and discussing puzzle progress, while *requesting feedback from others* is a conscious decision that fits that game environment less well.

**TAGC.15**: I was assigned specific tasks according to my abilities. *<Team Planning>*

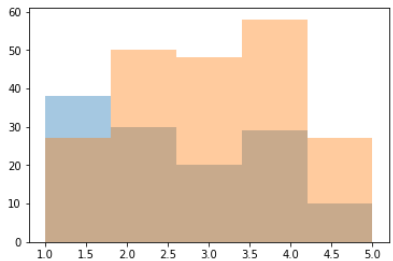

The differences here are (MarketPlace Live mode = 4, Daedalus mode = 1). The most interesting single outcome about this question is the fact that *MarketPlace Live* scores so centrally: The *MarketPlace Live* game explicitly makes people select certain 'play roles', and while that has no mechanical effect on which players can perform which tasks (all players can do anything), it would not be strange to expect that players followed these roles. It is possible that the phrase '*I was assigned*' in this question throws off that expectation, as players may have *taken on* tasks in accordance with their abilities instead. In *Daedalus*, every player had to complete every task in order to complete the game; the only real possible 'task division' could be 'which players solves which puzzle first'. *Daedalus*' different puzzles did task different skill sets, but from chat data we saw that most teams had one or several 'dedicated problem solvers' who blazed a trail for others to follow. Again, very little centralized 'task assigning' was seen.

**TAGC.16**: My team tried to resolve team conflicts by negotiating. *<Game Strategy>*

**Acknowledgement:** this research was developed with funding from the Defense Advanced Research Projects Agency (DARPA).

**Disclaimer:** The views, opinions and/or findings expressed are those of the author and should not be interpreted as representing the official views or policies of the Department of Defense or the U.S. Government.

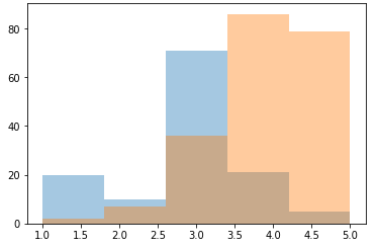

The differences here are (MarketPlace Live mode = 4, Daedalus mode = 3). The difference in score here is most likely because there were almost no observable conflicts in *Daedalus*. For most teams, the only observable 'conflict' was at the stage 5 prisoner's dilemma, which generally saw some negotiating about how all players should keep sharing (from the *Daedalus* slack chats):

- *But we need to do 3 rounds all voting share.*
- *Botkit updated and I shared :).*
- *Yes we're all sharing and if you don't you're a cop.*

While no team conflicts were seen in the Spring 2018 Slack chats, it is possible that players interpreted this question to mean 'conflicts in envisioned strategy', i.e. which approach they should be taking. Similar to TAGC.5 earlier, the fact that *MarketPlace Live* affords different strategies based on player interpretation of the game state would allow for these conflicts to arise, and give players both time and incentive to resolve them before the next quarter.

**TAGC.16**: My team articulated the best and worst strategies. <*Game Strategy*>

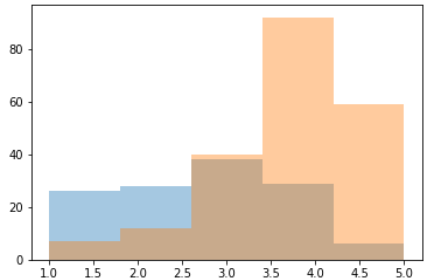

The differences here are (MarketPlace Live mode = 4, Daedalus mode = 3). Similar to TAGC.5/TAGC.8, the only real strategy in *Daedalus* was 'do everything correct as fast as you can'. Certain puzzle stages, particularly the stage 2 Tamagotchi puzzles, did enforce divergent strategies: Different players had to provide different emojis and were given different button layouts, generally forcing them to talk to each other to figure this out (*Daedalus* slack chat):

- *How did you just hatch it??*
- *I fed it :fire: emojis!*
- *But it looks like each egg wants a very specific temperature. Mine needed 85% exactly.*
- *They reset it for me, my snake preferred sun to fire, which is think is wise.*

Distribution Statement "A" (Approved for Public Release, Distribution Unlimited)      57
**Acknowledgement:** this research was developed with funding from the Defense Advanced Research Projects Agency (DARPA).

**Disclaimer:** The views, opinions and/or findings expressed are those of the author and should not be interpreted as representing the official views or policies of the Department of Defense or the U.S. Government.

In contrast, *MarketPlace Live* players would learn good and bad strategies based on inter-quarter feedback, with their results mostly depending on the results and choices of their competitors. This resulted in discussions like the following (a discussion fragment from one Spring 2018 Slack chat):

TeamMember-1
*yeah but i think compared to our competitors we overdid it a bit with the advertising*
*we should add bookkeeping software too, because our competitors all had that and we should have all the features or more that they do*

TeamMember-4
*we also need to do what our competitors are doing in the ads and brand because it is working for them so for our final price do you guys agree on $3300 for 100 rebate*

TeamMember-3
*Quarter 4 decisions: Two new brands one targeting both the mercedes group and the workhorse and one brand that targets mainly the workhorse. New brands are both lower priced than other in the industry but have the same amount of features, giving us a price advantage specifically to workhorse because we are the cheaper option. Split the sales people in half so we have people also selling to the workhorse target. Decided to open a new office in Paris because they have a larger workhorse market and they also have a large mercedes market.*

TeamMember-1
*im looking at the copmetitor profiles and im wondering if we should open a new office or not this quarter*

**TAGC.18**: I analyzed team member contributions to identify their errors. *<Feedback>*

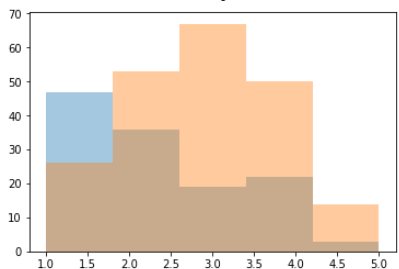

The differences here are (MarketPlace Live mode = 3, Daedalus mode = 1). This question is a counterpart to TAGC.13 (except asking players if they *gave* feedback instead of *received* it), with almost identical answer patterns. The same conclusions apply here.

**TAGC.23**: My team explicitly outlined everybody's responsibilities. *<Team Planning>*

**Acknowledgement:** this research was developed with funding from the Defense Advanced Research Projects Agency (DARPA).

**Disclaimer:** The views, opinions and/or findings expressed are those of the author and should not be interpreted as representing the official views or policies of the Department of Defense or the U.S. Government.

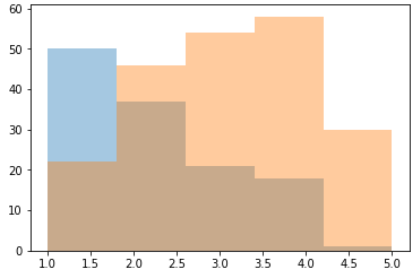

The differences here are (MarketPlace Live mode = 4, Daedalus mode = 1). This question combines elements of TAGC.11 ("*My team periodically updated a list of to-dos, needs, and objectives*") and TAGC.14 ("*I was assigned specific tasks according to my abilities*"). Similar conclusions can be drawn here: *Daedalus* simply did not support this level of careful planning and distribution of roles, while *MarketPlace Live* did.

**TAGC.22**: I accepted suggestions from teammates about how I could improve my performance. *<Feedback>*

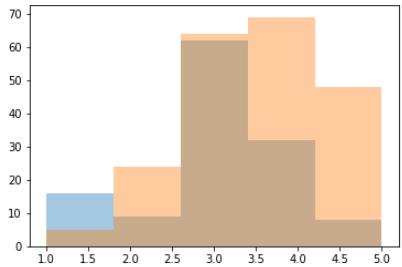

This question is almost identical in meaning to TAGC.13 ("*I received feedback from my teammates about my performance*"), except filtered through a lens of action and social obligation ('I received feedback' versus 'I put that feedback in action'). In that light, almost the most interesting outcome is how much more this question skews to 3-5 for both games, as opposed to TAGC.13's 1-4 distribution; nobody wants to outright say they *rejected* teammate suggestions.

**TAGC.25**: My team identified the key issues for improving its performance in Marketplace Live/Daedalus. *<Game Strategy>*

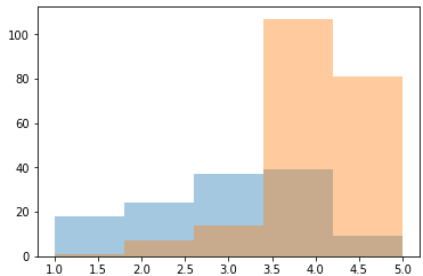

**Acknowledgement:** this research was developed with funding from the Defense Advanced Research Projects Agency (DARPA).

As established, *MarketPlace Live* has a range of success factors and potential issues that different teams will need to address, based on different play groups and the actions of their competitors, as well as their increasing understanding of the game logic and the behavior of the simulated world. The following Spring 2018 Slack chat demonstrates:

TeamMember-1

*Also an important think [sic] in designing the brand is to remember, while it is important to give the customer what they want, we also have to give them something else. Something they don't know that they want. That can distinguish us from the competition. Idk if you guys took Innovation yet, but it is important to remember that the "customer has no idea what they want. It is the company's job to deliver on things they didn't know they wanted."*

The centralized answer distribution for *Daedalus* could point to two things: Either players did not identify any 'key issues' beyond needing to solve puzzles more quickly, or players just did not perceive any issues at all. Since *Daedalus* frames its challenges as puzzles, it is likely that players did not see their problems as 'issues' as much as 'fun challenges to overcome'.

- *Does our adventuring team have any insight into which book/page/line might tell you how to hatch an egg?*
- *If there's still an egg on the table, try to hatch it before you go to the book table!! We still need either <@UC6S8M2EP> or <@UC69RPFJL> to hatch one for a +10h time bonus!!*

**TAGC.28**: When offered feedback, I was able to identify what would help me improve my performance. *<Game Strategy>*

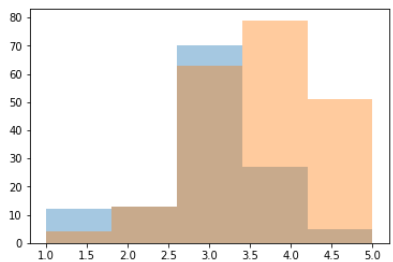

The differences here are (MarketPlace Live mode = 4, Daedalus mode = 3). The similar question text and similar answer pattern to TAGC.22 suggests that this question falls prey to similar social desirability effects: Nobody want to admit they did not understand their teammates' feedback. Alternatively, the centralized skew for *Daedalus* could mean that most players intended to answer this 'neutrally', i.e. getting and interpreting feedback was not really a factor in play. In contrast, *MarketPlace Live* tests a number of different economic skills and understanding of different fields of theory; feedback from other players with a better understanding of the field you are working on could very well be useful.

Distribution Statement "A" (Approved for Public Release, Distribution Unlimited)          60
**Acknowledgement:** this research was developed with funding from the Defense Advanced Research Projects Agency (DARPA).
**Disclaimer:** The views, opinions and/or findings expressed are those of the author and should not be interpreted as representing the official views or policies of the Department of Defense or the U.S. Government.

### 2.3.4. Histogram Investigations for Factor Scores between Daedalus and MPL

We also ran similar analysis on the factors scores for *MarketPlace Live* Spring 2018 (again shown in orange) and *Daedalus* (in blue). Below we discuss the details of this analysis.

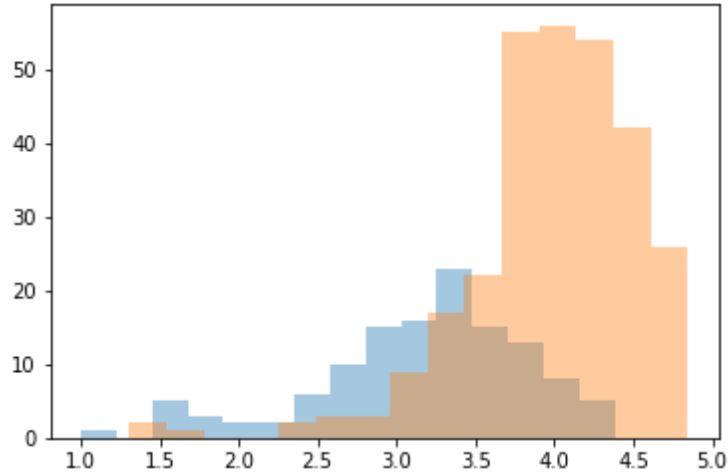

**Figure 2.15.** Histogram of TAG Factor: Game Strategy

This histogram (Figure 2.15) shows that Game Strategy skews strongly towards the right for *MarketPlace Live* while being more centralized for *Daedalus*. The Game Strategy factor encompasses elements, such as explicitly articulating objectives (TAGC.1), making plans for the future (TAGC.5), negotiating as a team (TAGC.15), and identifying causes of failure and success (TAGC.22). All of these elements are much more prevalent in *MarketPlace Live* than they are in *Daedalus*.This echoes the recurring findings throughout this analysis that *MarketPlace Live* required careful planning, data interpretation, and analysis, while *Daedalus* prioritized speed and fast throughput.

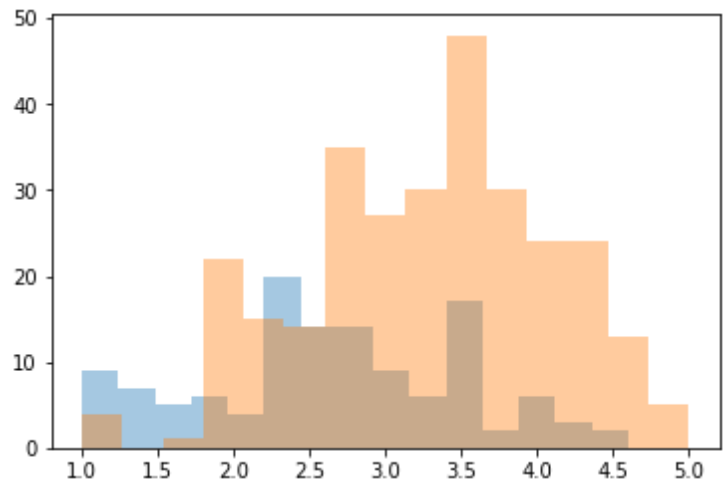

**Acknowledgement:** this research was developed with funding from the Defense Advanced Research Projects Agency (DARPA).

**Disclaimer:** The views, opinions and/or findings expressed are those of the author and should not be interpreted as representing the official views or policies of the Department of Defense or the U.S. Government.

**Figure 2.16.** Histogram of TAG Factor: Team Planning

The Team Planning factor histogram shown in Figure 2.16 is more scattered for both games, but we see that the *MarketPlace Live* score is more centralized and leaning towards the right. As addressed in the individual examples above, the design of *Daedalus* did not strongly encourage players to work together in game mechanics or to diversify roles; instead, the faster players would generally act as 'trailblazers' solving and explaining the puzzles for the slower players. See for instance the discussion at TAGC.15. Conversely, *MarketPlace Live* encouraged players to take on different roles and responsibilities and engage with different game mechanics. Even if there were no mechanical systems forcing players apart (it was possible for one player to make all decisions and, in fact, play the game alone) we noticed that players would commonly organize themselves along 'role' lines. Team Planning also incorporates elements like preemptive planning (TAGC.11) and contingency planning (TAGC.19), which exist in both games, but much more prevalently in *MarketPlace Live*.

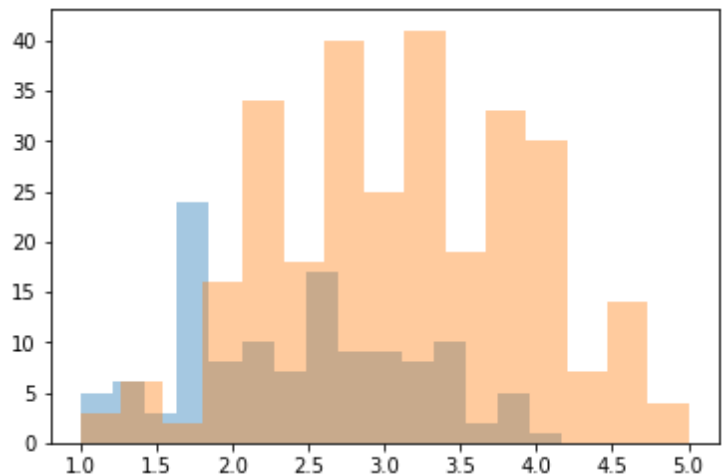

**Figure 2.17.** Histogram of TAG Factor: Feedback

The Feedback factor shown in Figure 2.17 is skewed towards the left for *Daedalus*, but all over the place for *MarketPlace Live*. This reinforces the finding that *Daedalus* did not really incorporate 'feedback' as an element. The goal was to solve puzzles, and to get the entire team to complete puzzles, meaning there was both little room and little incentive to spend time giving performance feedback. See for instance the discussion around TAGC.8, TACG.14, and TAGC.18. *MarketPlace Live* did incentivize feedback, but not all teams made use of this possibility. For instance, compare the following answers on TAGO.4, "*How did your team decide which actions to take?*":

**Acknowledgement:** this research was developed with funding from the Defense Advanced Research Projects Agency (DARPA).

**Disclaimer:** The views, opinions and/or findings expressed are those of the author and should not be interpreted as representing the official views or policies of the Department of Defense or the U.S. Government.

- *One of my team members was very assertive, so they would share what they thought was best. The other members would either agree or show why they disagreed. This was perfect because we always had an idea to work off of and some great feedback and dialogue.*
- *We had discussions with anyone's ideas and usually came to a quick conclusion on whether it would achieve our goals.*
- *Some of us are more experienced in this and thus they would normally lead the game while listening to ideas from all group members.*

The comparative factor analysis here clearly shows us that feedback is an issue with a range of possibilities in *MarketPlace Live*, while being mostly a non-issue in *Daedalus*.

## 2.3.5. Reliability and Validity of TAG

We calculated Cronbach's α for all three studies. Table x? shows the α-values and confidence intervals for all game factors. For the Fall 2017 study, Cronbach's α-values and 95% confidence intervals were: **TAG Factor 1: Game Strategy:** .92 [.90, .95], **TAG Factor 2: Team Planning:** .76 [.67, .84], and **TAG Factor 3: Feedback:** .80 [.73, .87]. For Spring 2018, Cronbach's α values were: **TAG Factor 1: Game Strategy:** .84 [.82, .87] This score is a little lower than the Fall 2017 score, with similar non-overlapping confidence intervals, **TAG Factor 2: Team Planning:** .73 [.69, .78] This score is a little lower than the Fall 2017 score, with a narrower confidence interval, and **TAG Factor 3: Feedback:** .84 [.81, .86]. For Summer 2018 (on *Daedalus*), we got the following scores: **TAG Factor 1: Game Strategy:** .86 [.82, .89]. This score is almost exactly identical to the Version 2 scores, and differs in the same way from Version 1. **TAG Factor 2: Team Planning** .73 [.66, .80]. This score falls between the Version 1 and Version 2 scores, and confidence intervals overlap in both directions. This factor seems consistent in strength and application, even in the face of the concept's poor fit within the *Daedalus* context. **TAG Factor 3: Feedback:** .76 [.69, .82]. This score is lower than both Versions 1 and 2 and does not fall within Version 2's confidence interval. This suggests that while the Feedback concept worked in *MarketPlace Live*, it was so poorly applicable to *Daedalus* that measurement accuracy has suffered as a result.

We next ran significance analysis to compare these reliability values and found that there is a significant effect between version 1 (Fall 2017) and the two applications of version 2 (Spring 2018 & Daedalus) of the Game Strategy factor, with p-values 0.001 (for Fall 2017-Spring 2018), and 0.16 (for Fall 2017-Daedalus). There was also a significant effect between Spring 2018 and Daedalus (p=0.011). This means that the version 2 survey has lowered this reliability significantly, although the effect size here is very small. The reason for this lowered reliability

**Acknowledgement:** this research was developed with funding from the Defense Advanced Research Projects Agency (DARPA).
**Disclaimer:** The views, opinions and/or findings expressed are those of the author and should not be interpreted as representing the official views or policies of the Department of Defense or the U.S. Government.

here may be the exclusion of the questions we decided to remove, or the rewording. However, since the effect size is so small, we deemed the reliability to be consistent and satisfiable high for such a survey.

**Table 2.5.** Comparing the three versions for reliability scores

|  | TAG1 | TAG2 | TAG3 |
|---|---|---|---|
| α (v1) - Fall 2017 | 0.92 [0.90-0.95] | 0.76 [0.67-0.84] | 0.80 [0.73-0.87] |
| α (v2) - Spring 2018 | 0.84 [0.82-0.87] | 0.73 [0.69-0.78] | 0.84 [0.81-0.86] |
| α (v2) - Summer 2018 | 0.86 [0.82-0.89] | 0.73 [0.66-0.80] | 0.76 [0.69-0.82] |

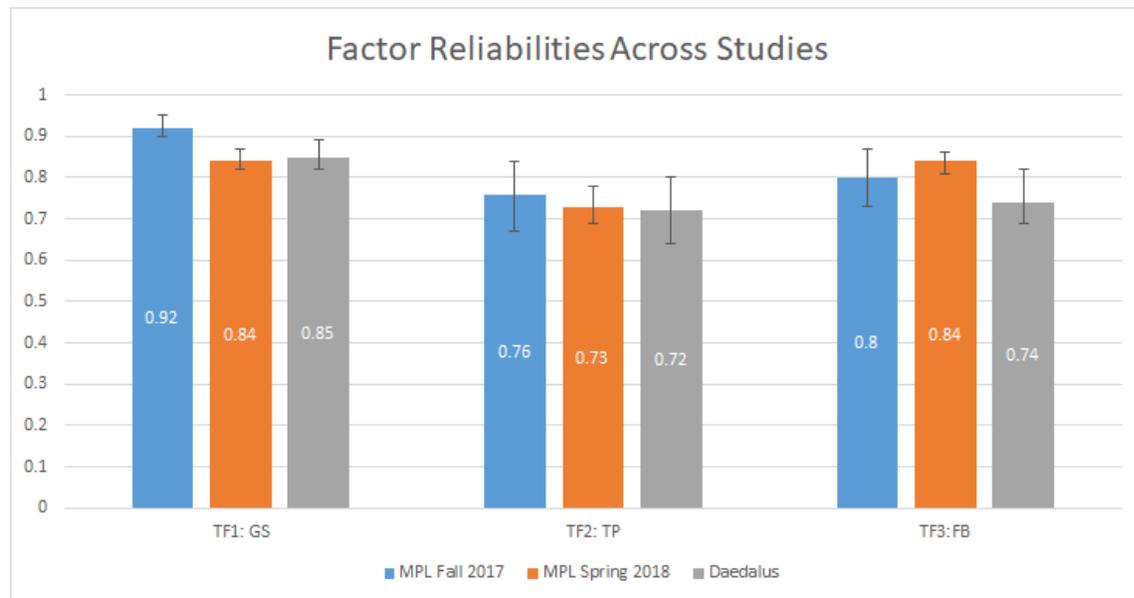

**Figure 2.18.** Factor reliability scores and confidence intervals

For our PCA, we followed the traditional procedures for validity. First, in order to see if a factor analysis was even feasible, we checked for Kaiser-Meyer-Olkin (KMO) Measure of Sampling Adequacy for high value (close to 1.0 and not less than 0.50) and conducted Bartlett's test of sphericity to test the hypothesis that the correlation matrices are an identity matrix. The values

**Acknowledgement:** this research was developed with funding from the Defense Advanced Research Projects Agency (DARPA).

**Disclaimer:** The views, opinions and/or findings expressed are those of the author and should not be interpreted as representing the official views or policies of the Department of Defense or the U.S. Government.

for all three studies satisfied these criteria. We then looked at the scree plots to determine the number of factors and the factor loadings (> 0.50) and communalities (> 0.40) to determine what items load on what factors. All scree plots suggest a possible 2-factor solution, but more likely a 3-factor or 4-factor one. However, the 4-factor solution led to only one or two items loading on a single factor, and we discarded it as a feasible solution. From the factor loadings we observed that, except for one item, the first factor ("Game Strategy") remained fairly consistent across the three survey implementations. However, the third factor ("Feedback") got subsumed into the second factor ("Team Planning") for the Spring 2018 study, and both the second and third factor got mixed up for the Daedalus study. This suggests that the second and third factor need to be further looked into; this echoes our own experience that it is difficult to conceptually tell these factors apart reliably. Our final step involved the consideration of several model fit indices (RMSR, RMSEA, TLI). While some values were deemed broadly acceptable, others were only 'just about' or close to being acceptable. This means that while there are some promising results, including consistency across implementations, some care is needed in interpreting results based on the factors. For this reason, we have decided to consider both the individual questionnaire items and the factors scores in our analyses.

Given the results above, we think that the TAG survey can be seen as a good novel contribution for measuring team adaptation processes within games. The analysis of open questions (as discussed above) can also allow us to characterize and further describe the strategies players have used, which can augment the analysis on the quantitative game log side by giving us better *a priori* ideas which analyses to try. In addition, we've applied the instrument to two different environments and got similar reliability: This shows that the instrument can reliably be used in different gaming environments, with questions context-independent or easily changeable (as shown in Appendix B). Consequently, we believe that the TAG is already in a position where other researchers could make good use of it, and that continued application (in different domains) will serve to increase its strength as an instrument to measure adaptation and teamwork processes in games.

## 2.4. Slack Data Coding Method

Our second new measurement approach to measure non-behavioral parts of the Rosen et al. model was qualitative coding of Slack chat data. We included qualitative data, because we wanted to gain more insight into the exact adaptation processes and behavior occurring in *MarketPlace Live* and *Daedalus*. Chat data have the advantage of providing us with rich, multi-faceted data that can contain insights and realizations we ourselves did not think of. However, time- and effort-intensive data preparation is often required to find interpretable, generalizable results, particularly if the data are intended to be used in mixed-method analysis. Our use of

**Acknowledgement:** this research was developed with funding from the Defense Advanced Research Projects Agency (DARPA).
**Disclaimer:** The views, opinions and/or findings expressed are those of the author and should not be interpreted as representing the official views or policies of the Department of Defense or the U.S. Government.

qualitative coding here was intended to explore the possibilities and develop methodological approaches that maximize the benefits of this kind of data while minimizing the drawbacks.

We were specifically interested in coding the chat data in a way that would support later quantitative analysis of behavioral data. The idea was this: Each Slack message that we coded had a UTC timestamp. Behavioral data drawn from both games also has these timestamps. By coding Slack messages for interesting but hard-to-observe data, we could correlate the messages to the behaviors that immediately follow them. This would allow us to see, for instance, if people taking on the responsibilities for a certain task then actually did the task (*role differentiation*), or if people that set a deadline for task completion actually managed to keep that deadline (*coordination*).

### 2.4.1 Applying the Approach to Daedalus

We started our coding process with the Slack chat data from *Daedalus* (section 3.2). We used the Rosen et al. (2011) model to code data, to support the aforementioned mixed-method analysis. However, because no extant qualitative code book based on this model could be found, we had to develop, iterate on, and evaluate one from scratch.

For purposes of length and readability, a complete overview of this process can be found in Appendix A. The coders' inter-rater reliability (IRR), calculated via the Fleiss-Kappa (k) measure, remained low throughout the coding processes, increasing from a score of .254 for the whole code book after the first iteration to a score of .381 for the whole code book on the fifth and final iteration. A short summary is presented here.

Our initial code book consisted of every Rosen et al. process descriptor, along with codes for user identification, *Daedalus* puzzles and cues, and pragmatic codes to track player organization and decision making. The three coders carried out a first iteration of coding with this code book on *Daedalus* Slack chat data, in order to ascertain their inter-rater reliability (IRR). A high enough IRR would indicate a (mostly) non-ambiguous coding of the data, which would support the follow-up quantitative analysis (which relies on the assumption that a 'true' objective coding of the chat data exists).

The first iteration of coding revealed low IRR (0.254), and several problems with applying the Rosen et al. codes to *Daedalus* Slack chat data. Four major change iterations were carried out: Every iteration consisted of the coders brainstorming about the issues they encountered, altering the code book to better reflect their new understanding, and trying another coding pass.

**Acknowledgement:** this research was developed with funding from the Defense Advanced Research Projects Agency (DARPA).
**Disclaimer:** The views, opinions and/or findings expressed are those of the author and should not be interpreted as representing the official views or policies of the Department of Defense or the U.S. Government.

Iteration 2 resulted in a small expansion of the code book. Iteration 3 resulted in a significant distillation: Several codes were either condensed into a single overarching code, or deleted entirely. Iterations 4 and 5 again resulted in small changes to or expansions to the code book. An IRR of .381 was achieved for the whole code book at the end of iteration 5; restricting the code book to only the eight most important codes raised this to .553, the highest value achieved in the work. This is addressed in significantly more detail in Appendix A.

While the IRR of the three coders went up after each iteration (to .381 for the whole code book / .553 in the restricted code book in the final iteration), it was still not high enough after five iterations to support the quantitative analysis: Coders could not convincingly argue that either of the three was able to code 'correctly'. Due to the time limits, we could not explore this further. We will be revisited in the future.

### 2.4.2. Applying the Approach to MPL

We used the Rosen et al. model code book on the Slack chat data from the Spring 2018 *MarketPlace Live* study (no Slack chat data was collected for the Fall 2017 study), with minor alterations to the code book to reflect the different context (potentially alongside the *MarketPlace Live* code book developed in section 5.2.1). However, the issues encountered during the *Daedalus* coding have put this idea on indefinite hold. An early attempt to code this chat data using the *MarketPlace Live* code book showed significant difficulty in applying a model created for open questions to Slack chat data, and as described in section 5.2.1, **we now no longer believe this specific type of coding (applying the open questions coding model to Slack data) to be very valuable.** In the interest of time, the decision was made to prioritize the *Daedalus* data and the Rosen et al. model code book. Consequently, no findings about coding the *MarketPlace Live* Slack chat data can be reported.

### 2.4.3. Reliability and Validity of the Approach

The initial *Daedalus* coding results have cast significant doubt on the assumption that we can meaningfully and unambiguously code Slack chat data to support quantitative analysis. Two main issues can be identified.

First, it is unclear if a 'ground truth' state of Rosen et al. model-coded adaptation behavior exists. In several recurring cases (see Appendix A), coders could not determine the 'correct' application of several similar codes, i.e. if the one code should be applied, the other code, or both. A lack of clear coding assumptions is partially at fault here, and coders making these assumptions explicit over time improved IRR scores. However, certain problems persisted to the degree that the existence of a 'true' coding must be called into question.

**Acknowledgement:** this research was developed with funding from the Defense Advanced Research Projects Agency (DARPA).
**Disclaimer:** The views, opinions and/or findings expressed are those of the author and should not be interpreted as representing the official views or policies of the Department of Defense or the U.S. Government.

Second, chat data was often highly ambiguous and hard to parse. Participants would speak in abbreviations and lingo, refer to each other by real names (instead of Slack identifiers) or nicknames (one team constantly referred to one person as 'mom'), and talk about puzzles and clues in ways that the coders could not easily follow. In theory, it could be possible to ask participants to restrict their speech to unambiguous identifiers, which would resolve this issue. In practice, this would likely limit their expression, and the argument can be made that at that point we are no longer coding 'real' chat data. This is assuming that participants would even keep these restrictions mind, which we know they do not: *Daedalus* players were asked to limit their communication exclusively to Slack, but several chats indicate or outright state that players met up in person or communicated in different ways.

Consequently, while the unpredictable and freeform nature of qualitative chat data can be very valuable to particular kinds of projects, our results seem to indicate that a code book and coding approach based on the Rosen et al.'s model cannot serve as a reliable basis for quantitative analysis of behavioral game data. In the future, we aim to look into the use of other coding techniques and theories.

## 3. Effect of Individual differences on Team Performance and Adaptation

Based on the scores above, we were able to verify the hypotheses we had articulated for this project. For convenience of the reader, the hypotheses investigated were:

- **H1:** Degree of **openness** will be highly predictive of **adaptability** and **Performance**. This is based on previous work by (Burke et al., 2006; Elaine D. Pulakos et al., 2002) who postulated that individuals who rate high on openness display high tolerance and curiosity when they confront unpredictable situations
- **H2:** Degree of **neuroticism** will be inversely predictive of **adaptability** and **Performance**. This is based on previous study by (Burke et al., 2006; Huang et al., 2014; Pulakos et al., 2002) who postulated that individuals with high neuroticism do not have the emotional stability to deal with unpredictable situations.
- **H3:** Degree of **cognitive flexibility** will be highly predictive of **adaptability** and **Performance**. This is based on previous study (Burke et al., 2006), which postulates that high cognitive flexibility will increase chances that individuals display adaptability in their behavior.

Since we also added the following hypotheses since we added new measures such as situational assessment and TAG Factors:

**Acknowledgement:** this research was developed with funding from the Defense Advanced Research Projects Agency (DARPA).

- **H4: TAG Factors** will correlate to **Situation Assesement Measures**, **Performance** and **Adaptability**.
- **H5: TAG Factors** will correlate to **Cognitive Flexibility**, **Neuroticism** and **Openness**.

## 3.1. Personality and Cognitive Flexibility and its effect on team work in MarketPlace Live

Our results are as follows:

For **H1**: there is **no effect** of **openness** on **performance** and **Adaptation**. **Openness** had a significant correlation with **conscientiousness** (r=0.27 and p=0.05).

For **H2**: there is an effect of **neuroticism** on **performance** and **Adaptation** (r=-0.325, p=0.02) and (r=-0.26, p=0.06).

For **H3**: there is **no effect** of **cognitive flexibility** on **performance** and **adaptation**.

For **H4**:

- **TAG Factor: Game Strategy** had a correlation with **Performance** (r=0.38, p = 0.026) **but not** **Adaptation**

- **TAG Factor: Team Planning** had a correlation with **Performance** (r=0.39, p = 0.026) **but not** **Adaptation**

- **Situation Awareness** had a correlation with **Performance**: Q4_SA (r=0.339, p = 0.015) and Q6_SA (r=0.438, p= 0.001).
- **Situation Awareness** had a correlation with **Adaptation**: Q3_SA (r=0.27, p=0.05), Q4_SA (r=0.26, p = 0.06) , Q5_SA (r=0.39, p=0.005 ), and Q6_SA (r=0.28, p= 0.048).

Other interesting effects found:

- **Performance** and **Adaptation** (r=0.38, p= 0.006)
- **Performance** and **Extroversion** (r=0.35, p=0.01).
- **Adaptation** and **Agreeableness** (r= 0.24, p=0.05).

**Acknowledgement:** this research was developed with funding from the Defense Advanced Research Projects Agency (DARPA).
**Disclaimer:** The views, opinions and/or findings expressed are those of the author and should not be interpreted as representing the official views or policies of the Department of Defense or the U.S. Government.

## 3.2. Personality and Cognitive Flexibility and its effect on team work in Daeldalus

### 3.2.1. Effect on Daedalus Performance

We developed a performance measure for Daedalus as described in Section 1.2.7. We used this measure to find correlation values and test significance with several surveys: TAG survey, IPIP-NEO (5 Personality Factors), and Cognitive Flexibility. We also tested correlation with our Situation Awareness score we developed and discussed in Section 2.2.

Our results are as follows:

For **H1**: there is **no effect** of **openness** on **performance**. **Openness** had a significant correlation with **cognitive flexibility** ($r=0.26$ and $p=0.0005$) and **TAG Feedback** ($r=0.185$, $p=0.051$).

For **H2**: there is **no effect** of **neuroticism** on **performance**.

For **H3**: there is **no effect** of **cognitive flexibility** on **performance**. **Cognitive flexibility** had a somewhat significant correlation with **TAG Game Strategy** ($r=0.1746$ and $p=0.0654$)

For **H4**: TAG Factors: **Game Strategy** had a significant correlation with **Situation Awareness** ($r=0.227$ and $p=0.0158$) and **Team Planning** had a significant correlation with **Situation Awareness** ($r=0.201$, $p=0.0337$). Team feedback had no effect which is not surprise since Daedalus did not have a feedback element.

For **H5**: **TAG Game Strategy** correlated with **Cognitive Flexibility** ($r=0.175$, $p = 0.065$). No other effects were found.

Other interesting effects found:

- **Performance** and **Situation Assessment** ($r=0.51$, $p=7.9e-09$).
- **Situation Awareness** and **TAG Factor: Game Strategy** ($r= 0.227$, $p= 0.0158$) and **TAG Factor: Team Planning** ($r=0.201$, $p=0.0337$).
- **Openness** and **TAG Factor: Team Feedback** ($r=0.185$, $p=0.05$)
- **Conscientiousness** and **TAG Factor: Game Strategy** ($r= 0.1937$, $p= 0.0407$)

**Acknowledgement:** this research was developed with funding from the Defense Advanced Research Projects Agency (DARPA).

| | PerformanceScore | Neuroticism | Extraversion | Openness | Agreeableness | Conscientiousness | Cog_Flexibility |
|---|---|---|---|---|---|---|---|
| PerformanceScore | 1.00 | 0.01 | -0.15 | 0.00 | -0.16 | -0.11 | -0.11 |
| Neuroticism | 0.01 | 1.00 | -0.27 | 0.04 | -0.22 | -0.35 | -0.42 |
| extraversion | -0.15 | -0.27 | 1.00 | 0.12 | 0.18 | 0.29 | 0.49 |
| Openness | 0.00 | 0.04 | 0.12 | 1.00 | 0.07 | 0.00 | 0.26 |
| Agreeableness | -0.16 | -0.22 | 0.18 | 0.07 | 1.00 | 0.08 | 0.16 |
| Conscientiousness | -0.11 | -0.35 | 0.29 | 0.00 | 0.08 | 1.00 | 0.53 |
| Cog_Flexibility | -0.11 | -0.42 | 0.49 | 0.26 | 0.16 | 0.53 | 1.00 |
| TAG_GameStrategy | 0.09 | -0.05 | 0.03 | 0.14 | -0.05 | 0.19 | 0.17 |
| TAG_TeamPlanning | 0.08 | 0.04 | 0.03 | 0.02 | -0.10 | 0.12 | 0.01 |
| TAG_Feedback | 0.08 | 0.00 | 0.14 | 0.19 | -0.12 | 0.17 | 0.12 |
| SA_Score | 0.51 | -0.12 | -0.18 | 0.04 | -0.01 | -0.03 | -0.08 |
| Age | -0.08 | -0.34 | -0.01 | 0.10 | 0.16 | 0.09 | 0.24 |

**Figure 3.1.** Correlation Analysis on Performance and Personality, Cognitive Flexibility as well as TAG Factors and Situation Awareness and Age (cont'd)

| | TAG_GameStrategy | TAG_TeamPlanning | TAG_Feedback | SA_Score | Age |
|---|---|---|---|---|---|
| PerformanceScore | 0.09 | 0.08 | 0.08 | 0.51 | -0.08 |
| Neuroticism | -0.05 | 0.04 | 0.00 | -0.12 | -0.34 |
| extraversion | 0.03 | 0.03 | 0.14 | -0.18 | -0.01 |
| Openness | 0.14 | 0.02 | 0.19 | 0.04 | 0.10 |
| Agreeableness | -0.05 | -0.10 | -0.12 | -0.01 | 0.16 |
| Conscientiousness | 0.19 | 0.12 | 0.17 | -0.03 | 0.09 |
| Cog_Flexibility | 0.17 | 0.01 | 0.12 | -0.08 | 0.24 |
| TAG_GameStrategy | 1.00 | 0.71 | 0.71 | 0.23 | 0.19 |
| TAG_TeamPlanning | 0.71 | 1.00 | 0.56 | 0.20 | 0.04 |
| TAG_Feedback | 0.71 | 0.56 | 1.00 | 0.08 | 0.15 |
| SA_Score | 0.23 | 0.20 | 0.08 | 1.00 | 0.07 |
| Age | 0.19 | 0.04 | 0.15 | 0.07 | 1.00 |

**Figure 3.2.** Correlation Analysis on Performance and Personality, Cognitive Flexibility as well as TAG Factors and Situation Awareness and Age (cont'd)

### 3.2.2. Effect on Daedalus Adaptation

We developed an adaptation score for Daedalus as described in Section 2.1. We used this measure to find correlation values and test significance with several surveys: TAG survey, IPIP-NEO (5 Personality Factors), and Cognitive Flexibility. We also tested correlation with our Situation Awareness score we developed and discussed in Section 2.2.

Our results are as follows:

For **H1**: there is **no effect** of **openness** on **adaptation**. **Openness** had a significant correlation with **cognitive flexibility** (r=0.26 and p=0.005).

**Acknowledgement:** this research was developed with funding from the Defense Advanced Research Projects Agency (DARPA).
**Disclaimer:** The views, opinions and/or findings expressed are those of the author and should not be interpreted as representing the official views or policies of the Department of Defense or the U.S. Government.

For **H2**: there is an effect of **neuroticism** on **Adaptation** (r=-0.192, p=0.04).

For **H3**: there is __no effect__ of **cognitive flexibility** on **Adaptation**.

For **H4**: **Situation Awareness** correlated with **Adaptation** (r=0.41 and p=4.1e-6).

| | Adaptation | neuroticism | extraversion | openness | agreeablene | conscientiousness | Cognitive Flexibility |
|---|---|---|---|---|---|---|---|
| Adaptation | 1 | | -0.069424 | 0.08170158 | 0.06424509 | -0.000640205 | 0.028227548 |
| neuroticism | -0.19155433 | 1 | -0.2742018 | 0.04447616 | -0.2137482 | -0.348692285 | -0.423112039 |
| extraversion | -0.069423959 | -0.2742018 | 1 | 0.11093554 | 0.18079064 | 0.294305996 | 0.49477265 |
| openness | 0.081701581 | 0.04447616 | 0.11093554 | 1 | 0.07246635 | -0.00174714 | 0.258910306 |
| agreeableness | 0.06424509 | -0.2137482 | 0.18079064 | 0.07246635 | 1 | 0.079946167 | 0.157229318 |
| conscientiousness | -0.000640205 | -0.3486923 | 0.294306 | -0.0017471 | 0.07994617 | 1 | 0.527748401 |
| Cognitive Flexibility | 0.028227548 | -0.423112 | 0.49477265 | 0.25891031 | 0.15722932 | 0.527748401 | 1 |

**Figure 3.3.** Adaptation, Personality and Cognitive Flexibility

| | Adaptation | TAG Factor 1: | TAG Factor 2: | TAG Factor 3: F | SA_score |
|---|---|---|---|---|---|
| Adaptation | 1 | 0.005092076 | -0.06074016 | 0.006069389 | 0.41355753 |
| TAG Factor 1: | 0.00509208 | 1 | 0.654464472 | 0.662001768 | 0.0071643 |
| TAG Factor 2: | -0.0607402 | 0.654464472 | 1 | 0.493753138 | 0.06003936 |
| TAG Factor 3: | 0.00606939 | 0.662001768 | 0.493753138 | 1 | -0.0891185 |
| SA_score | 0.41355753 | 0.007164305 | 0.06003936 | -0.089118536 | 1 |

**Figure 3.4.** Adaptation, TAG and Situation Awareness

# 5. Appendix A: Slack Data Coding

## Introduction

In order to examine individuals' and teams' adaptability during the ARG gameplay process, chat data was collected from the teams' Slack channels, anonymized, and organized for qualitative analysis. Three coders underwent an iterative coding process using *Rosen codes* derived from the four phase Rosen et al. adaptation model. The initial step was translating the individual factors of the model into codes, and assigning meaning to those codes based on their definitions within the literature. The coders also defined their own *pragmatic codes*, in order to track decision-making processes during play. These codes were applied in three situations: Users making proposals, users accepting proposals from other users, or users rejecting proposals from other users. The code book also included *user codes*, used to indicate which users were speaking or were being referenced in messages, *puzzle codes*, used to indicate that users were talking about certain puzzles, and *cue codes*, used to indicate that users were talking about or ascribing meaning to certain cues.

__Acknowledgement:__ this research was developed with funding from the Defense Advanced Research Projects Agency (DARPA).

__Disclaimer:__ The views, opinions and/or findings expressed are those of the author and should not be interpreted as representing the official views or policies of the Department of Defense or the U.S. Government.

Five coding iterations were carried out by the same three coders. As the Rosen et al. model was derived in a business/office context, the coders (throughout the iterations) discovered that pieces of the code book needed to be adjusted in order to better fit the context of the ARG in which the chat data was produced. Both the code definitions and the actual codes needed adjustments. The first code book consisted of 101 codes, of which 29 were identified as Rosen et al codes and Pragmatic codes. The three coders independently coded a contiguous sample of 219 chat excerpts, taken from the first team to complete the Daedalus game. For the Rosen et al codes, coders used the established definitions to determine when each code should be applied; each coder read the definitions and made their own judgment. For the Pragmatic codes, coders used Proposal whenever they identified a user making a proposal, and Agreement/Disagreement whenever they identified a user reacting to a proposal. The User/Puzzle/Cue codes were used whenever coders identified references to particular users, puzzles, or cues, again left to the coder's own judgment. The first and last versions of the code book are shown at the end of this document.

## Adaptation of the Rosen Model

The coders' inter-rater reliability (IRR), calculated via the Fleiss-Kappa (k) measure, remained low throughout the coding processes, increasing from a score of .254 for the whole code book after the first iteration to a score of .381 for the whole code book on the fifth and final iteration; eliminating a number of codes that the coders deemed unnecessary or problematic from the scoring process raised the score to .553 for , the highest achieved IRR score across the iterations. A number of causes for low scoring were diagnosed throughout the iterations and the coders met regularly to discuss their process and attempt to adapt the code book to address these issues.

### *Interpretative Definitions*

One main cause (identified after the first iteration) was that Rosen et al. codes had lengthy, complicated, and often vague definitions, leaving them vulnerable to subjective interpretation. These made it difficult for the three coders to agree upon the exact meaning of each code, resulting in inconsistent code application: Often one or two coders would believe that a code 'obviously' applied to an excerpt, but the remaining coder would understand the code to mean something else, and therefore not apply. In order to address this, the coders worked together to derive new definitions for each code, based on the original Rosen description, but shortened and adapted to better fit the context of the ARG. When these shorter definitions were not enough to alleviate the differences in interpretation, new codes were derived. An example is the Rosen et al. '*Coordination*' code. In the original definition, *Coordination* dealt with the orchestration of sequence and timing of interdependent actions. Within the context of the ARG, *Coordination*

**Acknowledgement:** this research was developed with funding from the Defense Advanced Research Projects Agency (DARPA).
**Disclaimer:** The views, opinions and/or findings expressed are those of the author and should not be interpreted as representing the official views or policies of the Department of Defense or the U.S. Government.

was applied to such excerpts as "*I will be a couple of minutes late*" and "*9 am Pacific?*" unanimously by all coders. However, excerpts such as "*Now we are in room 3 with safari animals, are you there also?*" and "*also, can we do a quick roll call/check in -- who is present and who has watched the video that comes from solving the safari puzzle?*" were only seen as *Coordination* by two of the three coders, who saw these types of messages as being inherently related to sequences of events, while the third coder did not believe that they were. The coders worked together to streamline *Coordination's* definition within the code book, and decided to create the code '*Status Update*' to differentiate 'check-in' messages from other messages dealing with timing/sequences and reduce such ambiguity.

### Inapplicable codes

A second cause, identified in a later iteration, was the fact that some sections of the Rosen code book did not apply heavily to the context of the Daedalus ARG; these codes were used sparingly or not at all. This was particularly apparent with regard to the Team Learning phase codes: Due to the game being a timed, competitive environment, in which players were encouraged to complete objectives as quickly as possible in order to win, there was little incentive to stop and reflect on one's actions. Similarly, no performance feedback was provided to the players that could give them an idea of whether they were doing well or poorly. Thus, it became apparent that the Team Learning phase as defined by Rosen et al. functionally did not occur within the context of the ARG, and as such the codes related to this phase did not appear in the chat excerpts. However, codes still felt that they had to keep all of these codes 'in mind', in case they did show up at any point. This caused mental overload and stress, resulting in coders forgetting or misapplying other codes. The coders chose to collapse the extant Team Learning codes into a single code called *Interpretation* in order to reduce this stress while still preserving the code concept for later use.

### Overlapping applications

After the second iteration, the coders analyzed their code applications and determined that several codes' applications were overlapping within the context of the ARG, despite having distinct definitions within the code book. This was determined to be due to the difference in context between the business/office environment that the model was initial derived in and the ARG environment in which the coders were working. This is illustrated most clearly by the codes *Goal Specification*, *Mission Analysis*, and *Systems Monitoring*. While these three codes were distinct in the business/work environment context that the Rosen et al. model was developed in, within the context of the ARG, it became apparent that their meanings were too close to reliably tease apart in the players' chat excerpts. *Goal Specification* was defined as "The

**Acknowledgement:** this research was developed with funding from the Defense Advanced Research Projects Agency (DARPA).
**Disclaimer:** The views, opinions and/or findings expressed are those of the author and should not be interpreted as representing the official views or policies of the Department of Defense or the U.S. Government.

explicit identification, articulation, and prioritization of the team's goals in the context of the mission". *Mission Analysis* was "The multilevel process of interpreting and evaluating the characteristics of a team's mission including tasks, subtasks, and available resources". And *Systems Monitoring* was "Tracking of team resources and environmental conditions as they relate to the mission". While these definitions are all distinct, coders disagreed on their application. For instance, the excerpt "*welp, looks like you guys gotta go on a Safari downtown*" was coded as *Goal Specification* by one coder, *Mission Analysis* by a second, and as none of the three by the third, and "*there's something in Node 2 to solve now :smile:*", which was coded as *Mission Analysis* by one coder, *Systems Monitoring* by a second coder, and none of the three by the third. This illustrates that, within the context of the ARG and specifically within the conversations of the players, these acts were not occurring in a way that could be consistently identified as distinct from one another. Thus, these three codes were collapsed into a single code called '*Game Analysis*'. While *Mission Analysis* and *Goal Specification* were originally a part of the Plan Formulation phase in the original model, and *Systems Monitoring* was a part of Plan Execution, we found that these two stages were not apparently distinct within the context of the ARG: Players rarely stopped to formulate a concrete plan before taking action, and more often went straight into interaction with the game environment with the hopes of learning and developing a plan as they went. Thus, *Game Analysis* was considered a Plan Formulation code as two of the codes that were collapsed into it were originally a part of this phase.

Another example of this phenomenon can be seen with *Back-Up Behavior* and *Mutual Monitoring*. *Back-Up Behavior* is defined as "back-up behavior can take many forms such as providing a teammate verbal feedback, assisting a teammate behaviorally in carrying out actions, or completing a task for a teammate"; *Mutual Monitoring* is defined as "a cognitive process in which team members observe their fellow team members' actions and behaviors in order to catch and correct performance discrepancies. Formally defined, it is the "team members' ability to keep track of fellow team members' work while carrying out their own… to ensure that everything is running as expected and… to ensure that they are following procedures correctly". These two codes were also difficult to differentiate within the ARG chat data, despite being distinct in the original model. Examples of difficulty in applying these codes consistently can be seen in excerpts such as "*Now we are in a room 3 with safari animals, are you there also?*", which was coded as *Mutual Monitoring* by only one coder, and *Back-Up Behavior* by a second, as well as "*can you give us some examples of the starting state, what you click, and the end state for your button colors?*" which was coded as *Back-Up Behavior* by only one coder, and *Mutual Monitoring* by the remaining two. Through analysis and discussion, it was determined that, while it was apparent to all three coders when players were taking action to assist or check in on each other, it was not consistently apparent whether the actions taking place were *Back-Up Behavior* or *Mutual Monitoring* as defined by Rosen et al. This is likely due to the ARG environment

 75

**Acknowledgement:** this research was developed with funding from the Defense Advanced Research Projects Agency (DARPA).

encouraging specific types of such collaboration that do not map directly to the work environment the model was developed in, such as the acts of walking people through puzzle solutions, or requesting status updates for the sake of coordination. Thus, these codes were collapsed into a single new code called '*Collaboration*'.

One final example of codes from the Rosen model that were adapted due to this reasoning is *Strategy Formulation: Contingency Planning* and *Strategy Formulation: Deliberate Planning*. While these codes are distinct, contingency planning was rarely-if-ever seen in the ARG Slack chat data (that was used for IRR calculations). This is likely due to the nature of the way in which players approached the game: Most tended to jump immediately into the puzzles and try everything until something works, which did not leave much room for planning in general, let alone contingency planning. Within the excerpts examined and coded in iteration 2, *Contingency Planning* was only coded for the excerpt "*Yeah, I just need to scroll down in the document. I'm pondering the sticky note now. Zach won't be able to join us until around 6 so I may hop on his slack account if we need to get to the same place together*", and only by a single coder, as the remaining two did not immediately recognize the excerpt as an example of contingency planning. Due to difficulty differentiating between the two codes, and the lack of contingency planning overall, it was decided to collapse the two codes into a single new code '*Preemptive Strategy Planning*', that would be considered distinct from *Reactive Strategy Planning*.

### Further refinement needed

After adjusting a number of codes from the Rosen model to better fit the ARG environment, the coders performed additional iterations of the coding process and discovered that a number of the adjusted codes required further refinement and organization in order to remove unnecessary ambiguity in meaning and application. One such example is with the code *Meaning Ascription*. In iteration 3, it became apparent to the coders that the new code *Game Analysis* was very close in concept to the Rosen code *Meaning Ascription*. It should be noted that at this point in time, the coders were not coding *Meaning Ascription* itself but rather its subcodes (*Correct Cue and Correct Meaning*, *Correct Cue but Incorrect Meaning*, and *Uncertain / Process*). However, they frequently saw that those subcodes tended to be applied to the same excerpts that *Game Analysis* was applied to: An example is the excerpt "*looks like some subliminal message pictures are stuck into that video*", which was coded as *Game Analysis* by two coders, and which also had *Correct Cue and Correct Meaning* applied to it. Similarly, it appeared that concern with regard to applying too many codes to any given excerpt was causing coders to choose between applying *Game Analysis* or the cue codes, but not both, leading to inconsistent applications. Thus, it was

**Acknowledgement:** this research was developed with funding from the Defense Advanced Research Projects Agency (DARPA).
**Disclaimer:** The views, opinions and/or findings expressed are those of the author and should not be interpreted as representing the official views or policies of the Department of Defense or the U.S. Government.

decided to remove the unused *Meaning Ascription* parent code and move the sub-codes under *Game Analysis*.

Another code that required further refinement was the newly defined code, *Collaboration*. This code was created by the coders after iteration 2 in order to encapsulate *Back-Up Behavior* and *Mutual Monitoring*, It became apparent after iterations 3 and 4, however that the code was too broad. This can be seen in such excerpts as "*I fed my baby dragon and he told me to look at the 9th line of a page of a book. Hatch some babies so we can figure out which page and book*" which was coded as *Collaboration* by two coders, but not the third, and which also encompasses a status update, data exchange, meaning ascription (as a part of *Game Analysis*), and reactive strategy formulation being communicated to the rest of the team. Another excerpt coded by all three as collaboration is "31st page of a book", which is merely the exchange of game data. Additionally, the code was still applied to excerpts that dealt with players helping one another, such as "*you have to get the light orb first from entering one of the green/red/black codes we posted earlier*", which was coded as *Collaboration* by all three coders. The coders were concerned that the broad reach of the code was causing meaning to be lost in the data, as it was not able to capture specific elements of player behavior through its varied applications. Thus, *Collaboration* was broken down into *Game Data Exchange*, which captured all instances of players sharing actual game data with each other, and *Assistance* (with subcodes *Provide* and *Request*), which captured all instances of players asking for or providing help.

Additionally, there continued to be inconsistencies between the coders with regard to when to apply the *Preemptive Strategy Planning* and *Reactive Strategy Planning* codes vs. *Coordination*. As an example, the excerpt "*I fed my baby dragon and he told me to look at the 9th line of a page of a book. Hatch some babies so we can figure out which page and book*" (Also discussed above) was marked as *Reactive Strategy Planning* by two coders, and *Coordination* by a third. Through discussion, it became apparent to the coders that that it was difficult to discern when players were performing actual strategy planning vs. simply coordinating with each other, both due to the nature of the game not encouraging the players to take time to plan before taking action and due to the short and simplistic sequences of actions needed to solve most of the puzzles. In addition, the coders' varied backgrounds made it difficult to settle on a unanimous definition of strategy planning within the context of the ARG. Additional discussion also drew attention to the fact that it was difficult to discern when players were performing reactive strategy planning vs. preemptive. All three codes were kept in the code book for the final iteration of coding, and the coders decided that *Preemptive Strategy Planning* would be any planning that occurred prior to receiving feedback on a puzzle, and *Reactive Strategy Planning* would be planning that occurred in response to such feedback. It was also decided that the planning codes would be applied to higher level and more complex sequences of

**Acknowledgement:** this research was developed with funding from the Defense Advanced Research Projects Agency (DARPA).

**Disclaimer:** The views, opinions and/or findings expressed are those of the author and should not be interpreted as representing the official views or policies of the Department of Defense or the U.S. Government.

actions to take, while the *Coordination* code would refer to more individual actions. Unfortunately, despite several discussions and refinement of definitions, the inconsistencies detailed above persisted in the last iteration.

## Slack as a medium

The final reason for the low IRR scores was concluded, by the coders, to lie within the nature of the chat based platform that the codes were being applied to. Slack (or similar chat based communication media) alone may not be enough to adequately apply the code book, as this style of analysis relies on the reporting of the users to be thorough enough as to be able to understand their actions. This can be seen very well in the difficulty the coders had applying the sub codes of *Meaning Ascription*. Codes to track whether the players had identified a correct cue and assigned the correct meaning were added to the code book, however after the first iteration it became apparent that it was not always obvious from what the players would say in their messages whether they were truly recognizing a cue or assigning meaning. Additionally, there were often scenarios where it was unclear what the players were referring to do to the lack of proper identifying nouns.

As an example, "*Did it ask you to heat it up?*", a message regarding the Tamagotchi puzzle, was coded by one coder as *Correct Cue and Correct Meaning*, but was not coded as any of the above subcodes by the remaining two coders. Similarly, a message chronologically prior to that one in the chat, "*but it looks like each egg wants a very specific temperature. Mine needed 85% exactly*", was coded as *Correct Cue and Correct Meaning* by two coders. Through discussion, it became apparent that it was also unclear whether cue recognition was something that persisted or should only be coded the first time it appeared in the chat for a particular cue, as the players would frequently keep discussing the cues and ascribing meanings for a number of messages. It was also discovered that it was inconsistent between coders what messages during that span were considered to be a part of the cue recognition/meaning ascription process. There was also disagreement between coders with regard to excerpts such as "*Well it looks like the next photo puzzle is at the Boston public garden*", in which the players simply stated information that had been given to them, should be considered cue recognition or meaning ascription, as the above excerpt was coded as *Correct Cue and Correct Meaning* by only one coder.

# Conclusion

The low IRR score indicates that the three coders were only able to agree on the meaning and applicability of a code less than 50% of the time. Above, we detail a number of reasons why this occurred, and present insight into the ways in which the Rosen et al. model would need to be adjusted in order to apply consistently to an ARG environment such as Daedalus, as well as an

**Acknowledgement:** this research was developed with funding from the Defense Advanced Research Projects Agency (DARPA).
**Disclaimer:** The views, opinions and/or findings expressed are those of the author and should not be interpreted as representing the official views or policies of the Department of Defense or the U.S. Government.

outline of a methodology through which such an adaptation could be carried out. The IRR score increase indicates that, given the time to perform numerous iterations, the Rosen et al. model could be converted into a comprehensive coding manual for measuring team adaptability in an ARG environment.

**Acknowledgement:** this research was developed with funding from the Defense Advanced Research Projects Agency (DARPA).
**Disclaimer:** The views, opinions and/or findings expressed are those of the author and should not be interpreted as representing the official views or policies of the Department of Defense or the U.S. Government.

# Code Book: Iteration 1

- MODEL: Rosen et al.
  - 1. Situation Assessment
    - Cue Recognition
    - Meaning Ascription
      - Correct Cue And Correct Meaning
      - Correct Cue But Incorrect Meaning
      - Incorrect Cue
    - Team Communication
  - 2. Plan Formulation
    - Goal Specification
    - Mission Analysis
    - Preemptive Conflict Management
    - Role Differentiation
    - Strategy Formulation: Contingency Planning
    - Strategy Formulation: Deliberate Planning
  - 3. Plan Execution
    - Affect Management
    - Back-Up Behavior
    - Coordination
    - Mutual Monitoring
    - Reactive Conflict Management
    - Strategy Formulation: Reactive Strategy Planning
    - Systems Monitoring
  - 4. Team Learning
    - Recap: Information Search and Structuring
    - Recap: Review Events
    - Reflection / Critique: Active Listening
    - Reflection / Critique: Framing / Convergent Interpretation
    - Reflection / Critique: Reframing / Divergent Interpretation
    - Reflection / Critique: Strength / Weakness Diagnosis
    - Summarize Lessons Learned: Accommodation / Integration
- PARTICIPANTS
  - Session: First Cohort, Team TC6DSCNBB
    - Admin User: UC74WV49K
    - Admin User: UC8AJQPBQ
    - Task Assigned To:

**Acknowledgement:** this research was developed with funding from the Defense Advanced Research Projects Agency (DARPA).

**Disclaimer:** The views, opinions and/or findings expressed are those of the author and should not be interpreted as representing the official views or policies of the Department of Defense or the U.S. Government.

- - TAT: UC694QBEC
  - - TAT: UC6S8M2EP
  - User: UC694QBEC
  - User: UC69RPFJL
  - User: UC6S8M2EP
  - User: UC83BCM55
  - User: UC8B6N50W
- PRAGMATIC CODES
  - Agreement
  - Disagreement
  - Proposal
- PUZZLE STRUCTURE
  - Stage 1
    - Puzzle 1.01: Few Dots Colour Code
      - Cue: The 9 coloured dots
    - Puzzle 1.02: Many Dots Colour Code
      - Cue: The 81 coloured dots
      - Cue: The text on the plaque
    - Puzzle 1.03: Cypher Wheel Instructions Colour Code
      - Cue: The coloured squares in the PDF
    - Puzzle 1.04: Glyph Cypher Puzzle
      - Cue: The completed cypher wheel
      - Cue: The symbol legend image
    - Puzzle 1.05: Safari Slideshow Colour Code
      - Cue: The coloured animals
    - Puzzle 1.06: Animal Pictures
      - Cue: The three statues
    - Puzzle 1.07: Safe Code
      - Cue: The safe code
  - Stage 2
    - Puzzle 2.01: Tamagotchi: Hatching
      - Cue: The feedback from the tamagotchi
      - Cue: The progress bar
    - Puzzle 2.02: Tamagotchi: Chessboard
      - Cue: The first set of buttons & effects
      - Cue: The second set of buttons & effects
    - Puzzle 2.03: Bookshelf Code

**Acknowledgement:** this research was developed with funding from the Defense Advanced Research Projects Agency (DARPA).
**Disclaimer:** The views, opinions and/or findings expressed are those of the author and should not be interpreted as representing the official views or policies of the Department of Defense or the U.S. Government.

- Cue: The bookshelf code(s)
- Stage 3
  - Puzzle 3.01: Cyber Garden
    - Cue: The statues needed for the puzzle
    - Cue: The timestamp(s) from the video
  - Puzzle 3.02: Locked Door Code
    - Cue: The door code
- Stage 4
  - Puzzle 4.01: Tangrams Mapping
    - Cue: The example-tangrams PDF
    - Cue: The instructions PDF
    - Cue: The layout PDF(s)
  - Puzzle 4.02: Symbols And Pictures Mapping
    - Cue: The Animal Channel video
    - Cue: The Photo Album pictures
    - Cue: The tangram-symbol mapping
  - Puzzle 4.03: TV Guide Mapping
    - Cue: The 3x3 word grid
    - Cue: The PDF with the symbols
    - Cue: The pigpen cypher
  - Puzzle 4.04: TV Channels Mapping
    - Cue: The symbol-word mapping
  - Puzzle 4.05: Telegraph Key Code
    - Cue: The 3x3 letter grid
    - Cue: The Morse code signs
  - Puzzle 4.06: Door Keypad Code
    - Cue: The nine-letter door keypad code
- Stage 5

**Acknowledgement:** this research was developed with funding from the Defense Advanced Research Projects Agency (DARPA).

**Disclaimer:** The views, opinions and/or findings expressed are those of the author and should not be interpreted as representing the official views or policies of the Department of Defense or the U.S. Government.

# Code Book: Iteration 5

- MODEL: Rosen et al.
    - 1. Situation Assessment
        - Meaning Ascription
    - 2. Plan Formulation
        - Game Analysis
            - Correct Cue And Correct Meaning
            - Correct Cue But Incorrect Meaning
            - Incorrect Cue
            - Uncertain / Process
        - Role Differentiation
        - Strategy Formulation: Preemptive Strategy Planning
    - 3. Plan Execution
        - Assistance
            - Provide
            - Request
        - Coordination
        - Game Data Exchange
        - Strategy Formulation: Reactive Strategy Planning
    - 4. Team Learning
        - Interpretation
        - Recap: Review Events
- PARTICIPANTS
    - Session: First Cohort, Team TC6DSCNBB
        - Admin User: UC74WV49K
        - Admin User: UC8AJQPBQ
        - User: UC694QBEC
        - User: UC69RPFJL
        - User: UC6S8M2EP
        - User: UC83BCM55
        - User: UC8B6N50W
        - User: UC8BVUW22
- PRAGMATIC CODES
    - Agreement
    - Disagreement
    - Proposal
- PUZZLE STRUCTURE

**Acknowledgement:** this research was developed with funding from the Defense Advanced Research Projects Agency (DARPA).

- Stage 1
    - Puzzle 1.01: Few Dots Colour Code
        - Cue: The 9 coloured dots
    - Puzzle 1.02: Many Dots Colour Code
        - Cue: The 81 coloured dots
        - Cue: The text on the plaque
    - Puzzle 1.03: Cypher Wheel Instructions Colour Code
        - Cue: The coloured squares in the PDF
    - Puzzle 1.04: Glyph Cypher Puzzle
        - Cue: The completed cypher wheel
        - Cue: The symbol legend image
    - Puzzle 1.05: Safari Slideshow Colour Code
        - Cue: The coloured animals
    - Puzzle 1.06: Animal Pictures
        - Cue: The three statues
    - Puzzle 1.07: Safe Code
        - Cue: The safe code
- Stage 2
    - Puzzle 2.01: Tamagotchi: Hatching
        - Cue: The feedback from the tamagotchi
        - Cue: The progress bar
    - Puzzle 2.02: Tamagotchi: Chessboard
        - Cue: The first set of buttons & effects
        - Cue: The second set of buttons & effects
    - Puzzle 2.03: Bookshelf Code
        - Cue: The bookshelf code(s)
- Stage 3
    - Puzzle 3.01: Cyber Garden
        - Cue: The statues needed for the puzzle
        - Cue: The timestamp(s) from the video
    - Puzzle 3.02: Locked Door Code
        - Cue: The door code
- Stage 4
    - Puzzle 4.01: Tangrams Mapping
        - Cue: The example-tangrams PDF
        - Cue: The instructions PDF
        - Cue: The layout PDF(s)

**Acknowledgement:** this research was developed with funding from the Defense Advanced Research Projects Agency (DARPA).
**Disclaimer:** The views, opinions and/or findings expressed are those of the author and should not be interpreted as representing the official views or policies of the Department of Defense or the U.S. Government.

- Puzzle 4.02: Symbols And Pictures Mapping
  - Cue: The Animal Channel video
  - Cue: The Photo Album pictures
  - Cue: The tangram-symbol mapping
- Puzzle 4.03: TV Guide Mapping
  - Cue: The 3x3 word grid
  - Cue: The PDF with the symbols
  - Cue: The pigpen cypher
- Puzzle 4.04: TV Channels Mapping
  - Cue: The symbol-word mapping
- Puzzle 4.05: Telegraph Key Code
  - Cue: The 3x3 letter grid
  - Cue: The Morse code signs
- Puzzle 4.06: Door Keypad Code
  - Cue: The nine-letter door keypad code
- Stage 5

# 6. Appendix B: TAG Versions

## TAG version 1

**Table B1. TAG closed questions (TAGC) and open questions (TAGO) used in Fall 2017 study. Scales for closed questions: Situation Assessment (SA), Plan Formulation (PF), Plan Execution (PE), Team Learning (TL). Open questions had no assigned scales.**

| Label | Item | Scale |
|-------|------|-------|
| TAGC.1 | My team explicitly articulated its objectives | PF |
| TAGC.2 | My team brainstormed strategies on how to play MarketPlace Live | PF |
| TAGC.3 | My team members passed information to one another in a timely and efficient manner | PE |
| TAGC.4 | My team defined potential problems it could encounter | SA |
| TAGC.5 | My team made plans for improving our performance in future quarters | TL |

**Acknowledgement:** this research was developed with funding from the Defense Advanced Research Projects Agency (DARPA).

**Disclaimer:** The views, opinions and/or findings expressed are those of the author and should not be interpreted as representing the official views or policies of the Department of Defense or the U.S. Government.

| TAGC.6 | My team came up with a list of tasks to accomplish its objectives | PF |
|---|---|---|
| TAGC.7 | My team identified the root causes of conflicts between its members | PE |
| TAGC.8 | I asked my teammates for feedback regarding my performance | TL |
| TAGC.9 | My team referred to past strategies when making decisions | TL |
| TAGC.10 | My team brainstormed new plans and strategies when it encountered problems | PF |
| TAGC.11 | My team maintained and updated a list of to-dos, needs, and objectives | PE |
| TAGC.12 | My team quickly noticed problems that could occur in the market | SA |
| TAGC.13 | I received feedback from my teammates for self-correction | PE |
| TAGC.14 | I was assigned specific tasks according to my abilities | PF |
| TAGC.15 | My team tried to resolve conflicts through negotiation | PE |
| TAGC.16 | My team articulated the best and worst strategies | PF |
| TAGC.17 | One or more of my teammates showed negative feelings such as stress, anger, depression, frustration or similar emotions that negatively impacted the team | PE |
| TAGC.18 | I analyzed team member contributions to identify their errors | TL |
| TAGC.19 | My team came up with a list of additional tasks in case a strategy did not work | PF |
| TAGC.20 | My team conversed without using sarcastic or non-constructive comments | TL |
| TAGC.21 | I provided feedback to teammates to facilitate self-correction | TL |
| TAGC.22 | My team identified why the project is successful or not successful thus far | SA |
| TAGC.23 | My team explicitly outlined everybody's responsibilities | PF |

**Acknowledgement:** this research was developed with funding from the Defense Advanced Research Projects Agency (DARPA).

**Disclaimer:** The views, opinions and/or findings expressed are those of the author and should not be interpreted as representing the official views or policies of the Department of Defense or the U.S. Government.

| TAGC.24 | I accepted suggestions from teammates about how I could improve my performance | TL |
| TAGC.25 | My team identified the key issues for improving its performance in Marketplace Live | TL |
| TAGC.26 | My team discussed errors and their causes | TL |
| TAGC.27 | My team attempted to provide opportunities for socializing | PE |
| TAGC.28 | When offered feedback, I was able to identify what would help me improve my performance | TL |
| TAGO.1 | Did your team have a specific strategy for how to play MarketPlace Live? If so, what was it? | |
| TAGO.2 | Did your team change its strategies based on results? If so, how? | |
| TAGO.3 | How did you collaborate with your teammates? | |
| TAGO.4 | How did your team members agree on how to pursue the team's goals in MarketPlace Live? | |
| TAGO.5 | What roles did your team assign to each of its members? Why? Did your team members fulfill this role and why yes or why not? | |

**TAG Factors deduced after running factor analysis.**

1. My team explicitly articulated its objectives [GS]
2. My team brainstormed strategies on how to play MarketPlace Live [GS]
3. My team members passed information to one another in a timely and efficient manner [GS]
4. My team defined potential problems it could encounter [GS]
5. My team made plans for improving our performance in future quarters [GS]
6. My team came up with a list of tasks to accomplish its objectives [TP]
7. My team identified the root causes of conflicts between its members <**unassigned**>
8. I asked my teammates for feedback regarding my performance [FB]
9. My team referred to past strategies when making decisions [kept separate]
10. My team brainstormed new plans and strategies when it encountered problems [GS]
11. My team maintained and updated a list of to-dos, needs, and objectives [TP]
12. My team quickly noticed problems that could occur in the market [GS]
13. I received feedback from my teammates for self-correction [FB]
14. I was assigned specific tasks according to my abilities [TP]

**Acknowledgement:** this research was developed with funding from the Defense Advanced Research Projects Agency (DARPA).

**Disclaimer:** The views, opinions and/or findings expressed are those of the author and should not be interpreted as representing the official views or policies of the Department of Defense or the U.S. Government.

15. My team tried to resolve conflicts through negotiation **[GS]**
16. My team articulated the best and worst strategies **[GS]**
17. One or more of my teammates showed negative feelings such as stress, anger, depression, frustration or similar emotions that negatively impacted the team **[GS]**
18. I analyzed team member contributions to identify their errors **[FB]**
19. My team came up with a list of additional tasks in case a strategy did not work **[TP]**
20. My team conversed without using sarcastic or non-constructive comments <**unassigned**>
21. I provided feedback to teammates to facilitate self-correction **[FB]**
22. My team identified why the project is successful or not successful thus far **[GS]**
23. My team explicitly outlined everybody's responsibilities **[TP]**
24. I accepted suggestions from teammates about how I could improve my performance **[FB]**
25. My team identified the key issues for improving its performance in Marketplace Live **[GS]**
26. My team discussed errors and their causes **[GS]**
27. My team attempted to provide opportunities for socializing <**unassigned**>
28. When offered feedback, I was able to identify what would help me improve my performance **[FB]**

# TAG Version 2

**Table B2. TAG closed questions (TAGC) and open questions (TAGO) used in Spring 2018. Rosen scales for closed questions: Situation Assessment (SA), Plan Formulation (PF), Plan Execution (PE), Team Learning (TL). TAG scales for closed questions: Game Strategy (GS), Team Planning (TP), Feedback (FB).**

| Label | Item | Rosen Scale | TAG Scale |
|-------|------|-------------|-----------|
| TAGC.1 | My team explicitly articulated its objectives | PF | GS |
| TAGC.2 | My team brainstormed strategies on how to play MarketPlace Live | PF | GS |
| TAGC.3 | My team members passed information to one another in an efficient manner | PE | GS |
| TAGC.4 | My team identified potential problems it could encounter while playing Marketplace Live | SA | GS |
| TAGC.5 | My team created strategies for improving our performance in future quarters | TL | GS |

**Acknowledgement:** this research was developed with funding from the Defense Advanced Research Projects Agency (DARPA).

**Disclaimer:** The views, opinions and/or findings expressed are those of the author and should not be interpreted as representing the official views or policies of the Department of Defense or the U.S. Government.

| TAGC.6 | My team came up with a list of tasks to accomplish its objectives | PF | TP |
|---|---|---|---|
| TAGC.7 | <deleted> | | |
| TAGC.8 | I asked my teammates for feedback regarding my performance | PE | FB |
| TAGC.9 | My team referred to past strategies when making decisions | TL | -- |
| TAGC.10 | My team brainstormed new plans and strategies when it encountered problems | PF | GS |
| TAGC.11 | My team periodically updated a list of to-dos, needs, and objectives | PE | TP |
| TAGC.12 | My team anticipated problems that could occur in the market | SA | GS |
| TAGC.13 | I received feedback from my teammates about my performance | PE | FB |
| TAGC.14 | I was assigned specific tasks according to my abilities | PF | TP |
| TAGC.15 | My team tried to resolve team conflicts by negotiating | PE | GS |
| TAGC.16 | My team articulated the best and worst strategies | PF | GS |
| TAGC.17 | One or more of my teammates showed negative feelings such as stress, anger, depression, frustration or similar emotions that negatively impacted the team | PE | GS |
| TAGC.18 | I analyzed team member contributions to identify their errors | TL | FB |
| TAGC.19 | My team came up with a list of additional ideas in case a strategy did not work | PF | TP |
| TAGC.20 | I provided feedback to teammates about their performance | TL | FB |
| TAGC.21 | <deleted> | | |
| TAGC.22 | My team was able to see why our strategies were or were not successful thus far | SA | GS |
| TAGC.23 | My team explicitly outlined everybody's responsibilities | PE | TP |

**Acknowledgement:** this research was developed with funding from the Defense Advanced Research Projects Agency (DARPA).

**Disclaimer:** The views, opinions and/or findings expressed are those of the author and should not be interpreted as representing the official views or policies of the Department of Defense or the U.S. Government.

| TAGC.24 | I accepted suggestions from teammates about how I could improve my performance | TL | FB |
|---|---|---|---|
| TAGC.25 | My team identified the key issues for improving its performance in Marketplace Live | TL | GS |
| TAGC.26 | My team discussed errors and their causes | TL | GS |
| TAGC.27 | <deleted> | | |
| TAGC.28 | When offered feedback, I was able to identify what would help me improve my performance | TL | FB |
| TAGO.1 | What was your team's strategy for playing Marketplace Live? | | |
| TAGO.2 | How did your team's strategies change during play? And why? | | |
| TAGO.3 | How would you describe your role in the team, and what were your tasks and responsibilities? | | |
| TAGO.4 | How did your team decide which actions to take? | | |
| TAGO.5 | What roles did your team assign to each of its members? Why? Did your team members fulfill this role and why yes or why not? | | |
| TAGO.6 | How did your team communicate? | | |

**Acknowledgement:** this research was developed with funding from the Defense Advanced Research Projects Agency (DARPA).
**Disclaimer:** The views, opinions and/or findings expressed are those of the author and should not be interpreted as representing the official views or policies of the Department of Defense or the U.S. Government.

**Acknowledgement:** this research was developed with funding from the Defense Advanced Research Projects Agency (DARPA).
**Disclaimer:** The views, opinions and/or findings expressed are those of the author and should not be interpreted as representing the official views or policies of the Department of Defense or the U.S. Government.